# ANALYSIS AND DESIGN OF NON-UNIFORM ARRAYS FOR DIRECTION FINDING

By

Elie Bou Daher

Dissertation
Submitted to the
College of Engineering
Villanova University
in partial fulfillment of the requirements
for the degree of

DOCTORATE OF PHILOSOPHY

In

Engineering

July, 2016

Villanova, Pennsylvania

# ANALYSIS AND DESIGN OF NON-UNIFORM ARRAYS FOR DIRECTION FINDING

By

Elie Bou Daher

Approved: ______________________________

*Dr. Fauzia Ahmad*
Associate Professor, ECE Department, Temple University
Primary Advisor

Approved: ______________________________

*Dr. Moeness G. Amin*
Professor, ECE Department, Villanova University
Co-Advisor

Approved: ______________________________

*Dr. Ahmad Hoorfar*
Professor, ECE Department, Villanova University

Approved: ______________________________

*Dr. Saleem A. Kassam*
Professor, ESE Department, University of Pennsylvania

Approved: ______________________________

*Dr. Gary Gabriele*
Dean of College of Engineering, Villanova University

# STATEMENT BY AUTHOR

This dissertation has been submitted in partial fulfillment of requirements for an advanced degree at the Villanova University.

Brief quotations from this dissertation are allowable without special permission, provided that accurate acknowledgment of source is made. Requests for permission for extended quotation from or reproduction of this manuscript in whole or in part may be granted by the head of the major department or the Associate Dean for Graduate Studies and Research of the College of Engineering when in his or her judgment the proposed use of the material is in the interests of scholarship. In all other instances, however, permission must be obtained from the author.

# ACKNOWLEDGMENTS

This dissertation is the result of my Ph.D. research at Villanova University. This work could not be completed without the help and support of many people. First, I would like to express my heartfelt gratitude to my advisor, Dr. Fauzia Ahmad, and co-advisor, Dr. Moeness Amin, for providing me with their support in terms of knowledge and patience during my research and for being forthcoming with novel ideas and suggestions. I would also like to thank the rest of my thesis committee, Dr. Ahmad Hoorfar and Dr. Saleem Kassam, for their invaluable support and guidance whenever needed. Thanks also go to Dr. P. P. Vaidyanathan, Dr. John Buck, Dr. Visa Koivunen, Dr. Yimin Zhang, and Dr. Aboulnasr Hassanien. Their invaluable comments and suggestions helped shape the content of this work. Special thanks go to Mrs. Janice Moughan for making administrative tasks seem trivial. I would also like to thank the Office of Naval Research for providing the financial support throughout my research. Finally, I would like to thank my parents, to whom I am ever indebted, my friends, my colleagues, and all those who contributed to the success of my work.

# TABLE OF CONTENTS

# LIST OF TABLES

# LIST OF FIGURES

# NOMENCLATURE

| | |
|---|---|
| BOMP | Block orthogonal matching pursuit |
| CMA-ES | Covariance matrix adaptation evolution strategy |
| CoSaMP | Compressive sampling matching pursuit |
| CPA | Co-prime array |
| CRB | Cramer-Rao bound |
| DCSC | Difference coarray of sum coarray |
| DFT | Discrete Fourier transform |
| DOA | Direction-of-arrival |
| DOF | Degree-of-freedom |
| EP | Evolutionary programming |
| ES | Evolution strategy |
| GA | Genetic Algorithm |
| Lasso | Least absolute shrinkage and selection operator |
| MCM | Mutual coupling matrix |
| MFO | Multi-frequency operation |
| MHA | Minimum hole array |
| MIMO | Multiple-input multiple-output |
| MRA | Minimum redundancy array |
| MUSIC | Multiple signal classification |
| NA | Nested array |
| OMP | Orthogonal matching pursuit |

PSO             Particle swarm optimization

RF              Radio frequency

RMIM            Receiving mutual-impedance method

RMSE            Root mean-square-error

SFO             Single-frequency operation

SINR            Signal-to-interference-plus-noise ratio

SNR             Signal-to-noise ratio

SVD             Singular value decomposition

ULA             Uniform linear array

# ABSTRACT

Antenna arrays are widely used in radar, sonar, and communication systems for direction finding and, more generally, target/source detection and localization. In these applications, the cost per sensor is typically significant because of the sensor itself and the associated electronics. As such, it is highly desirable to reduce the cost of the array by having fewer sensors. To this end, non-uniform arrays address the operational constraints on cost and hardware complexity by spanning large array apertures using far fewer elements than dictated by classical array theory. Numerous non-uniform array geometries have been introduced in the literature, including minimum redundancy arrays, minimum hole arrays, and more recently, nested and co-prime array configurations. Each configuration offers some advantages and disadvantages over the other configurations. The purpose of this research is to employ non-uniform arrays in different active and passive sensing applications for both narrowband and wideband operations, while providing a multitude of array processing methodologies that assist in dealing with the different encountered challenges.

The problem of direction-of-arrival (DOA) estimation using non-uniform arrays is considered in this research. The different challenges that are treated include the reduction of the available degrees-of-freedom (DOFs), the presence of coherent targets, and the mutual coupling effect in practical antenna arrays. Multi-frequency operation is exploited to increase the DOFs that are available for DOA estimation using both high-resolution subspace and sparse reconstruction techniques. In addition, a sparsity-based interpolation technique is presented to perform DOA estimation with increased DOFs. Moreover, a DOA estimation approach for a mixture of coherent and uncorrelated targets based on sparse reconstruction and active non-uniform arrays under narrowband signal platform is proposed. The aforementioned approaches deal with ideal operational scenarios. To address a more practical scenario, various methods for DOA estimation using non-uniform arrays in the presence of mutual coupling are presented. Extensive numerical simulations which validate the different proposed methods are also included.

# CHAPTER I

# INTRODUCTION AND MOTIVATION

Antenna arrays have found a wide range of signal processing applications due to their multitude of offerings including increased overall gain, diversity gain, interference cancellation, beam steering, and direction-of-arrival (DOA) estimation among others [1]. Generally speaking, the performance of an antenna array improves with an increasing number of elements in the array. This is due to the fact that the increased number of elements produces more degrees-of-freedom (DOFs). For instance, the overall gain of an antenna array with identical elements is the product of the element gain with the array factor. For a uniform linear array (ULA), the array factor is equal to the number of elements in the array. This means that a larger number of elements produces a larger gain. Another example is the number of resolvable sources in DOA estimation using a ULA. This number is tied to the number of elements in the array, and as such, for a larger number of elements, more sources can be estimated.

The offerings obtained by increasing the number of elements in the array usually come at the expense of increased cost, size, and hardware complexity. This has led to the introduction of non-uniform arrays, which have the ability to offer a comparable performance to that of a ULA, but with a reduced number of physical elements. Several non-uniform linear array configurations have been proposed in the literature [2-7]. These arrays have the ability to provide $O(N_A^2)$ DOFs using $N_A$ physical sensors. Minimum redundancy arrays (MRAs) constitute a class of non-uniform arrays that minimizes the redundancy in the difference coarray (the set of all spatial lags generated by the physical array [8]) for a given $N_A$, while ensuring that it has no missing elements or holes [2]. Minimum hole arrays (MHAs) form another class of non-uniform arrays that aims to reduce the redundancy in the coarray [3]. The corresponding difference coarray has no redundancy except at the zeroth lag; however, the set of coarray elements is not contiguous. Nested arrays constitute yet another class of non-uniform arrays which consists of two ULAs where one of the ULAs is spatially undersampled [4]. The corresponding coarray has no missing elements; however, its aperture is smaller than that of a MRA for given number of physical elements. Recently, a new structure of non-uniform linear arrays, known as co-prime arrays, has been proposed [5, 6]. A co-prime configuration consists of two spatially undersampled ULAs with co-prime number of elements and co-prime spatial sampling rates. The main objective of this research is to employ non-uniform arrays in different active and passive sensing applications for both single and multi-frequency operations, while devising effective array processing methodologies that deal with the different encountered challenges.

DOA estimation is a major application of array signal processing [9-11]. Non-uniform arrays provide the ability to resolve more sources than the number of physical sensors; however, this comes with some challenges. First, subspace-based DOA estimation techniques, such as MUSIC [12], require either complicated matrix completion processing [13-15] or are limited to the number of contiguous elements in the difference coarray. The latter results in a reduced number of the DOFs available for DOA estimation. This issue is relevant to co-prime arrays and MHAs as well since the difference coarrays of these configurations contain missing elements or holes. Second, the issue of mutual coupling comes into play when dealing with practical antenna arrays. Mutual coupling introduces a mismatch between the assumed model and the actual one, and results in DOA estimation errors. The majority of the proposed methods in the literature that deal with mutual coupling are limited to ULAs and cannot be applied directly to non-uniform arrays. Third, the presence of coherent sources or targets complicates the DOA problem due to the reduction of the rank of the noise-free covariance matrix. Spatial smoothing has been proposed to deal with this issue [16], but is only applicable to specific array structures and always results in reduction of the available DOFs.

In this research, several methods are proposed to address the aforementioned challenges in direction finding using non-uniform arrays. A high-resolution multi-frequency approach is utilized to fill in the missing elements of the difference coarray and permit DOA estimation with increased DOFs. In order to make use of all generated lags at the multiple frequencies, a sparsity-based multi-frequency approach is proposed. Alternatively, a sparsity-based interpolation technique is also proposed to generate the measurements at the missing elements in the difference coarray. Regarding the mutual coupling challenge in practical arrays, two methods which can be employed to perform DOA estimation using non-uniform arrays in the presence of mutual coupling are presented. Finally, a sparsity-based method which uses transmit/receive non-uniform arrays and allows direction finding of a mixture of coherent and uncorrelated targets is proposed.

## 1.1. Chapters Organization

The remainder of this dissertation is organized as follows. Chapter 2 provides a literature review about antenna arrays, non-uniform arrays, and the different DOA estimation methods that are used in this research. The major challenges that are encountered in direction finding using non-uniform arrays are also discussed at the end of this chapter. In Chapter 3, the two proposed multi-frequency DOA estimation techniques are presented. The sparsity-based interpolation technique is discussed in Chapter 4. Chapter 5 is divided into two main parts. In the first part, the effect of mutual coupling on the DOA estimation performance using different array configurations and different antenna types is investigated. In the second part, the two methods that allow DOA estimation using non-uniform arrays in the presence of mutual coupling are presented. In Chapter 6, the sparsity-based technique which allows direction finding of a mixture of coherent and uncorrelated targets is discussed. Supporting extensive numerical simulations, which validate the effectiveness of each of the proposed methods, are also included in the corresponding chapters. The major contributions of this research, specific to each topic, are summarized at the end of the relevant sections. Chapter 7 concludes the dissertation and provides future directions and recommendations.

## 1.2.    List of Publications

The following list of publications is the result of this research.

### 1.2.1.   Journal Articles

- E. BouDaher, F. Ahmad, and M. G. Amin, "Sparsity-based interpolation for DOA estimation using non-uniform arrays," *IEEE Trans. Signal Process.* (in preparation).

- E. BouDaher, F. Ahmad, M. G. Amin, and A. Hoorfar, "Mutual coupling effect and compensation in non-uniform arrays for direction-of-arrival estimation," *Digital Signal Processing Special Issue on Co-prime Sampling and Arrays* (in press).

- E. BouDaher, F. Ahmad, and M.G. Amin, "Sparsity-based direction finding of coherent and uncorrelated targets using active nonuniform arrays," *IEEE Signal Processing Letters*, vol. 22, no. 10, pp. 1628-1632, Oct. 2015.

- E. BouDaher, Y. Jia, F. Ahmad, and M. G. Amin, "Multi-frequency co-prime arrays for high-resolution direction-of-arrival estimation," *IEEE Trans. Signal Process.*, vol. 63, no. 14, pp. 3797-3808, Jul. 2015.

- E. BouDaher, F. Ahmad, and M. G. Amin, "Sparse reconstruction for direction-of-arrival estimation using multi-frequency co-prime arrays," *Eurasip Journal on Advances in Signal Processing Special Issue on Sparse Sensing in Radar and Sonar Signal Processing*, Nov. 2014.

### 1.2.2. Conference Papers

- E. BouDaher, F. Ahmad, M. G. Amin, and A. Hoorfar, "Effect of mutual coupling on direction-of-arrival estimation using sparse dipole arrays," in *IEEE AP-S/UNSC-URSI*, Fajardo, Puerto Rico, 2016.

- E. BouDaher, F. Ahmad, and M. G. Amin, "Sparsity-based extrapolation for direction-of-arrival estimation using co-prime arrays," in *SPIE Defense + Commercial Sensing Symposium*, Baltimore, MD, 2016.

- E. BouDaher, F. Ahmad, M. Amin, and A. Hoorfar, "DOA estimation with co-prime arrays in the presence of mutual coupling," in *23rd European Signal Processing Conference*, Nice, France, 2015, pp. 2830–2834.

- E. BouDaher, F. Ahmad, and M. G. Amin, "Sparsity-based DOA estimation of coherent and uncorrelated targets using transmit/receive co-prime arrays." in *SPIE Defense, Sensing, and Security Symposium, Compressive Sensing IV Conference*, Baltimore, MD, 2015.

- E. BouDaher, Y. Jia, F. Ahmad, and M. G. Amin, "Direction-of-arrival estimation using multi-frequency co-prime arrays," in *22nd European Signal Processing Conference*, Lisbon, Portugal, 2014.

# CHAPTER II

# BACKGROUND AND LITERATURE REVIEW

## 2.1.    Antenna Arrays

An antenna array is a collection of two or more spatially separated antennas arranged in a specific structure. The signals transmitted or received by these antennas are combined or processed in a certain way in order to achieve an improved performance over what would be obtained using the individual elements. An antenna array can increase the overall gain, provide diversity gain, cancel out interference from a set of directions, steer the beam in a particular direction, determine the DOA of incoming signals, and maximize the signal-to-interference-plus-noise ratio (SINR).

The array structure can assume many configurations including, but not limited to, linear, rectangular and circular as shown in Fig. 2.1. This research mainly focuses on one-dimensional linear arrays. The spacing between elements in a linear array is traditionally kept constant and set to half-wavelength at the operating frequency, resulting in uniform linear arrays.

The performance of an antenna array increases with the number of elements in the array. The main drawbacks or limitations are the increased cost and complexity. These drawbacks can be overcome by using non-uniform arrays. A non-uniform array can provide a performance similar to uniform array using a smaller number of physical elements. A non-uniform array has non-uniform spacing between consecutive elements in the array as shown in Fig. 2.2.

Before elaborating more on non-uniform arrays, the notion of coarrays should be discussed. The concept of coarrays will be used throughout the proposed research.

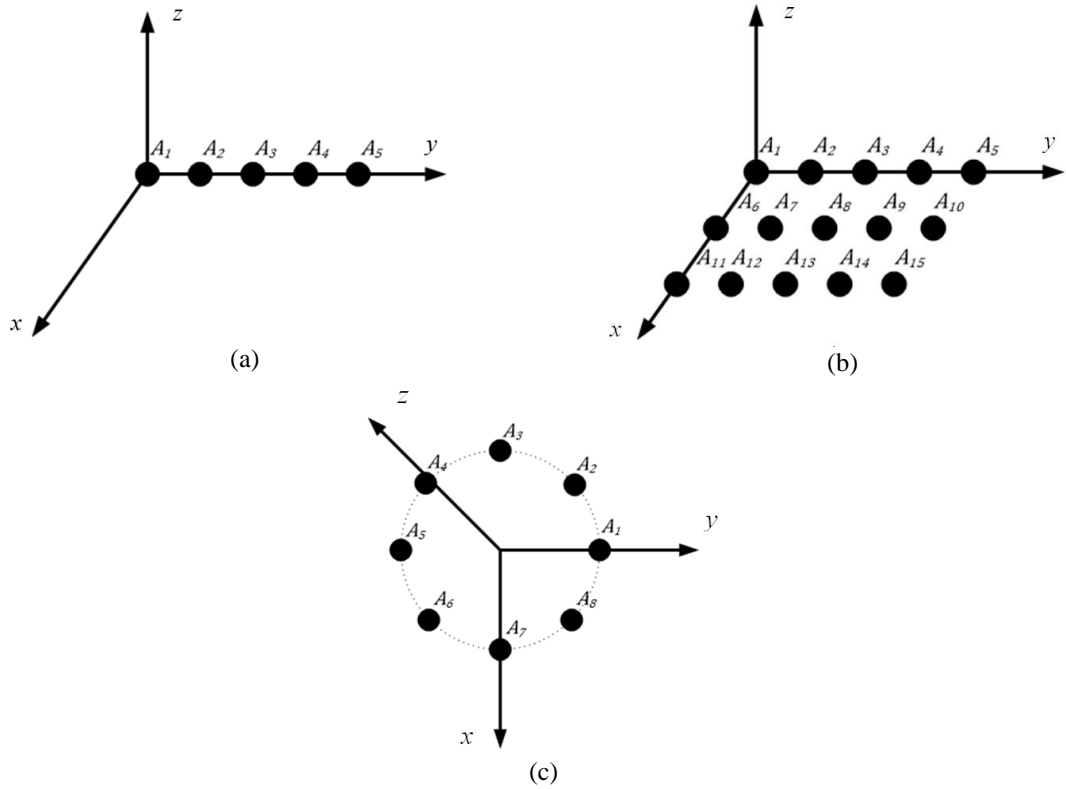

**Figure 2.1:** Antenna array structures (a) linear array (b) planar array (c) circular array.

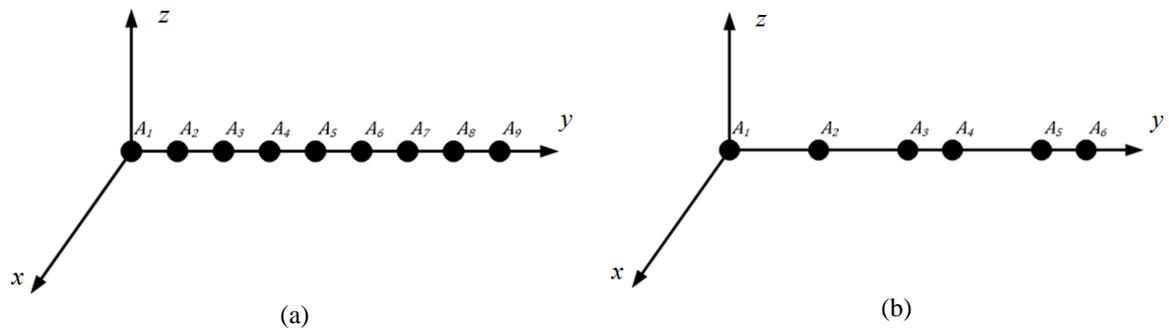

**Figure 2.2:** Array configurations (a) uniform linear array (b) non-uniform linear array.

### 2.1.1. Sum Coarray

The sum coarray is a virtual array that arises when dealing with active or transmit/receive arrays [8]. The elements of the sum coarray form the following set

$$S_{sum} = \{t_m + r_n\}, 1 \le m \le M_t \text{ and } 1 \le n \le N_r, \tag{2.1}$$

where $t_m$ is the position of the $m$th transmitter and $r_n$ is the position of the $n$th receiver. The transmit array consists of $M_t$ transmitters whereas the receive array consists of $N_r$ receivers. For illustration, a transmit/receive array consisting of three transmitters with positions $[0, 1, 2]d_0$ and three receivers with positions $[1, 4, 7]d_0$ is considered, where $d_0$ is the unit spacing. Following (2.1), the corresponding sum coarray consists of nine elements. Fig. 2.3 shows the transmit array, the receive array, and the corresponding sum coarray of this configuration.

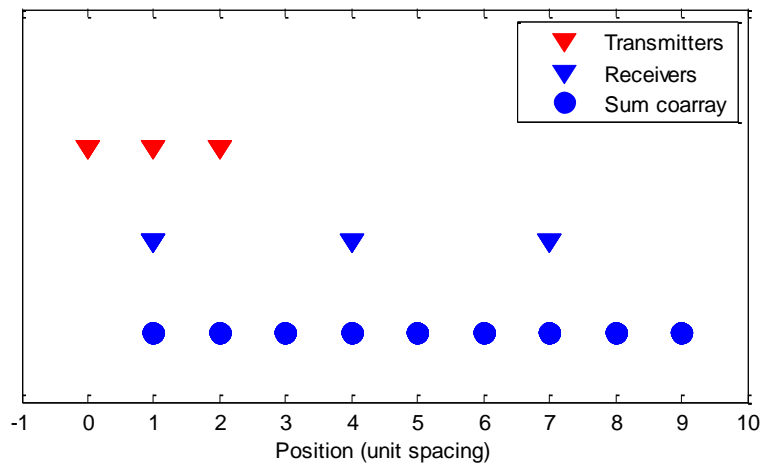

**Figure 2.3:** Transmit array, receive array, and corresponding sum coarray.

### 2.1.2. Difference Coarray

The difference coarray comes up when dealing with the second-order statistics of the received data. The difference coarray is defined as the set of all pairwise differences of array element locations, and, thus, it specifies the set of "lags" at which the spatial correlation function may be estimated [1, 8]. The difference coarray of a receive array with $N_A$ elements forms the following set

$$S_{diff} = \{x_m - x_n\}, 1 \leq m \leq N_A \ and \ 1 \leq n \leq N_A, \qquad (2.2)$$

where $x_m$ and $x_n$ denote the positions of the $m$th and $n$th elements, respectively. For illustration, a four-element array is considered. The sensor positions are given by $[0, 1, 3, 8]d_0$. The corresponding difference coarray consists of 13 elements. Fig. 2.4 shows the physical array and the corresponding difference coarray.

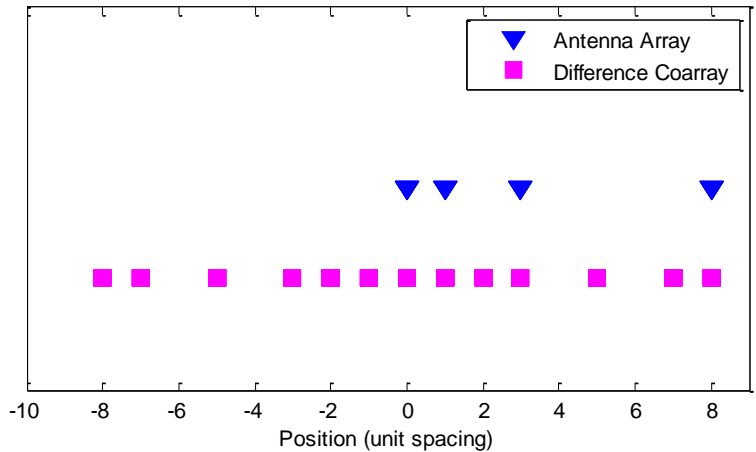

**Figure 2.4:** Receive array and corresponding difference coarray.

## 2.2. Non-Uniform Arrays

Non-uniform arrays are antenna arrays with non-uniform spacing between consecutive elements. The main motivation behind non-uniform arrays is that a similar performance to ULAs can be achieved with a smaller number of physical elements. Several non-uniform array configurations have been reported in the literature [2-6]. The most common ones are discussed below.

### 2.2.1. Minimum Redundancy Arrays

Minimum redundancy arrays are a class of sparse arrays which aims at maximizing the number of contiguous elements in the difference coarray for a given number of sensors [2]. The corresponding difference coarray contains the lowest possible redundancy without any missing lags or 'holes'. For a given number of physical sensors, MRAs require an exhaustive search through all possible combinations of the sensors to find the optimal design. A five-element MRA is considered. It has elements positioned at $[1, 2, 5, 8, 10]d_0$. The corresponding difference coarray contains 19 elements and is filled between $-9d_0$ and $9d_0$. The physical array and the corresponding difference coarray are shown in Fig. 2.5.

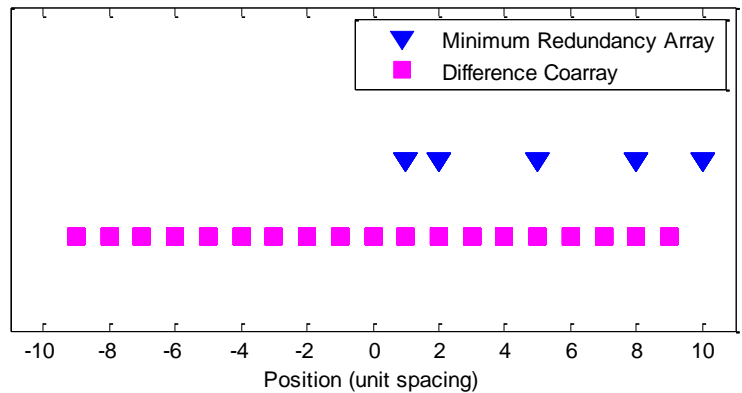

**Figure 2.5:** Five-element MRA with corresponding difference coarray.

### 2.2.2. Minimum Hole Arrays

Minimum hole arrays, also known as Golomb rulers, constitute another class of non-uniform arrays which aims to minimize the redundancy in the difference coarray [3]. The corresponding coarray contains the lowest possible number of holes with no redundancies except at the zeroth lag. Similar to MRAs, MHAs require an exhaustive search through the possible combinations to find optimal designs. As an illustration, a five-element MHA with elements at $[1, 2, 5, 10, 12]d_0$ is considered. Fig. 2.6 shows the array along with the corresponding coarray. The difference coarray contains 21 unique elements and extends from $-21d_0$ to $21d_0$. The coarray has two missing elements at $\pm 6d_0$. MHAs offer more DOFs than MRAs at the expense of the existence of holes in their difference coarray.

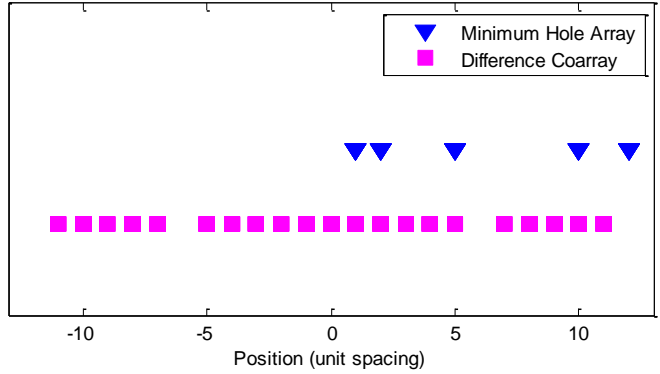

**Figure 2.6:** Five-element MHA with corresponding difference coarray.

### 2.2.3. Nested Arrays

Nested arrays are non-uniform arrays that can also increase the achievable DOFs [4]. In their basic configuration, nested arrays consist of a combination of two ULAs, where the inter-element spacing of the first array is equal to the unit spacing $d_0$ while the elements of the second ULA are separated by an integer multiple of $d_0$. The elements positions form the following set

$$S_{NA} = \{n_1 d_0 \cup (N_1 + 1) n_2 d_0\}, \tag{2.3}$$

where $1 \leq n_1 \leq N_1$ and $1 \leq n_2 \leq N_2$. $N_1$ and $N_2$ correspond to the number of elements in the first and the second ULAs, respectively. In order to maximize the DOFs for a given number of sensors $N_A$, the values of $N_1$ and $N_2$ are set as follows

$$\begin{cases} N_1 = N_2 = N_A/2, & N_A \; even \\ N_1 = \lfloor N_A/2 \rfloor, N_2 = \lceil N_A/2 \rceil, & N_A \; odd \end{cases} \tag{2.4}$$

where $\lfloor \cdot \rfloor$ is the floor operator and $\lceil \cdot \rceil$ is the ceil operator. The corresponding difference coarray is filled and contains no holes. The advantage of nested arrays over MRAs and MHAs is that the positions of the sensors and the achievable DOFs by nested arrays have closed-form expressions and do not require an exhaustive search. A five-element nested array along with its corresponding difference coarray is shown in Fig. 2.7. The five sensors are positioned at $[1, 2, 3, 6, 9]d_0$ and the corresponding difference coarray is filled between $-8d_0$ and $8d_0$.

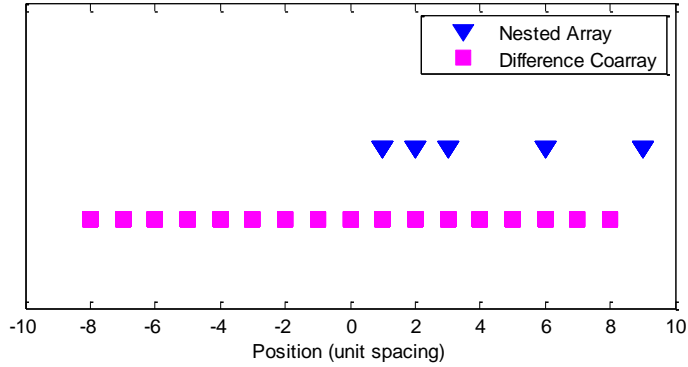

**Figure 2.7:** Five-element nested array with corresponding difference coarray.

### 2.2.4. Co-prime Arrays

A co-prime array comprises two spatially undersampled ULAs with co-prime spatial sampling rates [5, 6]. In the basic co-prime configuration, shown in Fig. 2.8 (a), the first array consists of $M$ elements with inter-element spacing $Nd_0$ and the second array has $N$ elements with spacing $Md_0$, with $M$ and $N$ being co-prime integers [5]. Without loss of generality, $M$ is assumed to be smaller than $N$. The elements positions form the following set

$$S_{CPA} = \{mNd_0 \cup nMd_0\}, \tag{2.5}$$

where $0 \le m \le M-1$ and $0 \le n \le N-1$. The element at the position 0 is shared by the two ULAs resulting in a total number of $(M+N-1)$ physical sensors. The corresponding difference coarray has $(MN+M+N-2)$ elements and is filled between $-(M+N-1)d_0$ and $(M + N − 1)d_0$. An extended co-prime array configuration was proposed in [6]. In this configuration, the number of elements in the first array is doubled as shown in Fig. 2.8 (b). The total number of physical elements is $(2M + N − 1)$, and the corresponding difference coarray has $(3MN + M − N)$ elements and is filled between $−(MN + M − 1)d_0$ and $(MN + M − 1)d_0$.

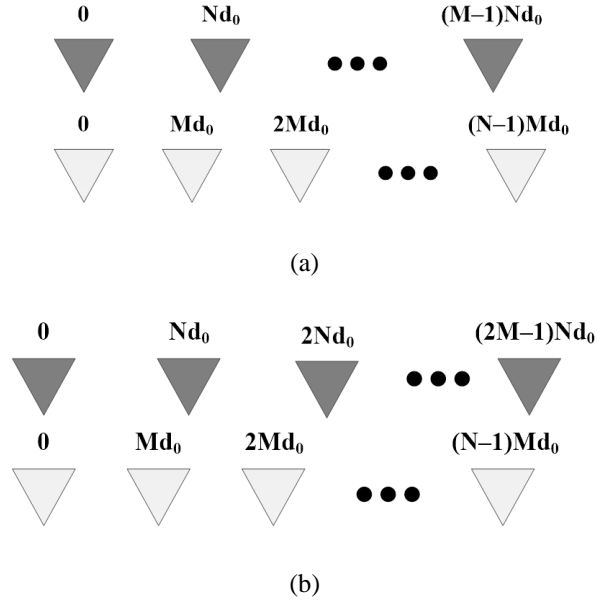

(a)

(b)

**Figure 2.8:** Co-prime array (a) basic configuration (b) extended configuration.

For illustration, a six-element co-prime with an extended configuration is considered. $M$ and $N$ are set to 2 and 3 respectively. The first uniform array consists of four elements with positions $[0, 3, 6, 9]d_0$ and the second array has three elements positioned at $[0, 2, 4]d_0$. The corresponding difference coarray comprises 17 elements and is filled between $−7d_0$ and $7d_0$. The physical co-prime array along with its difference coarray are shown in Fig. 2.9.

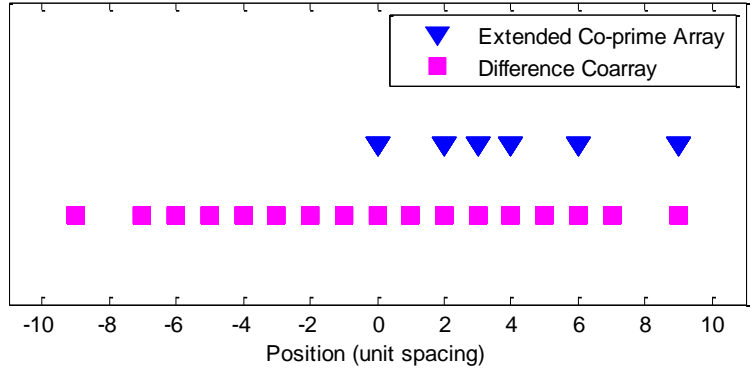

**Figure 2.9:** Six-element extended co-prime array with corresponding difference coarray.

Co-prime configurations offer some advantages over the other popular non-uniform arrays. As previously mentioned, for a given number of physical sensors, MRAs and MHAs require an exhaustive search through all possible combinations. This is not required for co-prime array configurations since the positions of the sensors constituting the co-prime configuration as well as the DOFs have closed-form expressions. Although the same is true with nested arrays, the elements of one of the arrays constituting the nested structure are closely separated, which may lead to problems due to mutual coupling between the sensors. Co-prime arrays reduce the mutual coupling between most adjacent sensors by spacing them farther apart [5]. Because of all of the aforementioned characteristics, co-prime arrays are finding broad applications in the areas of communications, radar, and sonar [17-23].

## 2.3. DOA Estimation Techniques

DOA estimation has received considerable research interest due to its applications in radar, sonar, and wireless communications [9-11]. Traditional high-resolution DOA techniques, such as MUSIC [12] and ESPRIT [24], can only estimate up to $(N_A - 1)$ sources when applied to an $N_A$- element ULA. Non-uniform arrays provide the ability to estimate the DOAs of more sources than the number of physical sensors.

In what follows, the signal model for DOA estimation is discussed and the major DOA estimation techniques, used in this research, are reviewed.

### 2.3.1. Signal Model

As a signal impinges on a linear array from a direction $\theta$, where $\theta$ is measured relative to broadside, the signal arrival at each antenna element encounters a propagation delay as it travels across the array. For a uniform linear array, shown in Fig. 2.10, the additional distance traveled by the wavefront between two consecutive elements is equal to $d_0 \sin \theta$, where $d_0$ is the unit spacing which is usually set to half-wavelength at the operating frequency. The propagation delay is a key element in array signal processing.

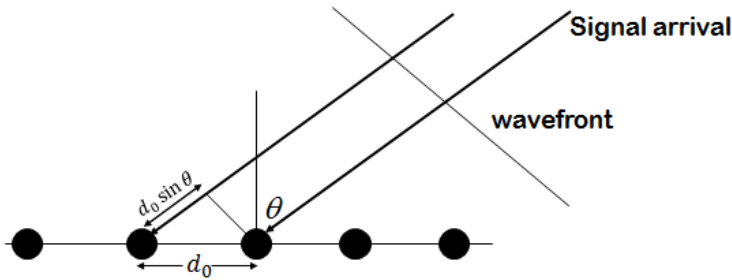

**Figure 2.10:** Propagation delay across the array.

A general $N_A -$ element linear array is considered. The elements positions are assumed to be integer multiples of the unit spacing, i.e., $x_i = n_i d_0, i = 1, \dots, N_A$, where $x_i$ is the position of the $i$th array element and $n_i$ is an integer. Assuming that $D$ narrowband sources with directions $\{\theta_1, \theta_2, \ldots, \theta_D\}$ and powers $\{\sigma_1^2, \sigma_2^2, \ldots, \sigma_D^2\}$ impinge on the array, the received data vector at snapshot $t$ can be expressed as

$$\mathbf{x}(t) = \mathbf{A}\mathbf{s}(t) + \mathbf{n}(t), \tag{2.6}$$

where $\mathbf{s}(t) = [s_1(t), s_2(t), \ldots, s_D(t)]^T$ is the $D \times 1$ source signal vector, $\mathbf{n}(t) = \left[n_1(t), n_2(t), \ldots, n_{N_A}(t)\right]^T$ is the $N_A \times 1$ noise vector, and $\mathbf{A}$ is the $N_A \times D$ array manifold matrix whose $(i, d)$th element is given by

$$[\mathbf{A}]_{i,d} = \exp(jk_0 x_i \sin \theta_d). \tag{2.7}$$

Here, $k_0$ is the wavenumber at the operating frequency and $\theta_d$ is the DOA of the $d$th source. The $d$th column in $\mathbf{A}$ corresponds to the steering vector of the array relative to direction $\theta_d$. Under the assumptions of uncorrelated sources and spatially and temporally white noise, the covariance matrix can be expressed as

$$\mathbf{R}_{xx} = E\{\mathbf{x}(t)\mathbf{x}(t)^H\} = \mathbf{A}\mathbf{R}_{ss}\mathbf{A}^H + \sigma_n^2\mathbf{I}, \tag{2.8}$$

where $E\{\cdot\}$ is the expectation operator, $\mathbf{R}_{ss} = diag\{\sigma_1^2, \sigma_2^2, \ldots, \sigma_D^2\}$ is the source covariance matrix, $\sigma_n^2$ is the noise variance, and $\mathbf{I}$ is an $N_A \times N_A$ identity matrix. If some of the sources are correlated or coherent, the corresponding off-diagonal terms in $\mathbf{R}_{ss}$ become nonzero. In practice, the covariance matrix is obtained as a sample average

$$\widehat{\mathbf{R}}_{xx} = \frac{1}{T}\sum_{t=1}^{T} \mathbf{x}(t)\mathbf{x}^H(t), \tag{2.9}$$

where $T$ is the total number of available snapshots. The objective of DOA estimation techniques is to estimate the unknown source directions $[\theta_1, \ldots, \theta_D]$ given the observations $\mathbf{x}(t), 1 \leq t \leq T$. The major DOA estimation techniques that are employed in this research are reviewed below.

### 2.3.2. Delay and Sum Processor (Conventional Beamforming Method)

In this method, the average power at the output of the delay and sum processor is computed for different steering vectors. For a direction $\theta$, the output average power can be computed using

$$P(\theta) = E\{|\mathbf{a}^H(\theta)\mathbf{x}(t)|^2\} = \mathbf{a}^H(\theta)\mathbf{R}_{xx}\mathbf{a}(\theta), \tag{2.10}$$

where $\mathbf{a}(\theta)$ is the steering vector of the array relative to $\theta$. The values of $\theta$ that produce peaks in $P(\theta)$ are taken as the DOA estimates.

For illustration, this method is applied to a 10-element ULA. Three sources with directions $[-20°, 10°, 50°]$ are considered. The signal-to-noise ratio (SNR) for all sources is set to 0 dB and the total number of snapshots is set to 500. Fig. 2.11 shows the estimated spectrum $P(\theta)$. The directions of the actual sources are shown with vertical dotted lines. It is clear that the estimated spectrum has three peaks at the actual source directions. This confirms that this method is successful in estimating the DOAs.

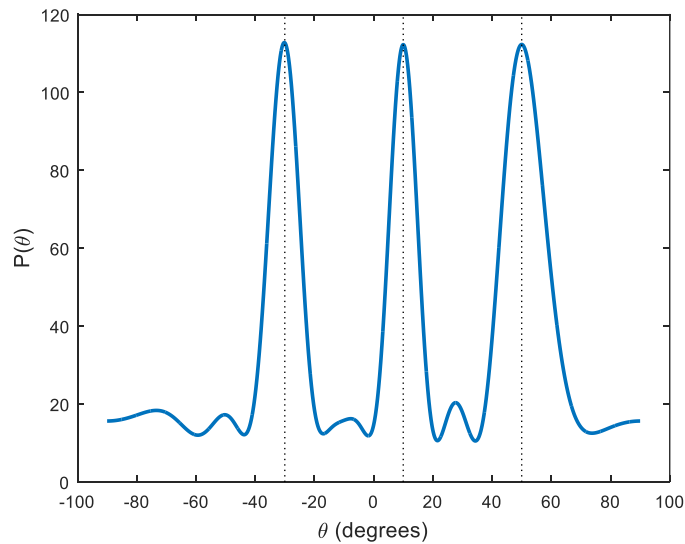

**Figure 2.11:** Delay and sum processor spectrum.

### 2.3.3. Multiple Signal Classification (MUSIC)

MUSIC is a high-resolution subspace-based DOA estimation technique [12]. MUSIC decomposes the covariance matrix into the signal subspace and the noise subspace. The first step in MUSIC is to perform eigenvalue decomposition of the covariance matrix

$$\mathbf{R}_{xx} = \mathbf{E}\mathbf{\Sigma}\mathbf{E}^H, \tag{2.11}$$

where $\mathbf{\Sigma} = diag\{\lambda_1^2, \dots, \lambda_D^2, \sigma_n^2, \dots, \sigma_n^2\}$ contains the eigenvalues and the columns of $\mathbf{E} = [\mathbf{E}_s \ \mathbf{E}_n]$ are the corresponding eigenvectors. The $D$ highest eigenvalues are associated with the powers of the actual sources and the corresponding eigenvectors $\mathbf{E}_s$ span the signal subspace. The remaining $(N_A - D)$ eigenvalues are associated with the noise and the corresponding eigenvectors $\mathbf{E}_n$ span the noise subspace. The MUSIC spectrum is computed as follows

$$P_{MUSIC}(\theta) = \frac{1}{\mathbf{a}^H(\theta)\mathbf{E}_n\mathbf{E}_n^H\mathbf{a}(\theta)}. \tag{2.12}$$

Since the signal and noise subspaces are orthogonal, the denominator in (2.12) will be equal to zero if $\theta$ is one of the source DOAs and the MUSIC spectrum will assume a peak. The DOA estimates are then obtained by localizing the $D$ highest peaks. It is clear that this method can only be applied to estimate up to $(N_A - 1)$ sources, and, therefore, cannot be directly applied to non-uniform arrays to estimate more sources than the number of sensors.

For illustration, MUSIC is applied to the same example that was discussed in the previous section. The obtained MUSIC spectrum is shown in Fig. 2.12. It is evident that the three sources are correctly estimated.

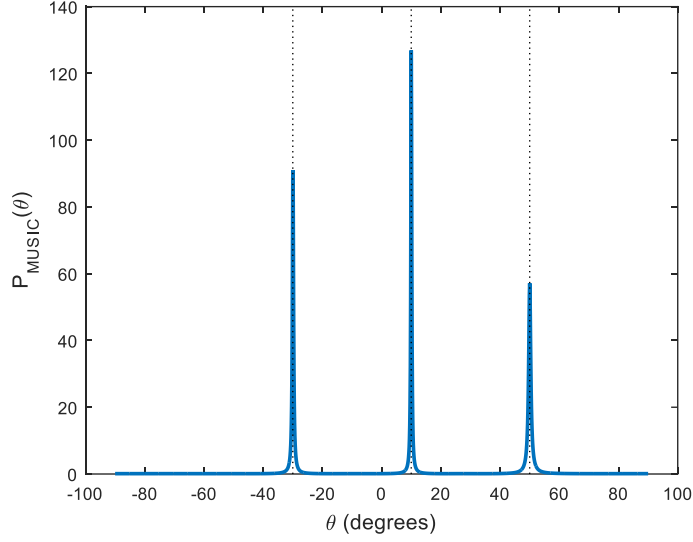

**Figure 2.12:** MUSIC spectrum.

### 2.3.4. Root-MUSIC

Root-MUSIC is a variation of the MUSIC algorithm that estimates the source DOAs by finding the roots of a polynomial [25]. In its original form, root-MUSIC was proposed for direction finding using uniform linear arrays. Starting with the MUSIC spectrum in (2.12), the denominator can be written as follows

$$P^{-1}(\theta) = \sum_{m=1}^{N_A} \sum_{n=1}^{N_A} \exp(-jk_0 m d_0 \sin\theta) \, [\mathbf{C}_{root}]_{m,n} \exp(jk_0 n d_0 \sin\theta), \qquad (2.13)$$

where $[\mathbf{C}_{root}]_{m,n}$ is the $(m,n)$th element of $\mathbf{C}_{root} = \mathbf{E}_n \mathbf{E}_n^H$. By letting $\ell = m - n$, (2.13) can be rewritten as

$$P^{-1}(\theta) = \sum_{\ell=-(N_A-1)}^{N_A-1} c_\ell \exp(-jk_0 \ell d_0 \sin\theta), \qquad (2.14)$$

where $c_\ell = \sum_{\ell=m-n} [\mathbf{C}_{root}]_{m,n}$ is the sum of entries of $\mathbf{C}_{root}$ along the $\ell - $th diagonal. A new polynomial $D(z)$ can be defined as

$$D(z) = \sum_{\ell=-(N_A-1)}^{N_A-1} c_\ell z^{-\ell}. \qquad (2.15)$$

Evaluating $P^{-1}(\theta)$ is equivalent to evaluating $D(z)$ on the unit circle. As a result, the peaks in the MUSIC spectrum can be attributed to the roots of $D(z)$ lying on the unit circle. In practice, the $D$ roots of $D(z)$ that are within the unit circle and that have the largest magnitudes are used for DOA estimation. The $d$th root $z_d$ can be associated with a source with direction

$$\theta_d = \sin^{-1}\left[\frac{1}{k_0 d_0} \arg(z_d)\right]. \qquad (2.16)$$

The same example, discussed in the previous two sections, is repeated with root-MUSIC. The three obtained roots which are closest to the unit circle are: $z_1 = 0.99 \exp(j0.5466)$, $z_2 = 0.9887 \exp(-j1.5681)$, and $z_3 = 0.985 \exp(j2.4069)$. The DOAs associated with these roots are $\theta_1 = \sin^{-1}\left[\frac{1}{\pi} 0.5466\right] = 10.01°$, $\theta_2 = \sin^{-1}\left[-\frac{1}{\pi} 1.5681\right] = -29.94°$, and $\theta_3 = \sin^{-1}\left[\frac{1}{\pi} 2.4069\right] = 50.01°$. Clearly, root-MUSIC is successful in estimating the source DOAs.

A modification of the polynomial $D(z)$ was introduced in [26] to make root-MUSIC applicable to non-uniform arrays.

### 2.3.5. $\ell_1 -$SVD

Starting with the signal model in (2.6), the receive data vectors at the $T$ snapshots can be combined in an $N_A \times T$ matrix as

$$\mathbf{X} = \mathbf{AS} + \mathbf{N}. \qquad (2.17)$$

where $\mathbf{X} = [\mathbf{x}(1), \mathbf{x}(2), \dots, \mathbf{x}(T)]$, the $D \times T$ source signal matrix is given by $\mathbf{S} = [\mathbf{s}(1), \mathbf{s}(2), \dots, \mathbf{s}(T)]$, and the $N_A \times T$ noise matrix is given by $\mathbf{N} = [\mathbf{n}(1), \mathbf{n}(2), \dots, \mathbf{n}(T)]$. Given the model in (2.17) and using the assumption that the sources are sparse in the spatial domain, sparse signal reconstruction can be employed to perform DOA estimation. The angular region of interest is discretized into a finite set of $K$ ($K \gg D$) grid points, $\{\theta_1^g, \theta_2^g, \dots, \theta_K^g\}$, with $\theta_1^g$ and $\theta_K^g$ being the limits of the search space. The sources are assumed to be located on the defined grid; however, several methods can be used to modify the model in order to deal with off-grid targets [20, 27, 28]. Then, (2.17) can be rewritten as

$$\mathbf{X} = \mathbf{A}^g \mathbf{S}^g + \mathbf{N}. \tag{2.18}$$

where the columns of the $N_A \times K$ matrix $\mathbf{A}^g$ are the steering vectors of the array corresponding to the defined angles in the grid, and the $K \times T$ matrix $\mathbf{S}^g$ holds the signals from potential sources on the defined grid. Although the signal of a particular source can change from one snapshot to another, it will occupy the same grid angle. As a result, the columns of $\mathbf{S}^g$ share a common support across the $T$ snapshots. That is, if a certain element in $\mathbf{S}^g$ has a nonzero value, the majority of the remaining elements in the same row should also be nonzero. The nonzero rows in $\mathbf{S}^g$ correspond to the signals from the actual sources, and by finding the nonzero rows, the directions of the sources can be determined. Fig. 2.13 shows an illustrative example where the number of grid points is set to six, the total number of snapshots is ten, and the actual number of sources is equal to two. In this example, the second and sixth rows have nonzero values and correspond directions of the actual sources. The common structure property suggests the application of a group sparse reconstruction to perform DOA estimation.

**Figure 2.13:** Sparse signal example, $T = 10$ snapshots, $K = 6$ potential directions, $D = 2$.

Group sparse reconstruction can be achieved by minimizing a mixed $\ell_{1,2} -$ norm where the elements of a particular row are combined using the $\ell_2 -$ norm. The $\ell_1 -$ norm is then applied to the obtained vector in order to encourage sparsity in the spatial domain. The drawback of this approach is that it depends on the number of snapshots, and for a large number of snapshots, the complexity of the model increases.

In order to solve this problem, singular value decomposition (SVD) can first be applied to the data matrix $\mathbf{X}$ in order to reduce its dimensionality [27]. By applying SVD, $\mathbf{X}$ can be expressed as

$$\mathbf{X} = \mathbf{U}_s \mathbf{\Lambda}_s \mathbf{V}_s^H + \mathbf{U}_n \mathbf{\Lambda}_n \mathbf{V}_n^H, \tag{2.19}$$

where the $N_A \times D$ matrix $\mathbf{U}_s$ and the $T \times D$ matrix $\mathbf{V}_s$ contain the left and right singular vectors corresponding to the largest $D$ singular values. The $N_A \times (N_A - D)$ matrix $\mathbf{U}_n$ and the $T \times (N_A - D)$ matrix $\mathbf{V}_n$ contain the left and right singular vectors corresponding to the $(N_A - D)$ remaining singular values. The $D$ largest singular values form the diagonal of the diagonal matrix $\mathbf{\Lambda}_s$, and the $(N_A - D)$ remaining singular values form the diagonal of the matrix $\mathbf{\Lambda}_n$. Multiplying (2.19) by $\mathbf{V}_s$ results in

$$\mathbf{X}_{SV} = \mathbf{X}\mathbf{V}_s = \mathbf{A}\mathbf{S}_{SV} + \mathbf{N}_{SV}, \tag{2.20}$$

where $\mathbf{S}_{SV} = \mathbf{SV}_s$, $\mathbf{N}_{SV} = \mathbf{NV}_s$, and the matrix $\mathbf{X}_{SV}$ is of size $N_A \times D$. As a result, the dimensionality of the data matrix is reduced, and the number of columns in $\mathbf{X}_{SV}$ is reduced from the number of snapshots $T$ to the number of sources $D$. Taking the discrete set of angles into account, (2.20) can be rewritten as

$$\mathbf{X}_{SV} = \mathbf{A}^g \mathbf{S}_{SV}^g + \mathbf{N}_{SV}, \tag{2.21}$$

where the nonzero rows of $\mathbf{S}_{SV}^g$ correspond to the actual sources. The group sparse solution is then obtained by minimizing the following mixed $\ell_{1,2} -$ norm cost function

$$\min \left\| \mathbf{X}_{SV} - \mathbf{A}^g \mathbf{S}_{SV}^g \right\|_F + \lambda \left\| \mathbf{S}_{SV}^g \right\|_{2,1}, \tag{2.22}$$

where $\|\cdot\|_F$ denotes the Frobenius norm of a matrix, $\lambda$ is a regularization parameter, and the mixed $\ell_{1,2} -$ norm $\left\| \mathbf{S}_{SV}^g \right\|_{2,1}$ is given by

$$\left\| \mathbf{S}_{SV}^g \right\|_{2,1} = \sum_{k=1}^{K} \left\| \left[ \mathbf{S}_{SV}^g \right]_{(k,:)} \right\|_2 \tag{2.23}$$

with $\left[ \mathbf{S}_{SV}^g \right]_{(k,:)}$ being the $k$th row of $\mathbf{S}_{SV}^g$. Since this method is applied in the data domain and the second order statistics are not needed, the performance is not affected by the correlation or coherence between the sources. The maximum number of resolvable sources is limited to the number of elements in the array [27].

For illustration, $\ell_1 -$ SVD is applied to the ten-element ULA with three sources at $[-20°, 10°, 50°]$. A grid of potential directions between $-90°$ and $+90°$ with a step size of 1° is considered. The same parameters as in the previous sections are used. The estimated spectrum is shown in Fig. 2.14. This figure confirms that $\ell_1 -$ SVD is successful in estimating the source directions.

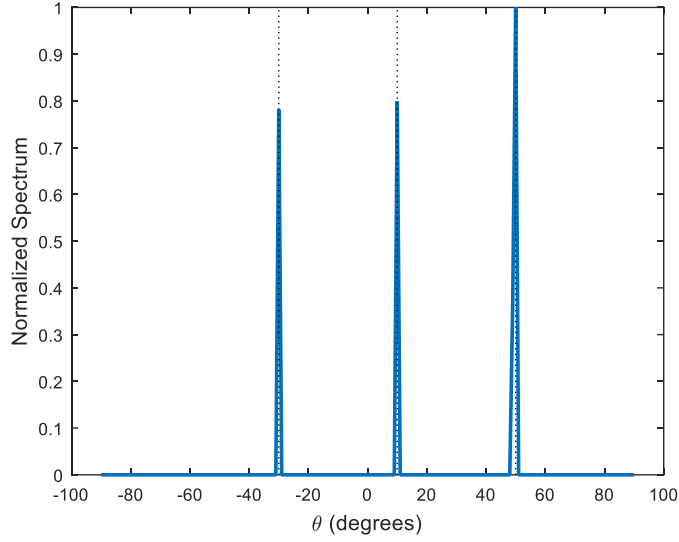

**Figure 2.14:** $\ell_1$ − SVD estimated spectrum.

The number of resolvable sources using the aforementioned techniques is limited by the number of physical sensors in the array. Several techniques can be applied to non-uniform arrays in order to estimate more sources than sensors. These techniques are reviewed below.

### 2.3.6. MUSIC with Spatial Smoothing

Starting with the model in (2.8), the vectorized covariance matrix can be expressed as

$$\mathbf{z} = vec\{\mathbf{R}_{xx}\} = \widetilde{\mathbf{A}}\mathbf{p} + \sigma_n^2\tilde{\mathbf{i}} = (\mathbf{A}^* \odot \mathbf{A})\mathbf{p} + \sigma_n^2\tilde{\mathbf{i}}, \tag{2.24}$$

where $\widetilde{\mathbf{A}}$ is the $N_A^2 \times K$ array manifold matrix corresponding to the difference coarray, $\mathbf{p} = [\sigma_1^2, \dots, \sigma_D^2]^T$ is the source powers vector, $\tilde{\mathbf{i}}$ is the vectorized identity matrix, and $\odot$ denotes the Khatri-Rao product [9]. The $N_A^2 \times 1$ vector $\mathbf{z}$ emulates observations at the difference coarray.

In (2.24), the sources are replaced by their powers and the noise is deterministic. As a result, the model in (2.24) is similar to that corresponding to a fully coherent environment. Spatial smoothing can be applied to restore the rank of the noise-free covariance matrix of $\mathbf{z}$ before proceeding with DOA estimation [6, 16]. However, due to the restrictions on the array geometries which are required by spatial smoothing, this method can only be applied to the filled part of the difference coarray. Assuming that the difference coarray has a contiguous part between $-Ld_0$ and $+Ld_0$, a new $(2L+1) \times 1$ vector $\mathbf{z}_f$, which comprises observations at the contiguous part of the difference coarray, is then formed as

$$\mathbf{z}_f = \widetilde{\mathbf{A}}_f \mathbf{p} + \sigma_n^2 \tilde{\mathbf{i}}_f, \tag{2.25}$$

where $\widetilde{\mathbf{A}}_f$ is the $(2L+1) \times D$ array manifold matrix corresponding to the contiguous part of the difference coarray and $\tilde{\mathbf{i}}_f$ is a $(2L+1) \times 1$ vector whose $(L+1)$th element is equal to one and all remaining elements are zeros. The contiguous part of the difference coarray is then partitioned into $(L+1)$ overlapping subarrays, each having $(L+1)$ elements. The received data vector at the $p$th subarray $(p = 1, 2, \dots, L+1)$ is denoted by $\mathbf{z}_{f,p}$ and holds observations at locations determined by the following set

$$\{(m+1-p)d_0, \quad m = 0, 1, \dots, L\}. \tag{2.26}$$

The overall spatially smoothed covariance matrix is then computed as

$$\mathbf{R}_{ZZ} = \frac{1}{L+1} \sum_{p=1}^{L+1} \mathbf{z}_{f,p} \mathbf{z}_{f,p}^H. \tag{2.27}$$

MUSIC can then applied to $\mathbf{R}_{ZZ}$ to estimate up to $L$ sources. As an example, the six-element extended co-prime array of Fig. 2.9 is considered. The total number of sources is set to seven and the sources are uniformly distributed between $-60°$ and $+60°$. The SNR for all sources is set to 0 dB and the total number of snapshots is set to 500. This example simulates a case where the number of sources is greater than the number of physical sensors. The difference coarray is filled between $-7d_0$ and $+7d_0$, which means that MUSIC after spatial smoothing can be applied to resolve up to seven sources. The estimated spectrum is shown in Fig. 2.15. This method correctly estimates all the source directions.

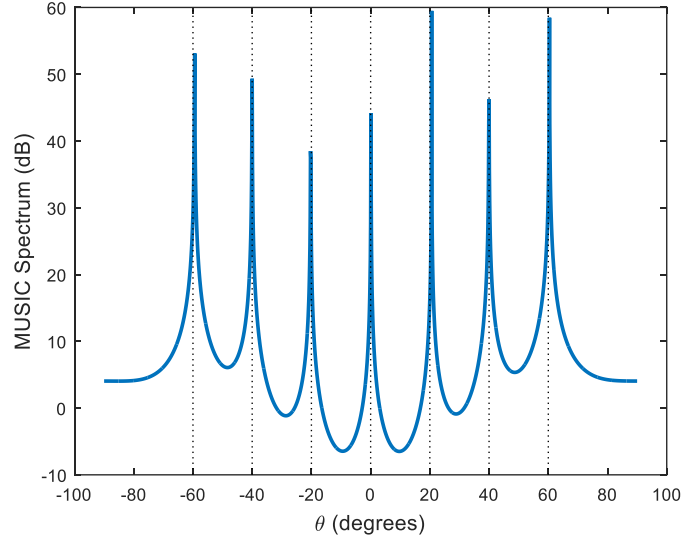

**Figure 2.15:** MUSIC with spatial smoothing spectrum.

### 2.3.7. Covariance Matrix Augmentation

Given the covariance matrix measurements, an augmented covariance matrix corresponding to a virtual ULA can be formed [13]. Assuming that the difference coarray has a contiguous part between $-Ld_0$ and $+Ld_0$, the virtual ULA has elements with positions $\{\ell d_0, 0 \leq \ell \leq L\}$. The augmented covariance has the following elements

$$\mathbf{R}_{Aug} = \begin{bmatrix} \mathbf{R}_{xx}\langle 0 \rangle & \mathbf{R}_{xx}\langle -d_0 \rangle & \cdots & \mathbf{R}_{xx}\langle -Ld_0 \rangle \\ \mathbf{R}_{xx}\langle +d_0 \rangle & \mathbf{R}_{xx}\langle 0 \rangle & \cdots & \mathbf{R}_{xx}\langle (-L+1)d_0 \rangle \\ \vdots & & \ddots & \vdots \\ \mathbf{R}_{xx}\langle +Ld_0 \rangle & \mathbf{R}_{xx}\langle (L-1)d_0 \rangle & \cdots & \mathbf{R}_{xx}\langle 0 \rangle \end{bmatrix}, \tag{2.28}$$

where $\mathbf{R}_{xx}\langle \ell \rangle$ denotes the element of $\mathbf{R}_{xx}$ at lag $\ell$. It should be noted that if two or more elements of $\mathbf{R}_{xx}$ have the same lag, their average value can be used in $\mathbf{R}_{Aug}$. After forming $\mathbf{R}_{Aug}$, MUSIC can be applied to estimate up to $L$ source directions. For illustration, a three-element non-uniform array is considered. The elements of the array are positioned at $[0, 1, 3]d_0$. The corresponding covariance matrix has elements

$$\mathbf{R}_{xx} = \begin{bmatrix} \mathbf{R}_{xx}\langle 0 \rangle & \mathbf{R}_{xx}\langle -d_0 \rangle & \mathbf{R}_{xx}\langle -3d_0 \rangle \\ \mathbf{R}_{xx}\langle d_0 \rangle & \mathbf{R}_{xx}\langle 0 \rangle & \mathbf{R}_{xx}\langle -2d_0 \rangle \\ \mathbf{R}_{xx}\langle 3d_0 \rangle & \mathbf{R}_{xx}\langle 2d_0 \rangle & \mathbf{R}_{xx}\langle 0 \rangle \end{bmatrix}. \tag{2.29}$$

The corresponding difference coarray is filled between $-3d_0$ and $+3d_0$. As a result, the augmented covariance matrix corresponds to a ULA with elements $[0, 1, 2, 3]d_0$ and has the following elements

$$\mathbf{R}_{Aug} = \begin{bmatrix} \mathbf{R}_{xx}\langle 0 \rangle & \mathbf{R}_{xx}\langle -d_0 \rangle & \mathbf{R}_{xx}\langle -2d_0 \rangle & \mathbf{R}_{xx}\langle -3d_0 \rangle \\ \mathbf{R}_{xx}\langle +d_0 \rangle & \mathbf{R}_{xx}\langle 0 \rangle & \mathbf{R}_{xx}\langle -d_0 \rangle & \mathbf{R}_{xx}\langle -2d_0 \rangle \\ \mathbf{R}_{xx}\langle +2d_0 \rangle & \mathbf{R}_{xx}\langle +d_0 \rangle & \mathbf{R}_{xx}\langle 0 \rangle & \mathbf{R}_{xx}\langle -d_0 \rangle \\ \mathbf{R}_{xx}\langle +3d_0 \rangle & \mathbf{R}_{xx}\langle +2d_0 \rangle & \mathbf{R}_{xx}\langle +d_0 \rangle & \mathbf{R}_{xx}\langle 0 \rangle \end{bmatrix}, \tag{2.30}$$

where $\left[\mathbf{R}_{Aug}\right]_{1,1} = \left[\mathbf{R}_{Aug}\right]_{2,2} = \left[\mathbf{R}_{Aug}\right]_{3,3} = \left[\mathbf{R}_{Aug}\right]_{4,4} = \frac{1}{3}\left\{[\mathbf{R}_{xx}]_{1,1} + [\mathbf{R}_{xx}]_{2,2} + [\mathbf{R}_{xx}]_{3,3}\right\}$. The remaining lags have no redundancies in $\mathbf{R}_{xx}$, which means that the same value is copied to the corresponding values in $\mathbf{R}_{Aug}$; e.g. $\left[\mathbf{R}_{Aug}\right]_{1,3} = \left[\mathbf{R}_{Aug}\right]_{2,4} = [\mathbf{R}_{xx}]_{2,3}$.

As an example, this method is applied to the six-element co-prime array of Fig. 2.9. The same seven sources from Section 2.3.6 are also used in this example and the same parameters are kept. Fig. 2.16 shows the estimated spectrum using MUSIC applied to the augmented covariance matrix. It is evident that this method is successful in estimating all the DOAs.

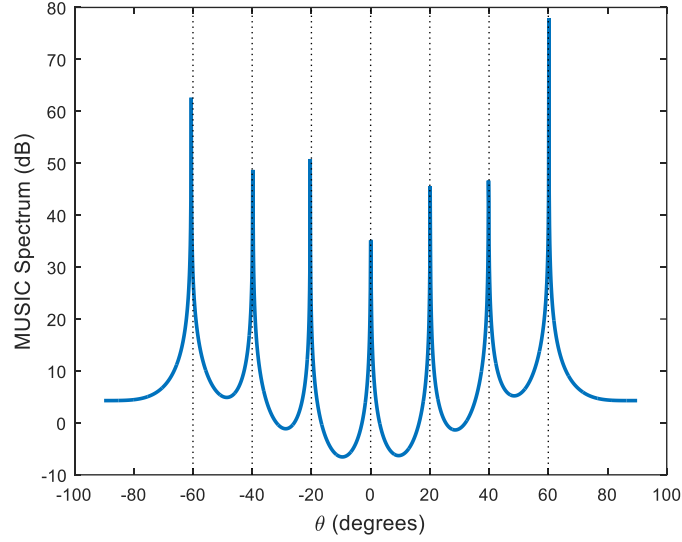

**Figure 2.16:** Augmented covariance matrix approach spectrum.

### 2.3.8. Sparsity-Based DOA Estimation Using Non-Uniform Arrays

Since MUSIC with spatial smoothing is limited to the contiguous part of the difference coarray, some of the available DOFs might not be exploited. This is the case where the difference coarray has missing elements or holes. Sparse reconstruction has been used to address this issue and allow the full exploitation of all available DOFs [29]. Using (2.24), a new vector, comprising the observations at the unique difference coarray elements, can be obtained as

$$\mathbf{z}_u = \widetilde{\mathbf{A}}_u \mathbf{p} + \sigma_n^2 \tilde{\mathbf{i}}_u. \tag{2.31}$$

The length of $\mathbf{z}_u$ is equal to $L_{sc}$ where $L_{sc}$ is the number of unique elements in the difference coarray. $\widetilde{\mathbf{A}}_u$ is the $L_{sc} \times D$ array manifold matrix corresponding to the difference coarray. $\tilde{\mathbf{i}}_u$ is a $L_{sc} \times 1$ vector with all zero elements except the $\frac{(L_{sc}+1)}{2}$th element , which assumes a unit value.

Sparse signal reconstruction can be applied based on the assumption that the sources are sparse in the spatial domain, i.e., only a small number of potential directions are occupied by sources. The angular region of interest is discretized into a set of $K$ ($K \gg D$) grid points, $\{\theta_1^g, \theta_2^g, \ldots, \theta_K^g\}$, with $\theta_1^g$ and $\theta_K^g$ being the limits of the search space. Eq. (2.31) can be rewritten as

$$\mathbf{z}_u = \widetilde{\mathbf{A}}_u^g \mathbf{p}^g + \sigma_n^2 \tilde{\mathbf{i}}_u, \tag{2.32}$$

where the columns of the $L_{sc} \times K$ array manifold matrix $\widetilde{\mathbf{A}}_u^g$ are steering vectors corresponding to the defined angles in the grid. $\mathbf{p}^g$ is a $D$-sparse source power vector of length $K$, with its $K$ nonzero elements corresponding to the powers of the actual sources. DOA estimation proceeds by solving the following minimization problem

$$[\hat{\mathbf{p}}^g; \hat{\sigma}_n^2] = \arg \min_{\mathbf{p}^g, \sigma_n^2} \left[ \frac{1}{2} \left\| \mathbf{z}_u - \widetilde{\mathbf{A}}_u^g \mathbf{p}^g - \sigma_n^2 \tilde{\mathbf{i}}_u \right\|_2 + \lambda \|\mathbf{p}^g\|_1 \right] \text{ s.t. } \mathbf{p}^g \succcurlyeq \mathbf{0}. \tag{2.33}$$

The constraint $\mathbf{p}^g \succcurlyeq \mathbf{0}$ is added to account for the fact that the source powers always assume positive values. The $\ell_2 -$ norm ensures data fidelity and the $\ell_1 -$ norm encourages sparsity in the reconstructed signal. $\lambda$ is a regularization parameter that controls the sparsity level of the reconstructed signal. For the sparse reconstruction approach, the number of resolvable sources is limited to the number of positive lags in the difference coarray, i.e., $(L_{sc} - 1)/2$.

For illustration, the same example from the previous two sections is repeated with this method. The regularization parameter is empirically set to 0.25 and the angular region of interest is divided into 181 grid points with step size 1°. The estimated spectrum using this method is shown in Fig. 2.17. Clearly, all source DOAs are correctly estimated.

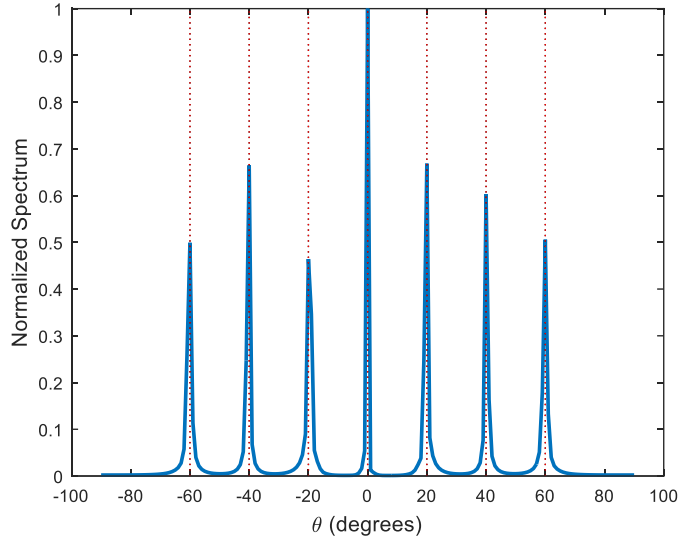

**Figure 2.17:** Sparsity-based estimated spectrum.

In the following example, the performance of the sparsity-based approach is compared to the high-resolution approach. An extended co-prime array, with $M = 3$ and $N = 5$, is considered for the comparison. The physical array consists of 11 elements with positions $[0, 3, 5, 6, 9, 10, 12, 15, 20, 25]d_0$. The corresponding difference coarray extends between $-25d_0$ and $+25d_0$ and has contiguous elements between $-17d_0$ and $+17d_0$. Two sources with varying source powers and varying DOAs are considered. The direction of the first source fixed to $u_1 = \sin\theta_1 = 0$, and the SNR of the second source fixed to 0 dB. The SNR of the first source is varied between 0 dB and 20 dB, and the direction of the second $u_2 = \sin\theta_2$ is varied between 0.03 and 0.20 to simulate different source separation scenarios. For each set of parameters, 1,000 Monte Carlo runs are used and the average root-mean-square error (RMSE) is computed. The number of snapshots is first set to 500. For the sparsity-based approach, the angular region of interest is divided into 181 bins, and the regularization parameter is set to 0.25. Fig. 2.18 compares the performance of MUSIC with spatial smoothing applied to the contiguous part of the coarray to that of the sparsity-based approach under multiple scenarios. In Fig. 2.18(a), the source separation $\Delta u = |\sin\theta_1 - \sin\theta_2|$ is set to 0.03 and the average RMSE is plotted as function of $\Delta SNR$. Clearly, the estimation error increases, for both methods, as $\Delta$SNR increases. Moreover, the sparse reconstruction approach outperforms the high-resolution approach. Fig. 2.18(b) and Fig. 2.18(c) show the RMSE plots for $\Delta u = 0.05$ and $\Delta u = 0.09$, respectively. It can be noticed that MUSIC fares better as the source separation begins to increase.

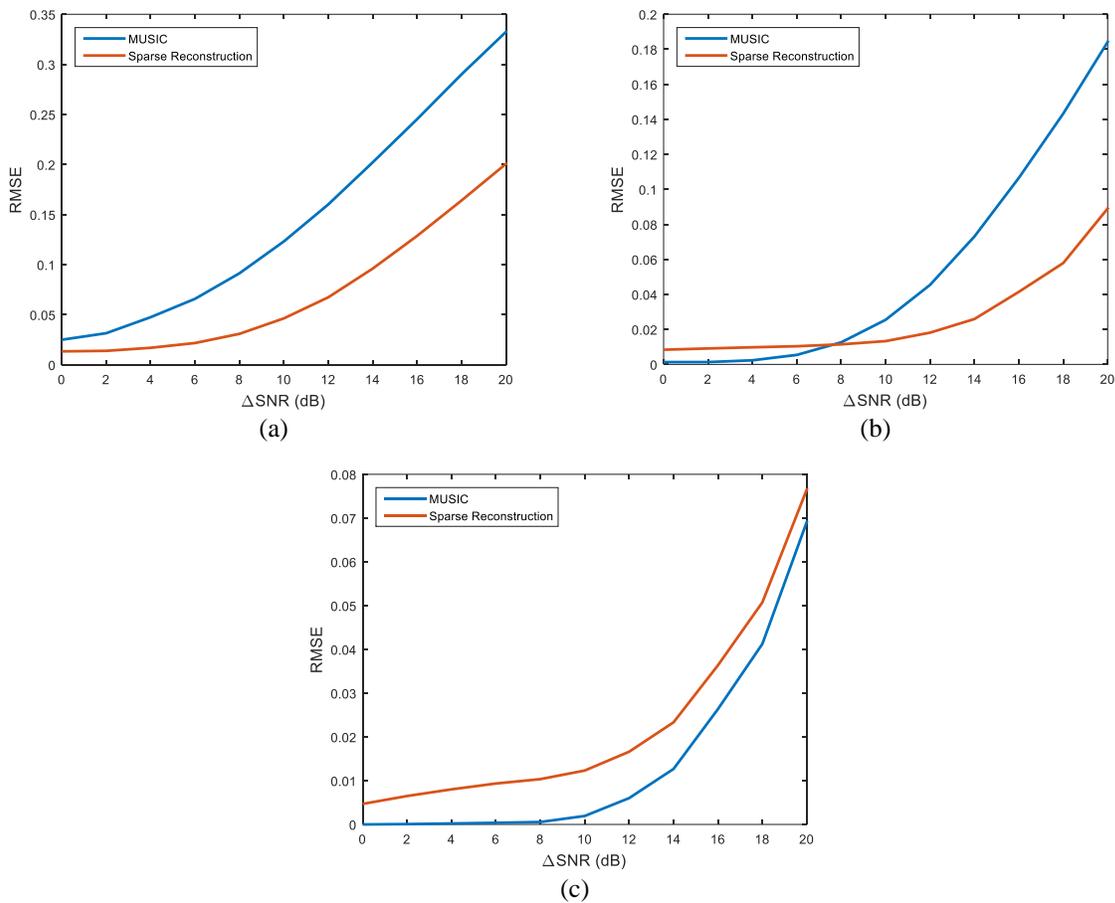

**Figure 2.18:** MUSIC vs. sparse reconstruction, $T = 500$ snapshots (a) $\Delta u = 0.03$: RMSE vs. $\Delta$SNR (b) $\Delta u = 0.05$: RMSE vs. $\Delta$SNR (c) $\Delta u = 0.09$: RMSE vs. $\Delta$SNR.

Fig. 2.19 shows the same results, but with a larger number of snapshots. The number of snapshots is increased to $T = 1,000$, and the remaining parameters are kept the same. Similar trends are observed in this figure. However, the high-resolution shows an improvement in the performance as compared to the previous case. Multiple conclusions can be made by observing these figures. First, the estimation accuracy improves for both methods as the source separation increases. Second, the estimation accuracy worsens as the separation between the source powers increases. Third, the sparsity-based approach provides a better performance when a low number of snapshots or when the sources are closely separated. The high-resolution approach provides an improved performance in the remaining cases.

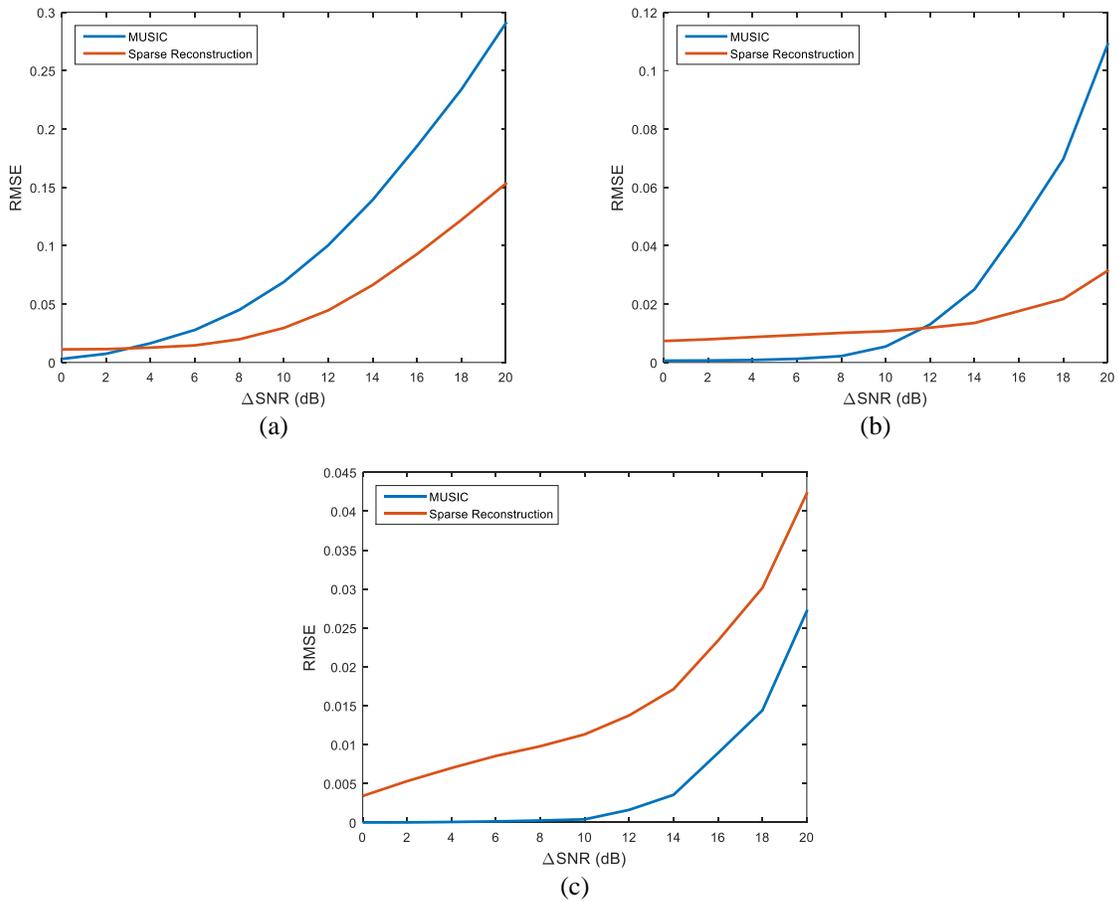

(a)

(b)

(c)

**Figure 2.19:** MUSIC vs. sparse reconstruction, $T = 1000$ snapshots (a) $\Delta u = 0.03$: RMSE vs. $\Delta$SNR (b) $\Delta u = 0.05$: RMSE vs. $\Delta$SNR (c) $\Delta u = 0.09$: RMSE vs. $\Delta$SNR.

## 2.4.    DOA Estimation Challenges

DOA estimation using non-uniform arrays comes with multiple challenges. Three challenges are considered in this dissertation and various techniques are proposed to solve them.

### 2.4.1.   Reduction of Available DOFs

Since the difference coarrays of co-prime and minimum hole arrays contain multiple missing elements, MUSIC with spatial smoothing employs only that part of the difference coarray which has contiguous elements with no holes. As such, only a subset of the total DOFs offered by the co-prime and minimum hole arrays can be utilized for high-resolution DOA estimation using the vectorized covariance matrix approach. The augmented covariance matrix approach, on the other hand, can exploit all the DOFs but at the expense of additional complicated matrix completion processing [15].

### 2.4.2.   Mutual Coupling

One further challenge occurs when dealing with practical antenna arrays due to mutual coupling between the physical elements. If unaccounted for, mutual coupling introduces a mismatch between the assumed model and the actual one, resulting in DOA estimation errors. In the literature, the majority of the DOA estimation methods that deal with the presence of mutual coupling have been developed for ULAs [30-32]. These methods take advantage of the special structure of ULAs and, therefore, cannot be applied to non-uniform arrays to estimate more sources than the number of sensors. An illustrative example which shows the effect of mutual coupling on DOA estimation is considered. The same six-element co-prime array is used in this example and the effect of mutual coupling is added. The modeling of mutual coupling will be discussed in Chapter V. Fig. 2.20 shows the estimated MUSIC spectrum without accounting for the mutual coupling effect. By comparing Fig. 2.20 to Fig. 2.15, it can be noticed that mutual coupling introduced errors in the DOA estimates.

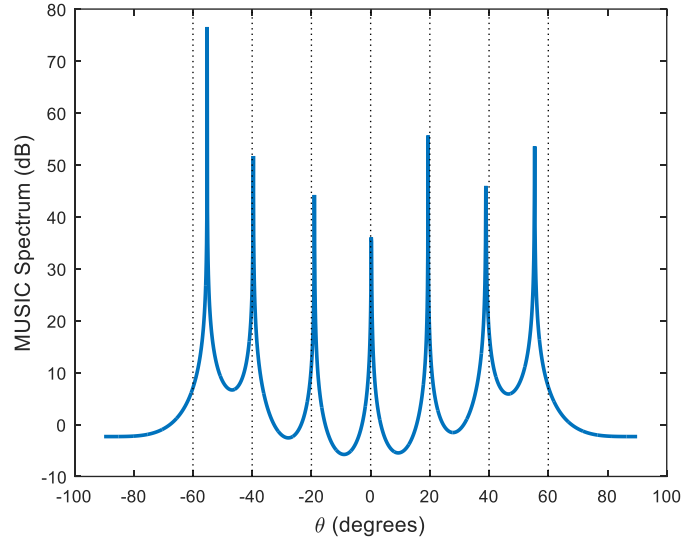

**Figure 2.20:** MUSIC spectrum in the presence of mutual coupling.

### 2.4.3. Coherent Environment

Another DOA estimation challenge rises when dealing with the presence of correlated or coherent sources. This could occur due to, for example, multipath propagation. In this case, the noise-free covariance matrix becomes rank deficient rendering traditional subspace-based DOA estimation techniques inapplicable. Fig. 2.21 shows the MUSIC spectrum applied to a ten-element ULA where two of the three sources are coherent. In this example, the coherent sources are shown with vertical red lines. It can be noticed that the estimated spectrum misses the two coherent sources and correctly estimates the third source.

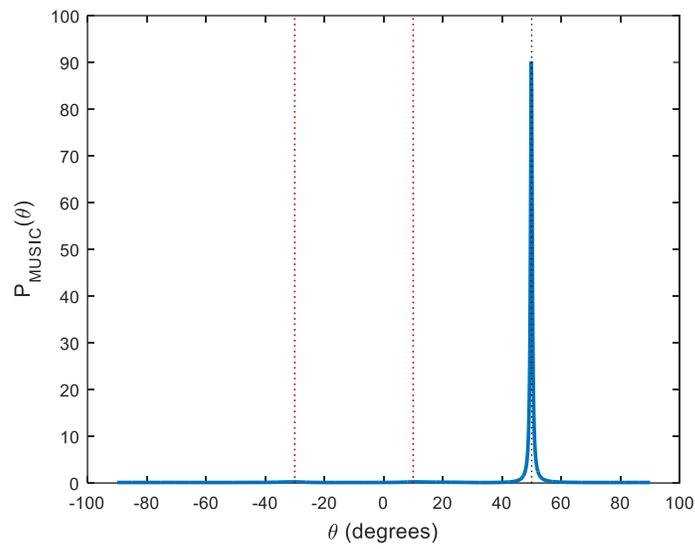

**Figure 2.21:** MUSIC spectrum with two coherent sources at −30° and 10°.

Spatial smoothing can be used to restore the rank of the covariance matrix [16]. However, it can only be applied to specific array structures and always results in reducing the DOFs that are available for DOA estimation.

# CHAPTER III

# MULTI-FREQUENCY DOA ESTIMATION USING NON-UNIFORM ARRAYS

In this chapter, two multi-frequency DOA estimation methods are presented to alleviate the issue of reduced DOFs when dealing with non-uniform arrays with missing elements in their difference coarray. The first method is a high-resolution method which utilizes multiple additional frequencies to fill in the missing elements and perform DOA estimation using the entire coarray aperture. The second method is a sparsity-based method that uses the entire set of observations that are generated at all additional frequencies in order to perform DOA estimation with increased DOFs. It should be noted that the aforementioned methods are initially proposed for use with extended co-prime arrays, but they can be readily applied to other non-uniform arrays. However, the closed-form expressions, presented in this chapter, are only specific to extended co-prime array configurations.

## 3.1. Subspace-Based High-Resolution Approach

In this approach, multi-frequency operation to utilize all of the DOFs for DOA estimation in co-prime arrays is considered. More specifically, a set of additional frequencies is employed to recover the missing lags through dilations of the coarray [33, 34]. The sources are assumed to have a bandwidth large enough to cover all specific frequencies required for filling the holes. Only the array elements involved in filling the missing holes in the difference coarray are required to be operated at one or more of the additional frequencies. The multi-frequency measurements are used to construct a virtual covariance matrix corresponding to an equivalent filled uniformly spaced coarray at a single frequency [35]. High-resolution subspace techniques, such as MUSIC, can then be applied to this virtual covariance matrix for DOA estimation. It is important to note that full utilization of the DOFs using multiple additional frequencies comes with a restriction on the sources' spectra. More specifically, the source spectra at all operational frequencies are required to be proportional. Deviations from this restriction can lead to higher DOA estimation errors.

Multiple frequencies have previously been used for alias-free DOA estimation of broadband sources [36, 37]. In [37], frequency diversity was exploited on a single spatial sampling interval to mitigate spatial aliasing in DOA estimation with a sparse non-uniformly spaced array. Ambiguities in the source location estimates were resolved by proper choice of chosen operational frequencies in [37] for arrays with periodic spatial spectra. Spatial sampling interval diversity at a single narrowband frequency was exploited in [5] to disambiguate aliased DOAs. Both spatial sampling and frequency diversity were exploited in [35] through multi-frequency coarray augmentation for high-resolution DOA estimation. However, no attempt was made therein to select the best number of employed frequencies or determine their best values. Multi-frequency coarray augmentation is effectively applied to co-prime arrays in this section. The main contribution lies in exploiting the specific structure of the coarray corresponding to co-prime configuration to determine the number and values of the additional frequencies required for recovering the missing lags. Closed-form expressions for the additional frequencies, which are 'best' in the sense of minimum operational bandwidth requirements, are provided. Exploitation of the redundancy in the coarray to reduce the system hardware complexity for multi-frequency co-prime arrays is also described. Further, the effects of noise and deviation from the proportional source spectra constraint on the DOA estimation performance of the multi-frequency co-prime arrays are investigated.

The remainder of this section is organized as follows. In Section 3.1.1, the multi-frequency approach for filling the missing elements in the coarray and utilizing all the DOFs offered by the co-prime configuration is described. Section 3.1.2 delineates the system bandwidth requirement for the multi-frequency operation, taking into account the specificities of the coarray structure corresponding to co-prime arrays. Coarray redundancy is also examined to reduce the number of antennas engaging in multiple frequency operation. A comparison between co-prime arrays and minimum hole arrays is also included in this section. In Section 3.1.3, performance of the proposed method is evaluated through extensive simulations under both proportional and non-proportional source spectra, and Section 3.1.4 summarizes the main contributions of this approach.

### 3.1.1. High-Resolution DOA Estimation with Multi-Frequency Co-Prime Arrays

The sources are assumed to have a bandwidth large enough to cover all frequencies required for filling the holes. Discrete Fourier transform (DFT) or filterbanks are used to decompose the array output vector into multiple non-overlapping narrowband components and extract the received signal at each considered frequency [38, 39]. The observation time is assumed to be sufficiently long to resolve the different frequencies.

The extended co-prime configuration of Fig. 2.8(b), where the unit spacing $d_0$ is assumed to be half-wavelength at the reference frequency $\omega_0$, is considered. Following (2.6), the received signal at $\omega_0$ can be expressed as

$$\mathbf{x}(\omega_0) = \mathbf{A}(\omega_0)\mathbf{s}(\omega_0) + \mathbf{n}(\omega_0), \tag{3.1}$$

where $\mathbf{x}(\omega_0)$, $\mathbf{A}(\omega_0)$, $\mathbf{s}(\omega_0)$, and $\mathbf{n}(\omega_0)$ have the same definitions as $\mathbf{x}(t)$, $\mathbf{A}$, $\mathbf{s}(t)$, and $\mathbf{n}(t)$, respectively. By operating the physical co-prime array at a different frequency, $\omega_q = \alpha_q \omega_0$, the received signal at $\omega_q$ has the form

$$\mathbf{x}(\omega_q) = \mathbf{A}(\omega_q)\mathbf{s}(\omega_q) + \mathbf{n}(\omega_q), \tag{3.2}$$

where $\mathbf{A}(\omega_q)$ is the $(2M + N - 1) \times D$ array manifold at $\omega_q$ with its $(i, d)$th element given by

$$\left[\mathbf{A}(\omega_q)\right]_{i,d} = e^{jk_q x_i \sin(\theta_d)}. \tag{3.3}$$

In (3.3), $k_q = \omega_q/c$ is the wavenumber at $\omega_q$. Since $k_q = \alpha_q k_0$ , (3.3) can be rewritten as

$$\left[\mathbf{A}(\omega_q)\right]_{i,d} = e^{jk_0 \alpha_q x_i \sin(\theta_d)}. \tag{3.4}$$

By comparing (2.7) and (3.4), it can be observed that the array manifold at $\omega_q$ is equivalent to the array manifold at $\omega_0$ of a scaled version of the physical co-prime array. The position of the $i$th element in the equivalent scaled array is given by $\alpha_q x_i$. This results in the difference coarray at $\omega_q$ to be a scaled version of the coarray at the reference frequency $\omega_0$ [40]. Values of $\omega_q$ higher than $\omega_0$ cause an expansion of the coarray, while the coarray contracts if $\omega_q$ is lower than $\omega_0$. In other words, operation at the additional frequency adds extra points at specific locations in the coarray. A suitable choice of additional operating frequencies will cause some of these extra points to occur at the locations of the holes in the difference coarray at $\omega_0$.

For illustration, an extended co-prime array with $M = 3$ and $N = 7$ is considered and the sensor positions given by $[0d_0\ 3d_0\ 6d_0\ 7d_0\ 9d_0\ 12d_0\ 14d_0\ 15d_0\ 18d_0\ 21d_0\ 28d_0\ 35d_0]$. The corresponding difference coarray at $\omega_0$ is shown in Fig. 3.1(a). Operating the array at frequency $\omega_1 = 8/7\omega_0$, which is larger than $\omega_0$, results in stretching the difference coarray of Fig. 3.1(a), as shown in Fig. 3.1(b). On the other hand, if the array is operated at a smaller frequency, $\omega_2 = 6/7\omega_0$, the difference coarray undergoes contraction as depicted in Fig. 3.1(c).

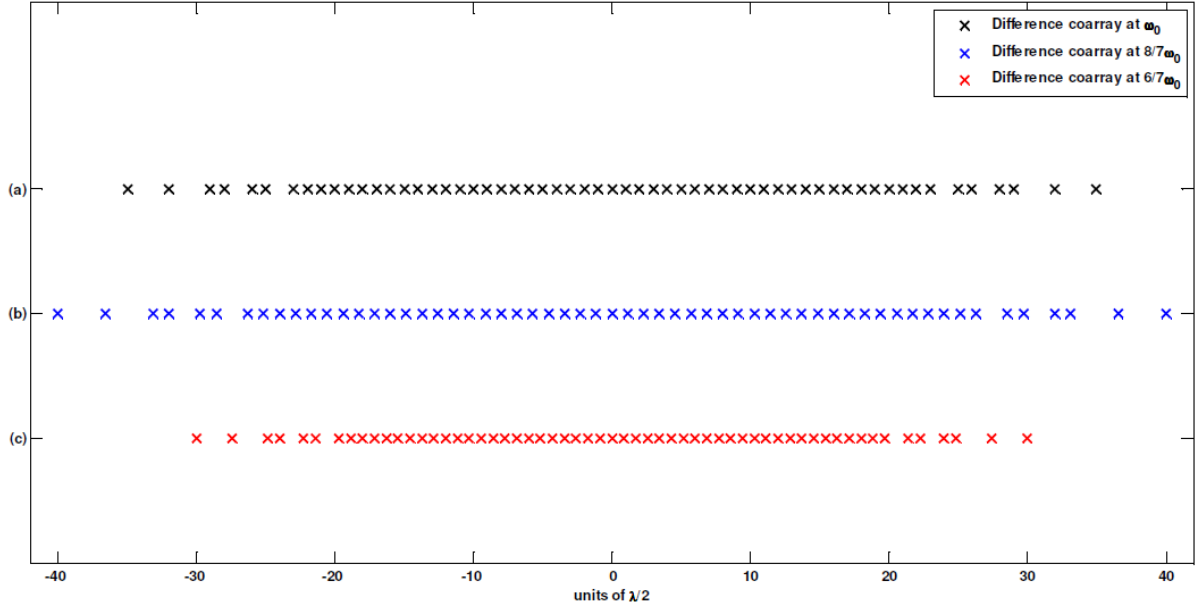

**Figure 3.1:** Difference coarray of extended co-prime array ($M = 3$, $N = 7$) at (a) $\omega_0$ (b) $\omega_1 = 8/7\,\omega_0$ (c) $\omega_2 = 6/7\omega_0$.

### A.    *Virtual Covariance Matrix Formation*

The total number of operational frequencies, including the reference, is assumed to be $Q$. As shown below, a virtual covariance matrix can be constructed using the multi-frequency measurements, which is equivalent to that of a ULA with $(2M-1)N+1$ elements operating at the reference frequency [35, 41]. This would allow DOA estimation of $(2M-1)N$ sources instead of $(MN+M-1)$ sources using the $(2M+N-1)$ physical sensors of the extended co-prime array.

A $(2M+N-1) \times (2M+N-1)$ support matrix $\mathbf{C}(\omega_q)$ is defined such that its $(i,j)$th element is given by [35, 41]

$$\left[ \mathbf{C}(\omega_q) \right]_{i,j} = \alpha_q x_i - \alpha_q x_j. \tag{3.5}$$

That is, the $(i,j)$th element of $\mathbf{C}(\omega_q)$ is the spatial lag or the coarray element position which is the support of the $(i, j)$th element of the covariance matrix $\mathbf{R}_{xx}(\omega_q)$

$$\mathbf{R}_{xx}(\omega_q) = E\{\mathbf{x}(\omega_q)\mathbf{x}^H(\omega_q)\} = \mathbf{A}(\omega_q)\mathbf{R}_{ss}(\omega_q)\mathbf{A}^H(\omega_q) + \sigma_n^2(\omega_q)\mathbf{I}, \qquad (3.6)$$

where $\mathbf{R}_{ss}(\omega_q) = diag\{\sigma_1^2(\omega_q),\ \sigma_2^2(\omega_q), ..., \sigma_D^2(\omega_q)\}$ is the source covariance matrix at frequency $\omega_q$. It should be noted that $\mathbf{C}(\omega_q) = \alpha_q \mathbf{C}(\omega_0)$, where $\mathbf{C}(\omega_0)$ is the support matrix at the reference frequency $\omega_0$. Let $\mathbf{C}_v(\omega_0)$ and $\mathbf{R}_v(\omega_0)$ be the support and the covariance matrices corresponding to the desired ULA with $(2M - 1)N + 1$ sensors operating at $\omega_0$. Given that the $Q$ operational frequencies are sufficient to fill all the holes in the difference coarray of the co-prime array, then

$$[\mathbf{C}_v(\omega_0)]_{i,j} = [\mathbf{C}(\omega_q)]_{p,r} \text{ for some } q, p, r \text{ and all } i, j \qquad (3.7)$$

The map that arranges selected elements of the multi-frequency support matrices, $\{\mathbf{C}(\omega_q)\}_{q=0}^{Q-1}$, into the desired virtual support matrix $\mathbf{C}_v(\omega_0)$ is denoted by $h$. Using the same map, the virtual covariance matrix $\mathbf{R}_v(\omega_0)$ corresponding to the equivalent ULA can then be constructed from the covariance matrices $\{\mathbf{R}_{xx}(\omega_q)\}_{q=0}^{Q-1}$ corresponding to the $Q$ operational frequencies [35].

For illustration, an extended co-prime array with $M = 2$ and $N = 3$ is considered. The sensor positions are the same as those shown in Fig. 2.9. The support matrix $\mathbf{C}(\omega_0)$ at the reference frequency takes the form

$$\mathbf{C}(\omega_0) = \begin{bmatrix} 0 & -2 & -3 & -4 & -6 & -9 \\ 2 & 0 & -1 & -2 & -4 & -7 \\ 3 & 1 & 0 & -1 & -3 & -6 \\ 4 & 2 & 1 & 0 & -2 & -5 \\ 6 & 4 & 3 & 2 & 0 & -3 \\ 9 & 7 & 6 & 5 & 3 & 0 \end{bmatrix} d_0. \qquad (3.8)$$

The difference coarray of this configuration is shown in Fig. 2.9. It has holes at $-8d_0$ and $8d_0$.

In order to fill these holes and form the virtual covariance matrix, an additional frequency $\omega_1 = 8/9\omega_0$ is required. With this choice of the second operational frequency, the support matrix at $\omega_1$ is given by

$$\mathbf{C}(\omega_1) = \begin{bmatrix} 0 & -16/9 & -8/3 & -32/9 & -16/3 & -8 \\ 16/9 & 0 & -8/9 & -16/9 & -32/9 & -56/9 \\ 8/3 & 8/9 & 0 & -8/9 & -8/3 & -16/3 \\ 32/9 & 16/9 & 8/9 & 0 & -16/9 & -40/9 \\ 16/3 & 32/9 & 8/3 & 16/9 & 0 & -8/3 \\ 8 & 56/9 & 16/3 & 40/9 & 8/3 & 0 \end{bmatrix} d_0. \tag{3.9}$$

The support matrix $\mathbf{C}_v(\omega_0)$ of the desired 10-element ULA, whose elements are positioned at $[0, 1, \dots, 9]d_0$, has the structure

$$\mathbf{C}_v(\omega_0) = \begin{bmatrix} 0 & -1 & -2 & \dots & -8 & -9 \\ 1 & 0 & -1 & \dots & -7 & -8 \\ 2 & 1 & 0 & \dots & -6 & -7 \\ \vdots & \vdots & \vdots & \ddots & \vdots & \vdots \\ 8 & 7 & 6 & \dots & 0 & -1 \\ 9 & 8 & 7 & \dots & 1 & 0 \end{bmatrix} d_0. \tag{3.10}$$

From (3.8)-(3.10), it can be noticed that several possibilities exist for constructing $\mathbf{C}_v(\omega_0)$ using $\mathbf{C}(\omega_0)$ and $\mathbf{C}(\omega_1)$, since several elements of $\mathbf{C}(\omega_0)$ and $\mathbf{C}(\omega_1)$ correspond to the same element of $\mathbf{C}_v(\omega_0)$. Either a single element or an average of all such elements can be used to specify the map for forming the desired virtual support matrix and, subsequently, the virtual covariance matrix $\mathbf{R}_v(\omega_0)$ [35, 41].

It should be noted that since the difference coarray at $\omega_0$ has two holes at $\pm 8d_0$, only those elements of $\mathbf{R}_{xx}(\omega_1)$ that correspond to these two lags are required to form $\mathbf{R}_v(\omega_0)$. This means that instead of operating the entire co-prime array at $\omega_1$, only the sensors that produce the $\pm 8d_0$ lags at $\omega_1$ should be operated at the additional frequency. For example, operating the two sensors with positions $[0, 9]d_0$ at $\omega_1$ produces the following reduced support matrix

$$\mathbf{C}_r(\omega_1) = \frac{8}{9}\mathbf{C}_r(\omega_0) = \frac{8}{9}\begin{bmatrix} 0 & -9 \\ 9 & 0 \end{bmatrix} d_0 = \begin{bmatrix} 0 & -8 \\ 8 & 0 \end{bmatrix} d_0. \tag{3.11}$$

The two support matrices $\mathbf{C}(\omega_0)$ and $\mathbf{C}_r(\omega_1)$ can then be combined to form $\mathbf{C}_v(\omega_0)$. This procedure results in reducing hardware complexity. A more detailed discussion in this regard is provided later.

### B.    Proportional Spectra Requirement

For multi-frequency DOA estimation, the normalized covariance matrices are employed instead of $\{\mathbf{R}_{xx}(\omega_q)\}_{q=0}^{Q-1}$. The $(i,j)$th element of the normalized covariance matrix $\overline{\mathbf{R}}_{xx}(\omega_q)$ at frequency $\omega_q$ can be expressed as [41]

$$\left[\overline{\mathbf{R}}_{xx}(\omega_q)\right]_{i,j} = \frac{E\left\{\left[\mathbf{x}(\omega_q)\right]_i\left[\mathbf{x}^*(\omega_q)\right]_j\right\}}{\frac{1}{N_A(\omega_q)}E\{\mathbf{x}^H(\omega_q)\mathbf{x}(\omega_q)\}}, \tag{3.12}$$

where $\left[\mathbf{x}(\omega_q)\right]_i$ is the $i$th element of the data vector at frequency $\omega_q$, and $N_A(\omega_q)$ is the number of sensors that are operated at $\omega_q$. This results in the source and noise powers in the covariance matrix representation of (2.8) being replaced by the normalized powers [35], which are given by

$$\bar{\sigma}_k^2(\omega_q) = \frac{\sigma_k^2(\omega_q)}{\sum_{d=1}^D \sigma_d^2(\omega_q) + \sigma_n^2(\omega_q)} \tag{3.13}$$

$$\bar{\sigma}_n^2(\omega_q) = \frac{\sigma_n^2(\omega_q)}{\sum_{d=1}^D \sigma_d^2(\omega_q) + \sigma_n^2(\omega_q)}, \tag{3.14}$$

where $\bar{\sigma}_k^2(\omega_q)$ is the normalized power of the $k$th source at frequency $\omega_q$ and $\bar{\sigma}_n^2(\omega_q)$ is the normalized noise power at the same frequency. The virtual covariance matrix $\mathbf{R}_v(\omega_0)$, constructed by using the normalized covariance matrices $\{\overline{\mathbf{R}}_{xx}(\omega_q)\}_{q=0}^{Q-1}$ following the procedure outlined in Section 3.1.1.A, must appear to have been generated by the virtual array as if it were the actual array operating at frequency $\omega_0$. However, some of the elements of the constructed virtual covariance matrix have contributions from frequencies other than $\omega_0$. The virtual covariance matrix will be exact provided that the normalized power of each source is independent of frequency,

$$\bar{\sigma}_k^2(\omega_q) = \sigma_k^2, \text{ for all } q \in \{0, 1, \dots, Q-1\} \text{ and all } k \in \{1, 2, \dots, D\}. \tag{3.15}$$

For a high SNR, a sufficient condition for the virtual covariance matrix to be exact is that the sources must have proportional spectra at the employed frequencies [41]. That is,

$$\frac{\sigma_k^2(\omega_q)}{\sigma_l^2(\omega_q)} = \beta_{k,l}, \tag{3.16}$$

where $\beta_{k,l}$ is a constant for each source pair $(k, l)$ over all frequencies $\omega_q$. This condition is satisfied, for example, when the $D$ sources are BPSK or chirp-like signals.

### 3.1.2. Frequency Selection For Multi-frequency Co-Prime Array Operation

In order to quantify the operational frequency set for filling the holes, the specific structure of the difference coarray corresponding to an extended co-prime configuration needs first to be examined. The difference coarray, corresponding to the extended co-prime array of Fig. 2.8(b), is shown in Fig. 3.2. The total number of filled and missing elements in the coarray equals $2(2M - 1)N + 1$, whereas the total number of holes is determined to be $(M - 1)(N - 1)$. As the coarray is symmetric, only the portion corresponding to the non-negative lags is taken into account. Several observations can be made by examining this portion. The portion of the coarray extending from $0$ to $(MN + M - 1)d_0$ is uniform and has no holes. The first hole appears at $(MN + M)d_0$, followed by another filled part from $(MN + M + 1)d_0$ to $(MN + 2M - 1)d_0$.

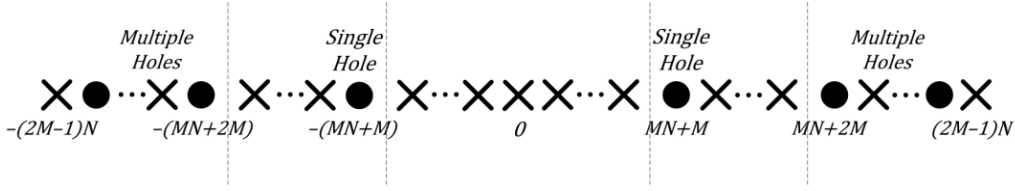

**Figure 3.2:** Difference coarray of the extended co-prime array.

The final part of the coarray from $(MN + 2M)d_0$ to $(2M - 1)Nd_0$ is non-uniform and contains $[(M-1)(N-1)/2 - 1]$ holes.

### A. One Additional Frequency (Dual-Frequency Operation)

The two holes at $-(MN + M)d_0$ and $(MN + M)d_0$ can be filled using only one additional frequency. The choice of the additional frequency is not unique. The value of $\omega_1$ that minimizes the separation between $\omega_0$ and $\omega_1$ is given by

$$\omega_1 = \alpha_1 \omega_0 = \frac{MN + M}{MN + M + 1}\omega_0, \tag{3.17}$$

where the numerator and the denominator of the scaling factor $\alpha_1$ correspond to the respective positions of the hole to be filled and the adjacent filled element to the right of the hole (considering the non-negative lags) that is used to fill it. It is to be noted that the value of $\omega_1$ in (3.17) is less than $\omega_0$. It can be readily shown that using neighboring elements other than the right adjacent one yields values of $\omega_1$, which result in a larger separation from $\omega_0$.

Filling the two holes at $\pm(MN + M)d_0$ causes the uniform part of the difference coarray to extend from $-(MN + 2M - 1)d_0$ to $(MN + 2M - 1)d_0$. As a result, the directions of up to $(MN + 2M - 1)$ sources can be estimated after forming the corresponding virtual covariance matrix. This implies that, compared to the single frequency operation, $M$ additional sources can be estimated using one additional frequency.

*B.     Multiple Additional Frequencies (Multiple-Frequency Operation)*

The remaining $(M-1)(N-1)-2$ holes in the difference coarray can also be filled through the use of additional frequencies. The exact number and values of the frequencies are tied to the non-uniformity pattern in the coarray beyond $\pm(MN+2M)d_0$, which varies from one co-prime configuration to the other. Assuming that each additional frequency is used to fill only two holes (one missing positive element and its negative counterpart), at the most $\frac{1}{2}((M-1)(N-1)-2) = (MN-M-N)/2$ additional frequencies are required to yield a filled uniform coarray extending from $-(2M-1)Nd_0$ to $(2M-1)Nd_0$.

*C.     Maximum Frequency Separation*

The maximum frequency separation from the reference frequency determines the required operational bandwidth of the antennas and receiver front end for the proposed multi-frequency approach. It is determined by the distance of the farthest hole from its nearest filled right neighbor and the location of the neighbor. The maximum number of consecutive holes in the difference coarray is $(M-1)$, and this pattern of $(M-1)$ consecutive holes repeats $\lfloor N/M \rfloor$ times at each end of the difference coarray, as shown in Fig. 3.3 for the non-negative lags. However, it is the first set of $(M-1)$ consecutive holes (those on extreme left in Fig. 3.3) that requires operational frequencies with the maximum separation from $\omega_0$ in order to be filled. The repeated hole patterns at larger lags yield smaller frequency separation values. The first missing element in the leftmost set of consecutive holes occurs at $\left[(2M-1)N-(M-1)-\left(\left\lfloor\frac{N}{M}\right\rfloor-1\right)M\right]d_0$, while the nearest right filled element is positioned at $\left[(2M-1)N-\left(\left\lfloor\frac{N}{M}\right\rfloor-1\right)M\right]d_0$. Therefore, the required frequency to fill this hole is given by

$$\widetilde{\omega} = \frac{(2M-1)N - (M-1) - \left(\left\lfloor \frac{N}{M} \right\rfloor - 1\right)M}{(2M-1)N - \left(\left\lfloor \frac{N}{M} \right\rfloor - 1\right)M}\,\omega_0.$$

(3.18)

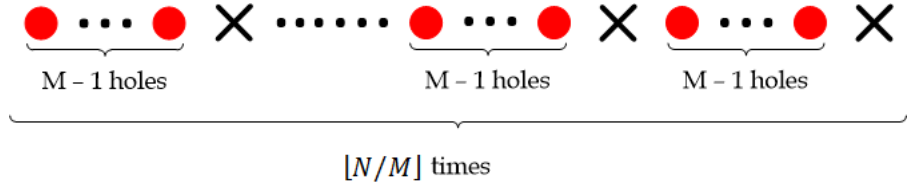

**Figure 3.3:** Positive end part of the difference coarray corresponding to the co-prime array.

The maximum frequency separation can, thus, be computed as

$$\Delta\omega_{max} = |\omega_0 - \widetilde{\omega}| = \left|\frac{1-M}{(2M-1)N - \left(\left\lfloor \frac{N}{M} \right\rfloor - 1\right)M}\right|\omega_0.$$

(3.19)

Table 3.1 shows the maximum frequency separation for different co-prime array configurations under two cases: i) when one additional frequency is used to fill the first pair of holes, and ii) when all holes are filled using multiple frequencies. For each of the aforementioned cases, the additional number of estimated sources compared to single frequency operation are also specified. Clearly, the maximum frequency separation decreases with increasing values of $M$ and $N$. This is because both the holes and the elements that are used to fill them occur at larger spatial lags for higher values of $M$ and $N$, which, in turn, implies a smaller value of the scaling factor in (3.19).

**Table 3.1:** Maximum frequency separation for dual and multi-frequency operations.

| $M$ | $N$ | Dual-frequency | | Multi-frequency | |
|---|---|---|---|---|---|
| | | Additional estimated sources | $\Delta\omega_{max}$ | Additional estimated sources | $\Delta\omega_{max}$ |
| 2 | 3 | 2 | 11.11% | 2 | 11.11% |
| 3 | 4 | 3 | 6.25% | 6 | 10.00% |
| 3 | 5 | 3 | 5.26% | 8 | 8.00% |
| 5 | 7 | 5 | 2.44% | 24 | 6.35% |
| 7 | 9 | 7 | 1.41% | 48 | 5.13% |

## D.    *Reduced Hardware Complexity*

Since only a few observations at each employed frequency other than $\omega_0$ are used for the proposed multi-frequency high-resolution DOA estimation scheme and the remaining observations are discarded, it is not economical to operate the entire physical array at each of the additional $(Q-1)$ frequencies. Therefore, only the receive elements that generate the desired spatial lags for filling the holes need to be operating at more than one frequency. As determined in the previous section, the bandwidth requirement for the multi-frequency operation is not that high, especially for larger values of $M$ and $N$. As such, only the multi-frequency receive elements require a DFT or a filterbank to extract the information at the different frequencies, leading to a significant reduction in system hardware complexity.

It becomes of interest to determine the smallest number of sensors that are required to operate at the additional frequency or frequencies. As the holes occur in symmetric pairs, the lags corresponding to each pair can be generated using only two sensors in the physical array. In case of redundancy in the difference coarray, there is more than one antenna pair that can generate the same spatial lag. In order to reduce the number of antennas engaging in multiple frequency processing, one should therefore seek and identify each sensor that participates in filling all the holes or at least many of them. This becomes important when there is flexibility in sensor participation choices implied by the redundancy property of the spatial lags. Clearly, only the redundant spatial lags occurring beyond the first symmetric hole pair at $\pm(MN + M)d_0$ need to be considered, since these are used to fill the holes in the difference coarray. It can be readily shown that there are a total of $2(M - 2)$ redundant lags beyond $\pm(MN + M)d_0$ at $\pm(MN + kN)d_0$ with weights given by

$$W(\pm(MN + kN)d_0) = M - k, \text{for } k = 1,2,\dots,M-2. \tag{3.20}$$

For illustration, an example where $M = 4$ and $N = 5$ is considered. The co-prime array consists of 12 elements positioned at $[0,4,5,8,10,12,15,16,20,25,30,35]d_0$. Fig. 3.4 shows the difference coarray weighting function corresponding to this array. The first hole pair in the coarray occurs at $\pm(MN + M)d_0 = \pm24d_0$. Beyond the first holes, $2(M - 2) = 4$ redundant lags exist. The first redundant lag pair occurs at $\pm(MN + N)d_0 = \pm25d_0$ with weight equal to $(M - 1) = 3$. The second redundant pair occurs at $\pm(MN + 2N)d_0 = \pm30d_0$ and has a weight of $(M - 2) = 2$. In order to minimize the maximum frequency separation, only the redundant lags that occur immediately to the right of the holes (considering the nonnegative lags) can be used. For the case where $mod(N, M) = 1$, all the redundant lags in the non-uniform part of the coarray occur immediately after the holes. This can be confirmed by observing the weighting function in Fig. 3.4. For the case where $mod(N, M) = M - 1$, none of the redundant lags are immediately to the right of the holes, as illustrated in Fig. 3.5 for the case where $M = 4$ and $N = 7$. For the remaining cases, only a subset of the redundant lags in the non-uniform part is immediately after the holes.

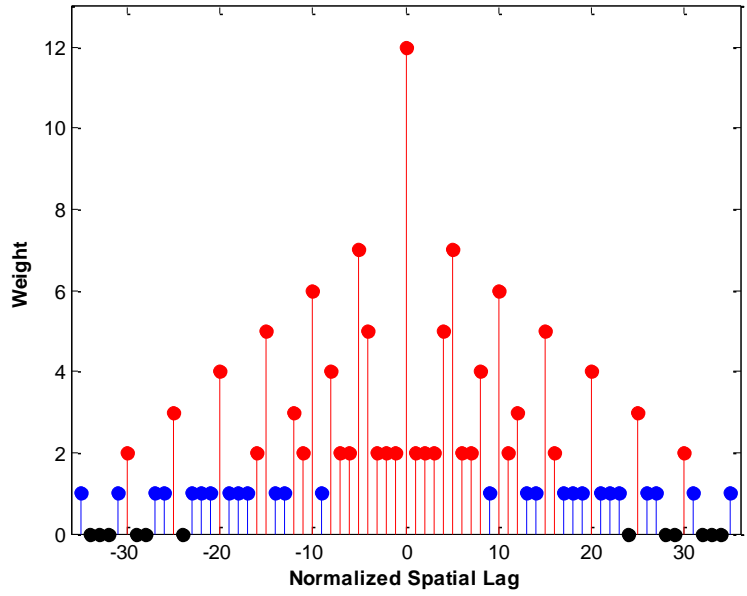

**Figure 3.4:** Difference coarray weight function: $M = 4$, $N = 5$.

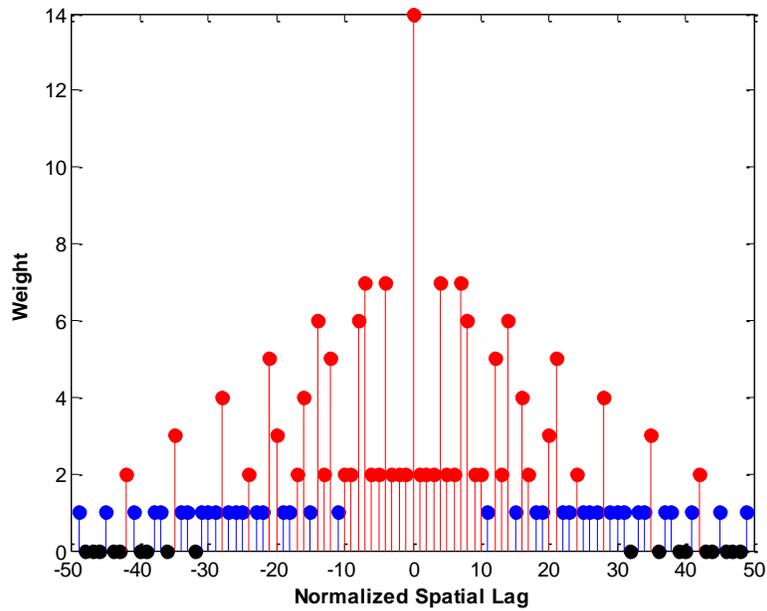

**Figure 3.5:** Difference coarray weight function: $M = 4$, $N = 7$.

For the illustration of the role of redundancy in reducing sensor engagement in hole filling, the following two examples are provided. Table 3.2 shows the additional frequencies and the corresponding sensor pairs that are required to fill all nine holes in the difference coarray for the case where $M = 4$ and $N = 7$. The corresponding physical array consists of 14 sensors at $[0, 4, 7, 8, 12, 14, 16, 20, 21, 24, 28, 35, 42, 49]d_0$. It is clear from Table 3.2 that only the six sensors located at $[0, 4, 8, 12, 16, 49]d_0$ are required to operate at more than one frequency in order to fill all the holes in the coarray. It should be noted that since $mod(N, M) = (M - 1)$ in this example, the redundant lags in the difference coarray cannot be used to further decrease the number of antennas that would operate at more than one frequency. Table 3.3 shows the required frequencies and the corresponding sensor pairs for the case where $M = 4$ and $N = 5$. Since $mod(N, M) = 1$, different sensor pairs can be used to fill the same holes. As shown in Table 3.3, the pairs that include common sensors at different frequencies are chosen in order to minimize the number of sensors that operate at more than one frequency. Table 3.4 shows the percentage of sensors that need to be operated at more than one frequency for different co-prime array configurations. It can be noticed that the number of sensors that need to be operated at multiple frequencies has a lower bound of one-third of the total number of sensors in the array, which is achieved for co-prime configurations with $N = (M + 1)$. It should be noted that the same choice of $N = (M + 1)$ also minimizes the total number of sensors in the co-prime arrays, as demonstrated in [18].

**Table 3.2:** Required frequencies and sensor pairs, $M = 4$, $N = 7$.

| Frequencies | Holes | Sensor Pairs |
|---|---|---|
| $\omega_1 = (32/33)\omega_0$ | $\pm 32d_0$ | $[16\ 49]d_0$ |
| $\omega_2 = (36/37)\omega_0$ | $\pm 36d_0$ | $[12\ 49]d_0$ |
| $\omega_3 = (39/41)\omega_0$ | $\pm 39d_0$ | $[8\ 49]d_0$ |
| $\omega_4 = (40/41)\omega_0$ | $\pm 40d_0$ | $[8\ 49]d_0$ |
| $\omega_5 = (43/45)\omega_0$ | $\pm 43d_0$ | $[4\ 49]d_0$ |
| $\omega_6 = (44/45)\omega_0$ | $\pm 44d_0$ | $[4\ 49]d_0$ |
| $\omega_7 = (46/49)\omega_0$ | $\pm 46d_0$ | $[0\ 49]d_0$ |
| $\omega_8 = (47/49)\omega_0$ | $\pm 47d_0$ | $[0\ 49]d_0$ |
| $\omega_9 = (48/49)\omega_0$ | $\pm 48d_0$ | $[0\ 49]d_0$ |

**Table 3.3:** Required frequencies and sensor pairs, $M = 4$, $N = 5$.

| Frequencies | Holes | Sensor Pairs | Chosen Pairs |
|---|---|---|---|
| $\omega_1 = (24/25)\omega_0$ | $\pm 24d_0$ | $[0\ 25]d_0$, $[5\ 30]d_0$, $[10\ 35]d_0$, | $[0\ 25]d_0$ |
| $\omega_2 = (28/30)\omega_0$ | $\pm 28d_0$ | $[0\ 30]d_0$, $[5\ 35]d_0$ | $[0\ 30]d_0$ |
| $\omega_3 = (29/30)\omega_0$ | $\pm 29d_0$ | $[0\ 30]d_0$, $[5\ 35]d_0$ | $[0\ 30]d_0$ |
| $\omega_4 = (32/35)\omega_0$ | $\pm 32d_0$ | $[0\ 35]d_0$ | $[0\ 35]d_0$ |
| $\omega_5 = (33/35)\omega_0$ | $\pm 33d_0$ | $[0\ 35]d_0$ | $[0\ 35]d_0$ |
| $\omega_6 = (34/35)\omega_0$ | $\pm 34d_0$ | $[0\ 35]d_0$ | $[0\ 35]d_0$ |

**Table 3.4:** Percentage of multi-frequency sensors for different co-prime arrays.

| $M$ | $N$ | Multi-frequency sensors |
|---|---|---|
| 2 | 3 | $2/6 = 33.3\%$ |
| 3 | 4 | $3/9 = 33.3\%$ |
| 3 | 5 | $4/10 = 40.0\%$ |
| 4 | 5 | $4/12 = 33.3\%$ |
| 4 | 7 | $6/14 = 42.8\%$ |
| 5 | 7 | $6/16 = 37.5\%$ |
| 6 | 7 | $6/18 = 33.3\%$ |

*E.      Comparison with Minimum Hole Arrays*

Even though the high-resolution approach was only discussed for application to co-prime arrays and the derived closed-form expressions are only specific to extended co-prime arrays, the same approach can be applied to other non-uniform arrays with missing elements in their coarrays [35]. MHAs constitute one class of non-uniform arrays with such property. MHAs do not have closed-form expressions for their sensor locations and their DOFs. The construction of a MHA requires an exhaustive search among all possible combinations. This requirement removes the ability to provide closed-form expressions for the number of frequencies and their values in order to fill all the holes in the difference coarray of a MHA. Similarly, the maximum frequency deviation cannot be computed without examining each individual MHA configuration and the location of the holes in the difference coarray.

Several MHA configurations have been reported in the literature [9, 11, 42]. For a given number of physical sensors, different arrays can yield a minimum hole configuration. These arrays have the same aperture, same coarray aperture, and the same number of holes in the coarray; they only differ by the location of the holes in the coarray. As a result, the number of resolvable sources using a MHA under single-frequency operation might change, for a given number of sensors $N_A$, depending on which configuration is used. In a similar fashion, the values of the additional frequencies, required to fill all the holes in the coarray, change with each configuration. Table 3.5 provides a comparison between different extended co-prime array and minimum hole array configurations for both single-frequency operation (SFO) and multi-frequency operation (MFO). The different MHA configurations are listed in Table 3.6. For a given number of elements, the number of resolvable sources using the different configurations is provided. The maximum frequency deviation, required to fill all the coarray holes, is also included. It should be noted that, similar to the case of co-prime arrays, only frequencies that are smaller than the reference frequency are considered for multi-frequency operation using MHAs.

**Table 3.5:** Multi-frequency operation: Co-prime arrays vs. minimum hole arrays.

| $N_A$ | Extended CPA | | SFO | MFO | | MHA | SFO | MFO | |
|---|---|---|---|---|---|---|---|---|---|
| | $M$ | $N$ | $D_{max}$ | $\Delta\omega_{max}$ | $D_{max}$ | | $D_{max}$ | $\Delta\omega_{max}$ | $D_{max}$ |
| 6 | 2 | 3 | 7 | 11.11% | 9 | $MHA_{6,1}$ | 13 | 12.50% | 17 |
| | | | | | | $MHA_{6,2}$ | 7 | 11.11% | 17 |
| | | | | | | $MHA_{6,3}$ | 13 | 12.50% | 17 |
| | | | | | | $MHA_{6,4}$ | 9 | 9.09% | 17 |
| 8 | 2 | 5 | 11 | 7.69% | 15 | $MHA_8$ | 15 | 7.14% | 34 |
| 9 | 3 | 4 | 14 | 10.00% | 20 | $MHA_9$ | 17 | 5.26% | 44 |
| 10 | 2 | 7 | 15 | 5.88% | 21 | $MHA_{10}$ | 35 | 10.00% | 55 |
| | 3 | 5 | 17 | 8.00% | 25 | | | | |

**Table 3.6:** Minimum hole array configurations.

| Configuration | Elements Positions |
|---|---|
| $MHA_{6,1}$ | $[0, 1, 4, 10, 12, 17]d_0$ |
| $MHA_{6,2}$ | $[0, 1, 4, 10, 15, 17]d_0$ |
| $MHA_{6,3}$ | $[0, 1, 8, 11, 13, 17]d_0$ |
| $MHA_{6,4}$ | $[0, 1, 8, 12, 14, 17]d_0$ |
| $MHA_8$ | $[0, 1, 4, 9, 15, 22, 32, 34]d_0$ |
| $MHA_9$ | $[0, 1, 5, 12, 25, 27, 35, 41, 44]d_0$ |
| $MHA_{10}$ | $[0, 1, 6, 10, 23, 26, 34, 41, 53, 55]d_0$ |

### 3.1.3. Numerical Results

In this section, DOA estimation results based on the MUSIC algorithm using multi-frequency co-prime arrays are presented. Both proportional and non-proportional source spectra cases are considered and performance comparison with single-frequency operation is provided. The contiguous part of the coarray is employed by the covariance matrix augmentation approach under single frequency operation. The RMSE in all examples in this section is based on a single realization, unless stated otherwise.

## A.    Proportional Spectra

A co-prime array configuration with six physical sensors, corresponding to $M = 2$ and $N = 3$, is first considered. The difference coarray of this configuration, shown in Fig. 2.9, has two holes at $\pm 8d_0$, which can be filled using an additional frequency $\omega_1 = (8/9)\omega_0$. Nine sources with proportional spectra, where $\sigma_d^2(\omega_1) = \sigma_d^2(\omega_0)$ for $d = 0, 1, \ldots, 8$ are considered. The sources are uniformly spaced between –0.95 and 0.95 in the reduced angular coordinate $\sin(\theta)$. A total number of 2,000 snapshots are used and the SNR is set to 0 dB for both frequencies. The estimated MUSIC spectrum, where only the reference frequency $\omega_0$ is used, is provided in Fig. 3.6. The elements in the covariance matrix corresponding to the holes in the difference coarray have been filled with zeros. This is equivalent to the case where the sources have zero powers at the additional frequency. The vertical lines in the figure indicate the true DOAs of the sources. Clearly, the single frequency approach fails to correctly estimate the DOAs of most of the sources. The RMSE is found to be 2.66°. Fig. 3.7 depicts the estimated MUSIC spectrum using the dual-frequency approach. The DOAs of all sources have been correctly estimated. In this case, the RMSE of the DOA estimates is equal to 0.17°.

It should be noted, that techniques, other than MUSIC, can also be applied to the augmented covariance matrix that is obtained from the observations at multiple frequencies. This is validated by applying root-MUSIC to the augmented covariance matrix of this example. The following table shows the estimated DOAs of the sources using root-MUSIC. Clearly, root-MUSIC is successful in estimating all DOAs and the corresponding RMSE is found to be 0.169°.

**Table 3.7:** Estimated DOAs using root-MUSIC

| $\theta$ | −71.81° | −45.44° | −28.36° | −13.74° | 0° | 13.74° | 28.36° | 45.44° | 71.81° |
|---|---|---|---|---|---|---|---|---|---|
| $\hat{\theta}$ | −71.39° | −45.48° | −28.27° | −13.74° | 0.02° | 13.55° | 28.52° | 45.44° | 71.70° |

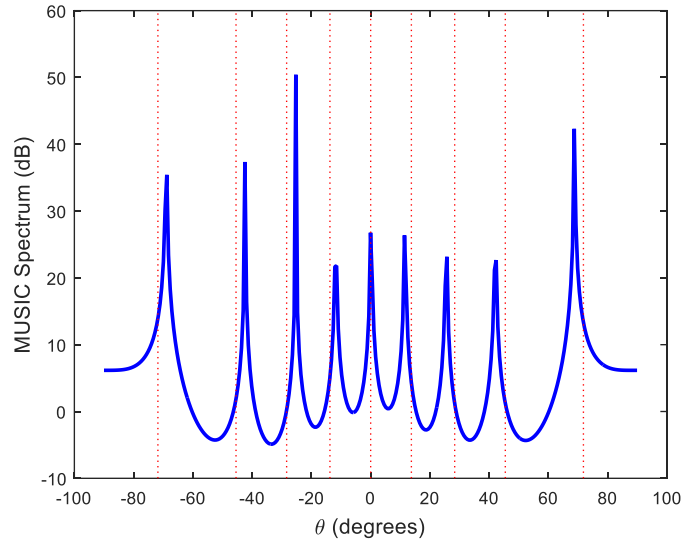

**Figure 3.6:** MUSIC spectrum using single frequency, $D = 9$ sources with proportional spectra.

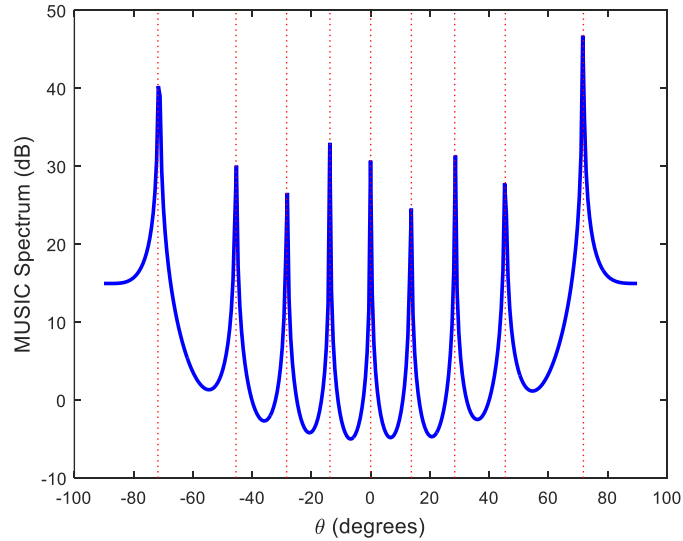

**Figure 3.7:** MUSIC spectrum using two frequencies, $D = 9$ sources with proportional spectra.

In the second example, a co-prime configuration with $M = 5$ and $N = 7$ is considered. The seven sensors of the first ULA are positioned at $[0, 5, 10, 15, 20, 25, 30]d_0$, and the second ULA has ten elements with positions $[0, 7, 14, 21, 28, 35, 42, 49, 56, 63]d_0$. The corresponding coarray extends from $-63d_0$ to $63d_0$ and has a total of 24 holes. The uniform portion of the coarray only extends from $-39d_0$ to $39d_0$. Thus, the single frequency operation can resolve a maximum of 39 sources. One additional frequency $\omega_1 = (40/41)\omega_0$ is first used to fill the holes at $\pm 40d_0$ in the coarray. As a result, the uniform part of the coarray now includes the lags from $-44d_0$ to $44d_0$, thereby increasing the maximum number of resolvable sources from 39 to 44. 44 sources with $\sin(\theta_d)$ uniformly distributed between –0.97 and 0.97 are considered. The sources are assumed to have identical power spectra at the two frequencies. A total of 2,000 snapshots are considered and the SNR is set to 0 dB for both frequencies. Fig. 3.8 shows the estimated spatial spectrum, wherein the DOAs of all 44 sources have been accurately estimated. The RMSE is determined to be $0.11°$ in this case. Next, 12 additional frequencies are employed to fill all 24 holes in the coarray. The additional frequencies and the corresponding holes they fill are listed in Table 3.8. It should be noted that the holes could have also been filled using only six additional frequencies. These frequencies are $\omega_1 = 5\omega_0$, $\omega_2 = 2\omega_0$ $\omega_3 = (47/49)\omega_0$, $\omega_4 = 3\omega_0$, $(\omega_5 = 59/63)\omega_0$, and $\omega_6 = (61/63)\omega_0$. However, this choice of frequencies results in a maximum frequency separation of $4\omega_0$, compared to $0.064\omega_0$ for the set of frequencies in Table 3.8. Fig. 3.9 shows the estimated spatial spectrum corresponding to 63 sources with $\sin(\theta_d)$ uniformly distributed between –0.97 and 0.97 and equal power spectra at the 12 frequencies. The SNR and the number of snapshots are taken to be the same as for Fig. 3.8. Again, the multi-frequency approach has estimated all sources accurately and the RMSE is $0.12°$.

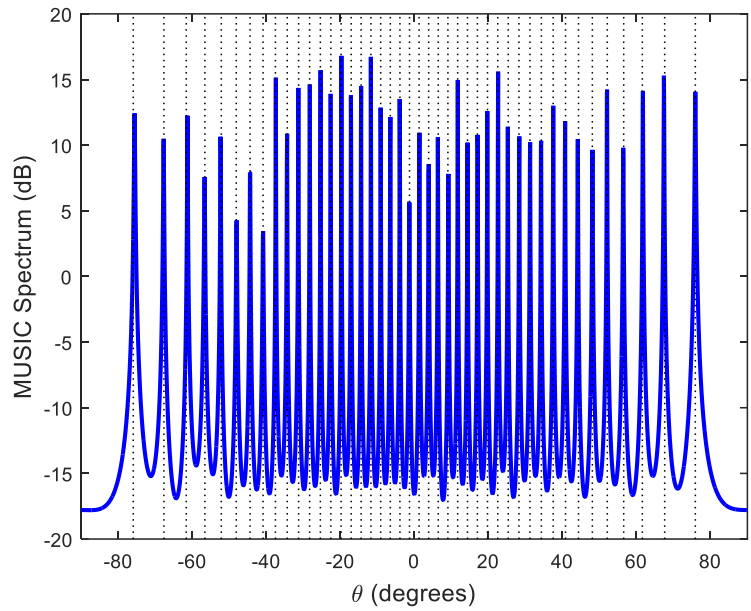

**Figure 3.8:** MUSIC spectrum with dual frequencies, $D = 44$ sources with proportional spectra.

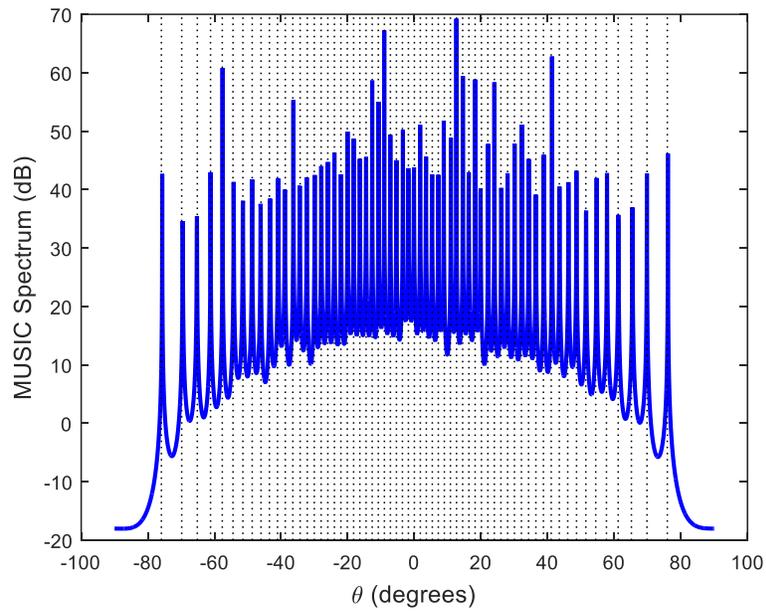

**Figure 3.9:** MUSIC spectrum with multiple frequencies, $D = 63$ sources with proportional spectra.

**Table 3.8:** Additional frequencies and corresponding holes, $M = 5$, $N = 7$.

| Frequency | Holes | Frequency | Holes |
|---|---|---|---|
| $\omega_1 = (40/41)\omega_0$ | $\pm 40d_0$ | $\omega_7 = (55/56)\omega_0$ | $\pm 55d_0$ |
| $\omega_2 = (45/46)\omega_0$ | $\pm 45d_0$ | $\omega_8 = (57/58)\omega_0$ | $\pm 57d_0$ |
| $\omega_3 = (47/48)\omega_0$ | $\pm 47d_0$ | $\omega_9 = (59/63)\omega_0$ | $\pm 59d_0$ |
| $\omega_4 = (50/51)\omega_0$ | $\pm 50d_0$ | $\omega_{10} = (60/63)\omega_0$ | $\pm 60d_0$ |
| $\omega_5 = (52/53)\omega_0$ | $\pm 52d_0$ | $\omega_{11} = (61/63)\omega_0$ | $\pm 61d_0$ |
| $\omega_6 = (54/56)\omega_0$ | $\pm 54d_0$ | $\omega_{12} = (62/63)\omega_0$ | $\pm 62d_0$ |

*B.    Non-proportional Spectra*

The DOA estimation performance of the multi-frequency co-prime arrays is evaluated when the condition of proportional source spectra is violated. In the first example, the same array and source configuration as in the first example in the previous section with $M = 2$ and $N = 3$ are considered. However, the nine sources are now assumed to have non-proportional spectra at $\omega_0$ and $\omega_1 = (8/9)\omega_0$. More specifically, the source powers at $\omega_0$ are assumed to be identical and equal to unity, whereas the source powers associated with $\omega_1$ are assumed to independently follow a truncated Gaussian distribution with a mean of 5.5 and a common variance. Two different values of 2.25 and 5.06 are considered for the variance. The variance controls the degree of non-proportionality. A higher variance increases the degree of non-proportionality of the source spectra, whereas a lower variance results in smaller variations in the source powers. Fig. 3.10 depicts the RMSE as a function of the variance and the SNR, averaged over 2,000 Monte Carlo runs. For comparison, the RMSE corresponding to both single-frequency operation and dual-frequency operation for the case when the sources have proportional spectra are also included. As expected, the single-frequency approach, wherein the elements of the virtual covariance matrix corresponding to the holes in the coarray are filled with zeros, provides the worst performance. Further, the RMSE corresponding to the multi-frequency approach for non-proportional spectra increases with increasing variance. This results in a degradation of the estimation performance. Finally, the multi-frequency approach works best when the spectra are proportional and the SNR is higher.

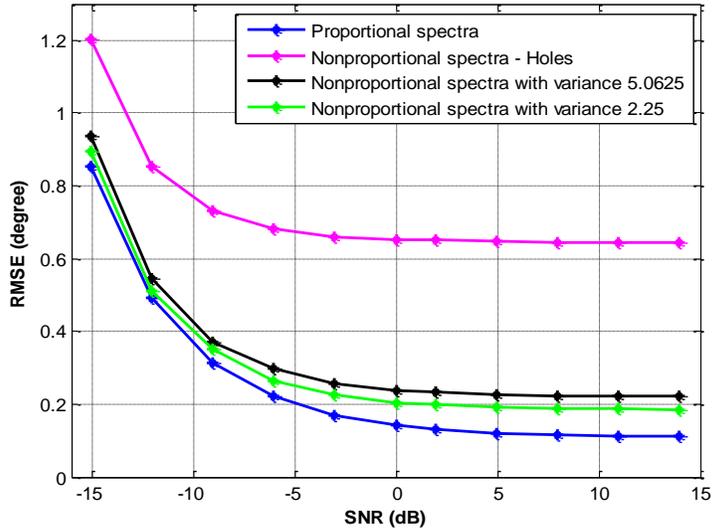

**Figure 3.10:** RMSE vs. SNR for $M = 2$, $N = 3$, $D = 9$.

In the following example, the performance of the multi-frequency approach is compared to single-frequency DOA estimation as a function of the assumed model order. The same array configuration with $M = 2$ and $N = 3$ is used. Two cases are considered in this example. The first case deals with sources with proportional spectra, while the second considers sources with non-proportional spectra. For the non-proportional case, the source powers associated with $\omega_0$ are assumed to be identical and equal to unity, and the source powers associated with $\omega_1$ follow a truncated Gaussian distribution with a mean of 5.5 and a variance 2. In both cases, the actual number of sources is set to four, and the assumed model order is varied between four and seven. 1,000 Monte Carlo are considered in this example. Fig. 3.11 shows the RMSE, averaged over 1,000 Monte Carlo runs, as a function of the assumed model order for both cases. In computing the RMSE, only the detected peaks that are closest to the actual source directions are considered. From Fig. 3.11, it can be noticed that the performance of the single-frequency approach is not affected by the non-proportionality of the source spectra, as expected. On the other hand, the multi-frequency DOA estimation exhibits superior performance for sources with proportional spectra compared to those with non-proportional spectra. Further, the multi-frequency approach is less sensitive to errors in model order as compared to the single-frequency approach.

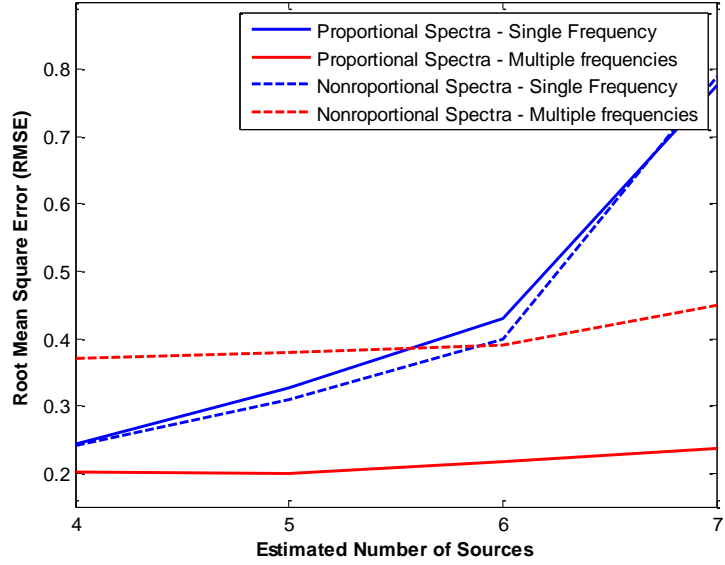

**Figure 3.11:** RMSE vs. assumed model order for $M = 2$, $N = 3$, $D = 4$.

The effect of the degree of non-proportionality on DOA estimation performance is next examined for the co-prime configuration of the second example in Section 3.1.3.A with $M = 5$ and $N = 7$ under both dual and multi-frequency operation. Again, the source powers at $\omega_0$ are assumed to be all equal to unity, whereas the source powers at additional frequencies follow a truncated Gaussian distribution with a mean of 5.5 and a common variance. Fig. 3.12 provides the RMSE, averaged over 2,000 Monte Carlo runs, as a function of SNR and variance under the dual-frequency operation for 44 sources. Similar observations to those in Fig. 3.10 can be made in this case as well. However, two differences can be noticed by comparing the RMSE plots in

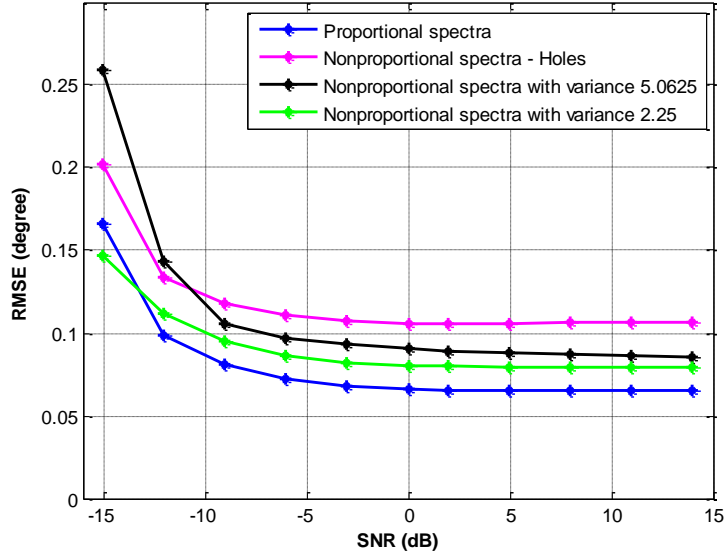

**Figure 3.12:** RMSE vs. SNR for $M = 5$, $N = 7$, $D = 44$

Figs. 3.10 and 3.12. First, the RMSE takes on lower values for all considered DOA estimation methods and variances for the co-prime configuration with $M = 5$ and $N = 7$. Second, the difference in performance between the single and dual frequency operations for the non-proportional spectra cases is much smaller at higher SNR values in this example. This is due to the fact that the ratio of the number of missing elements to the total number of elements in the filled part of the difference coarray is smaller in this example. This results in a smaller percentage of elements in the virtual covariance matrix to come from a different frequency or be filled with zeros for single frequency operation. The RMSE plots for the multi-frequency operation to fill all 24 holes are provided in Fig. 3.13, which corresponds to 60 sources with $\sin(\theta_d)$ uniformly distributed between –0.97 and 0.97. The performance difference between multi-frequency operation for sources with non-proportional spectra and those with proportional spectra is even less noticeable in this case, though the RMSE values themselves are slightly higher for high SNR. Also, the single-frequency operation exhibits a higher RMSE since a higher percentage of the virtual covariance matrix elements now have a zero value compared to that for Fig. 3.12.

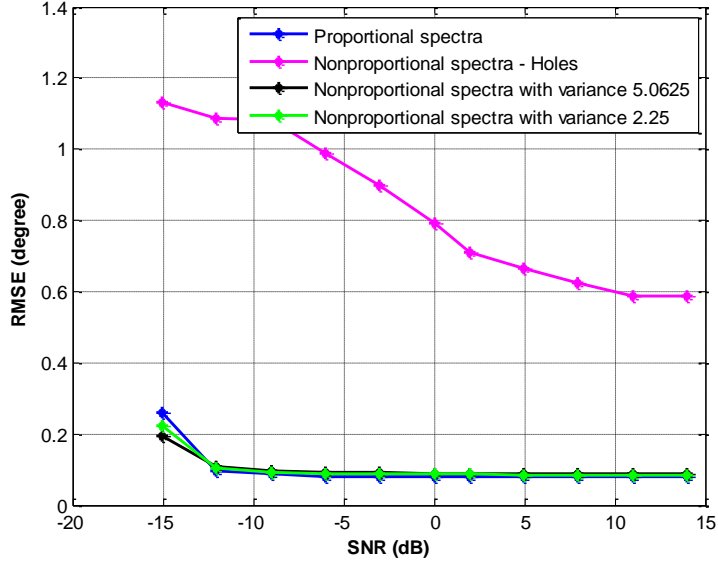

**Figure 3.13:** RMSE vs. SNR for $M = 5$, $N = 7$, $D = 60$.

The final example in this section examines the estimation performance for varying degree of non-proportionality of the source spectra for different values of $M$ and $N$ with the SNR fixed at 0 dB. Both dual-frequency operation for filling only the first hole pair and multi-frequency operation for filling all the holes are considered for each co-prime configuration. For each case, the maximum number of resolvable sources was used. A total of 2,000 Monte Carlo runs are considered in this example. The source powers associated with the reference frequency $\omega_0$ are identical and equal to unity. For the additional frequencies, the source powers follow a truncated Gaussian distribution with a mean of 5.5 and a common variance. The corresponding RMSE plots as a function of the variance of the source powers are depicted in Fig. 3.14. In order to have a fair comparison among co-prime arrays of different sizes, each RMSE plot is normalized by the Cramer Rao Bound (CRB) of an equivalent ULA with total number of elements equal to the number of contiguous nonnegative lags in the corresponding filled difference coarray. By examining Fig. 3.14, the following observations are in order. First, as expected, a decrease in the variance of the sources spectra results in a reduced estimation error. Second, by comparing the results of dual and multiple frequency operation for fixed $M$ and $N$, it can be noticed that, in general, the normalized RMSE error is smaller for the case when more than one additional frequencies are used.

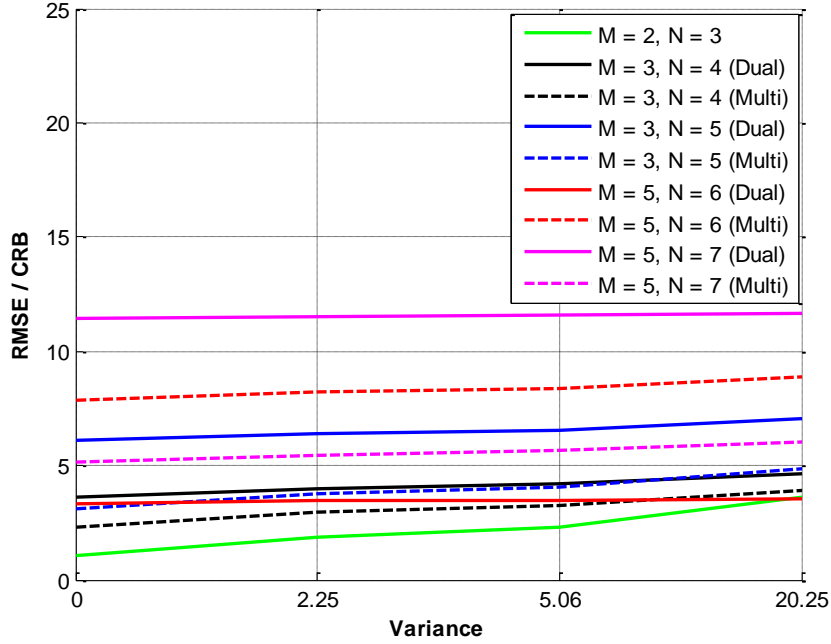

**Figure 3.14:** RMSE/CRB vs. variance, SNR = 0 dB.

## C. *Co-Prime Arrays vs. Minimum Hole Arrays*

In this section, the performance of a co-prime array is compared with the performance of a minimum hole array having the same number of physical sensors.

First, a six-element co-prime array with $M = 2$ and $N = 3$ is considered. For comparison, a six-element MHA with elements at $[0, 1, 4, 10, 12, 17]d_0$ is used. Fig. 3.15 shows the two arrays along with their corresponding difference coarrays. Fig. 3.16 compares the performance of these arrays for different source separations and different source powers. Two sources with proportional spectra are considered in this example. The direction of the first source is fixed to $u_1 = \sin \theta_1 = 0$, and the direction of the second source is varied. The SNR of the first source is varied between 0 dB and 30 dB, and the second source's SNR is set to 0 dB. For each set of parameters, the number of Monte Carlo runs is set to 100 and the average RMSE is calculated. The total number of snapshots is set to 1,000. By examining Fig. 3.16, it is clear that the MHA consistently outperforms the co-prime array even before filling the holes. In addition, applying the proposed technique results in improving the estimation performance for both arrays. For a fixed $\Delta$SNR and a large source separation $\Delta u$, the performance improvement is not noticeable.

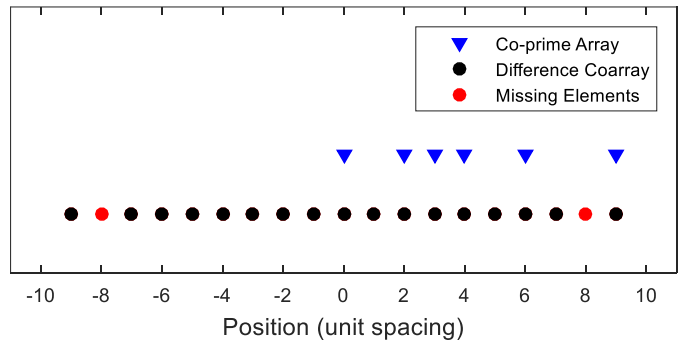

(a)

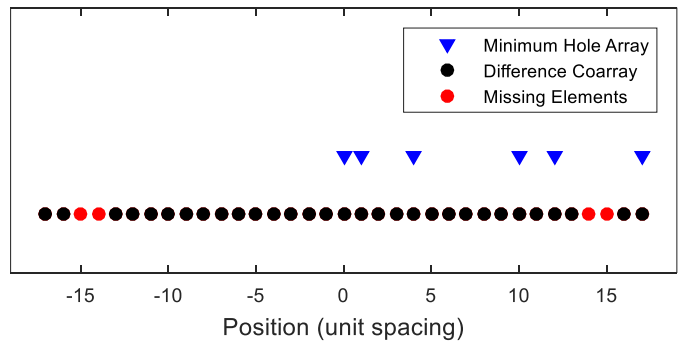

(b)

**Figure 3.15:** a) Extended co-prime array ($M = 2$, $N = 3$): physical array and coarray (b) six-element MHA ([0, 1, 4, 10, 12, 17]$d_0$): physical array and coarray.

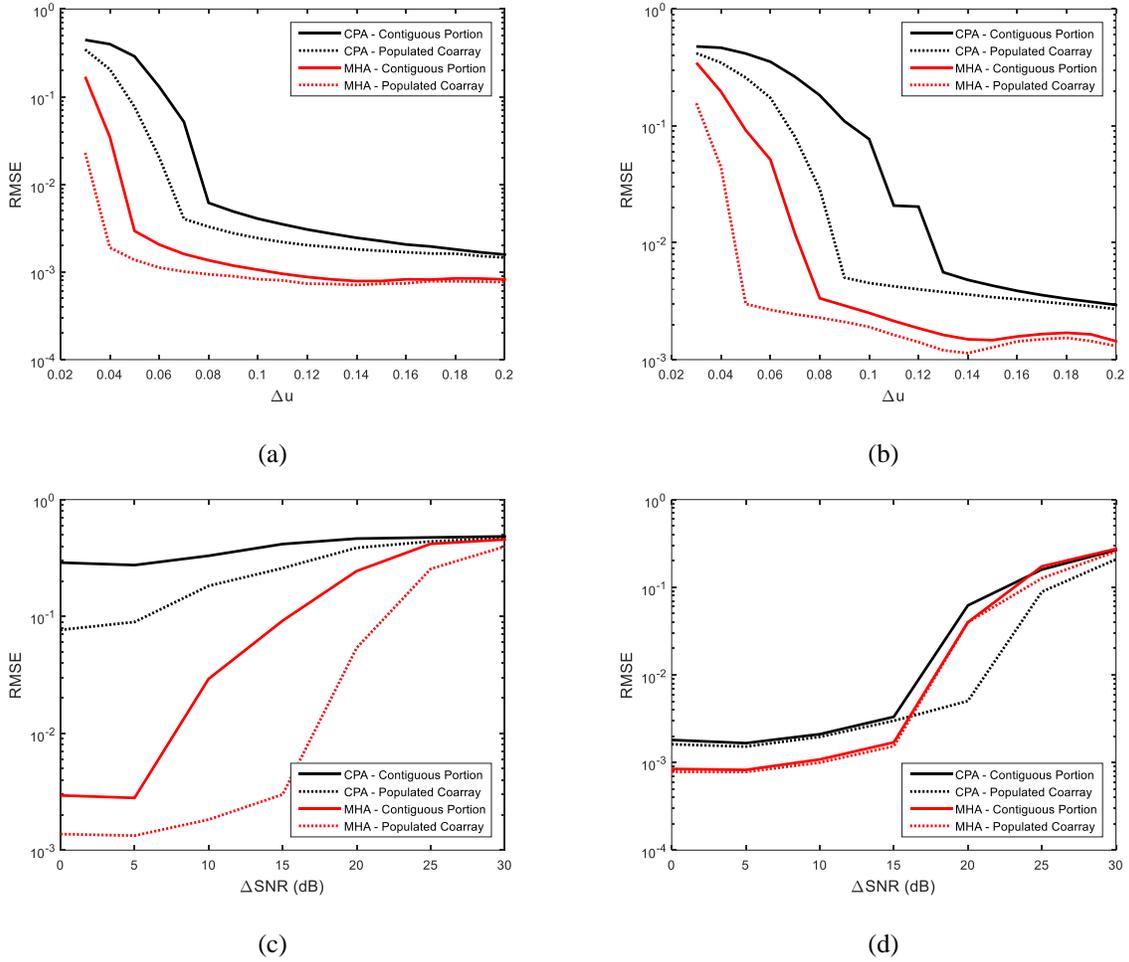

(a)

(b)

(c)

(d)

**Figure 3.16:** Extended co-prime array ($M = 2$, $N = 3$) vs. six-element MHA ([0, 1, 4, 10, 12, 17]$d_0$), $D = 2$ (a) $\Delta$SNR = 0dB: RMSE vs. $\Delta u$ (b) $\Delta$SNR = 15dB: RMSE vs. $\Delta u$ (c) $\Delta u$ = 0.05: RMSE vs. $\Delta$SNR (d) $\Delta u$ = 0.18: RMSE vs. $\Delta$SNR.

In the following example, the same six-element co-prime array is used; however, an MHA with a different configuration is considered. Fig. 3.17 shows the MHA configuration and the corresponding difference coarray. The coarray extends between $-17d_0$ and $17d_0$ and is filled between $-7d_0$ and $7d_0$. The contiguous part is similar to that of the co-prime array. Fig. 3.18 shows the different RMSE plots. The same simulation parameters as in the previous example are used. One notable difference between the results of Fig. 3.18 and those of Fig. 3.16 is that the co-prime array consistently outperforms the MHA before applying the proposed method. This is expected since the difference coarrays have the same contiguous part, and the difference coarray of the co-prime array has more redundancies than that of the MHA as shown in Fig. 3.19. Since the MHA is based on the concept of having no redundancies in the coarray, each coarray lag can only obtained with one pair of sensors. As a result, the covariance matrix estimates in the co-prime array have better estimates than those in the MHA and a better performance is expected from the co-prime array.

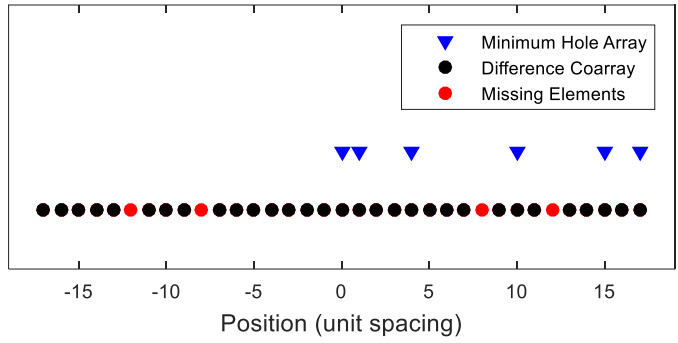

**Figure 3.17:** Six-element MHA ([0, 1, 4, 10, 15, 17]d₀): physical array and coarray.

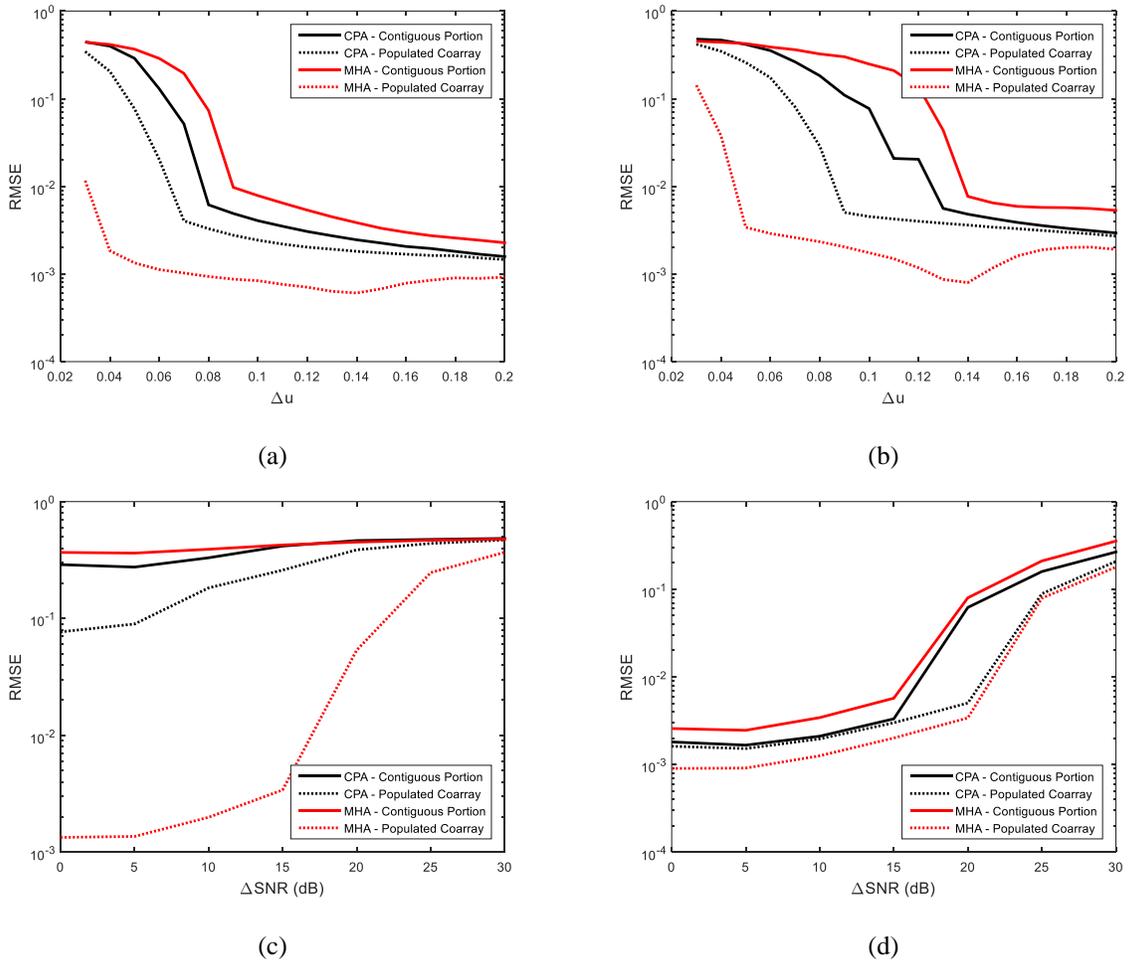

(a)

(b)

(c)

(d)

**Figure 3.18:** Extended co-prime array ($M = 2$, $N = 3$) vs. six-element MHA ([0, 1, 4, 10, 15, 17]d₀), $D = 2$ (a) $\Delta$SNR = 0dB: RMSE vs. $\Delta u$ (b) $\Delta$SNR = 15dB: RMSE vs. $\Delta u$ (c) $\Delta u = 0.05$: RMSE vs. $\Delta$SNR (d) $\Delta u = 0.18$: RMSE vs. $\Delta$SNR.

### 3.1.4. Contributions

The main contributions of this research are listed below.

1) Application of the multi-frequency approach to co-prime arrays in order to fill the holes in their difference coarray.

2) Exploiting the specific structure of the coarray to determine the number and values of the additional frequencies required for recovering the missing lags.

3) Determining closed-form expressions for the additional frequencies which are 'best' in the sense of minimum operational bandwidth requirements.

4) Exploiting redundancy in the coarray to reduce the system hardware complexity by employing a minimum number of antennas at additional frequencies.

5) Investigating the effects of noise and deviation from the proportional source spectra constraint on the DOA estimation performance.

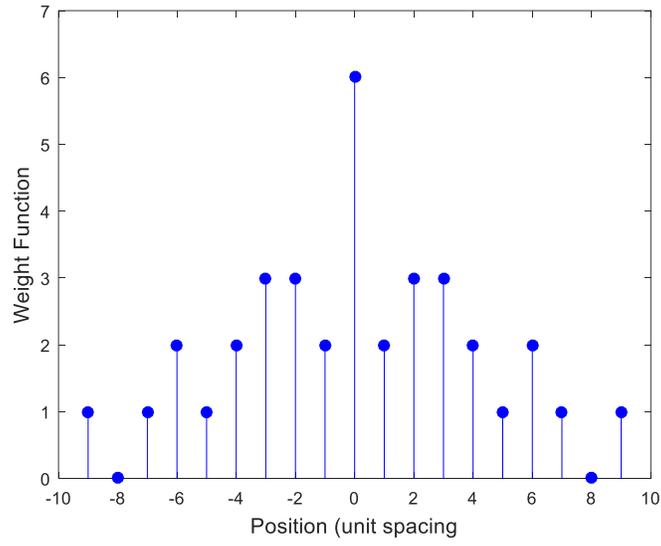

(a)

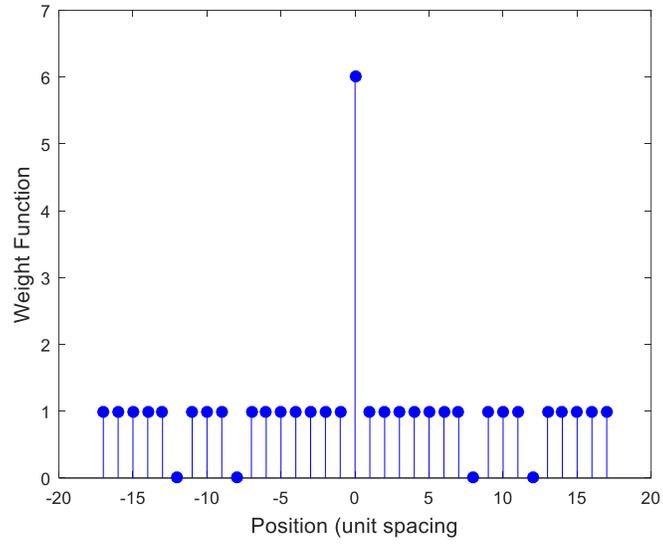

(b)

**Figure 3.19:** (a) Co-prime array weight function (b) MHA ($[0, 1, 4, 10, 15, 17]d_0$) weight function.

## 3.2. Sparsity-Based Approach

In the high-resolution approach, multiple frequencies were employed to exploit all of the DOFs of co-prime arrays, thus, increasing the number of resolvable sources. Measurements made at carefully chosen additional frequencies were used to fill in the missing elements in the difference coarray. In so doing, the filled part of the difference coarray is extended, which in turn, increases the maximum number of sources resolved by high-resolution DOA estimation techniques. If the entire array is operated at all frequencies, only a small portion of the additional measurements at frequencies other than the reference frequency are used; the rest are discarded.

In the sparsity-based, sparse reconstruction is considered to make use of the full measurement set corresponding to the multi-frequency operation for DOA estimation with co-prime arrays. This enhances the DOFs beyond those offered by single-frequency operation due to the additional virtual elements generated in the coarray under multi-frequency operation. For sources with proportional spectra, the observations at the different frequencies are cast as a single measurement vector model, which corresponds to a virtual array whose element positions are given by the union set of the difference coarrays corresponding to the multiple operational frequencies. Sparse reconstruction can then be applied for estimating the directions of signal arrivals. For the case where the sources have non-proportional spectra, the source signal vectors corresponding to the different frequencies have a common support, as the sources maintain their DOA even if their power varies with frequency. The common structure property of the sparse source vectors suggests the application of a group sparse reconstruction. It is noted that sparse recovery was previously applied for DOA estimation with co-prime arrays in [29, 43]; however, it was limited to single-frequency operation and did not consider enhancement of the DOFs of co-prime arrays through multi-frequency operation.

Performance evaluation of the proposed sparsity-based methods is conducted using numerical simulations. Three different cases for DOA estimation using sparse reconstruction at multiple frequencies are considered. In the first case, all sources are assumed to have the same bandwidth and all sensors operate at the same multiple frequencies. The second and third cases violate the above assumption with a subset of sensors only operating at multiple frequencies and the sources having nonidentical bandwidth but overlapping spectra.

The remainder of this section is organized as follows. In Section 3.2.1, the multi-frequency signal model for co-prime arrays is presented. In Section 3.2.2, the sparse reconstruction based DOA estimation for multi-frequency co-prime arrays under proportional spectra is discussed. The case of sources with non-proportional spectra is considered in Section 3.2.3 and the group sparsity based reconstruction is presented. The performance of the proposed methods is evaluated in Section 3.2.4 through numerical simulations, and Section 3.2.5 summarizes the contributions of this research.

### 3.2.1. Signal Model

It should be noted that, in this approach, it is not required for the reference frequency $\omega_0$ to be one of the $Q$ operational frequencies. If it is included in the operational frequency set, the corresponding $\alpha_q$ assumes a unit value. Starting with the received data vector of (3.2), the covariance matrix at frequency $\omega_q$ can be vectorized following (2.24) as

$$\mathbf{z}(\omega_q) = vec\left(\mathbf{R}_{xx}(\omega_q)\right) = \widetilde{\mathbf{A}}(\omega_q)\mathbf{p}(\omega_q) + \sigma_n^2(\omega_q)\widetilde{\mathbf{i}}, \tag{3.21}$$

where $\widetilde{\mathbf{A}}(\omega_q) = \mathbf{A}^*(\omega_q) \odot \mathbf{A}(\omega_q)$, $\mathbf{p}(\omega_q)$ is the sources powers vector at $\omega_q$, $\mathbf{p}(\omega_q) = \left[\sigma_1^2(\omega_q), \sigma_2^2(\omega_q), \dots, \sigma_D^2(\omega_q)\right]^T$, and $\widetilde{\mathbf{i}}$ is the vectorized form of $\mathbf{I}$. The vector $\mathbf{z}(\omega_q)$ behaves as the received signal vector at a longer virtual array with sensor positions given by the difference coarray at $\omega_q$ of the physical array.

As previously mentioned, for multi-frequency DOA estimation, the normalized covariance matrices at the $Q$ operational frequencies are employed. Therefore, the received vector $\mathbf{z}(\omega_q)$ of (3.21) is replaced by the vectorized form of the normalized covariance matrix

$$\bar{\mathbf{z}}(\omega_q) = vec\left(\bar{\mathbf{R}}_{xx}(\omega_q)\right) = \widetilde{\mathbf{A}}(\omega_q)\bar{\mathbf{p}}(\omega_q) + \bar{\sigma}_n^2(\omega_q)\bar{\mathbf{i}}, \qquad (3.22)$$

where $\bar{\mathbf{p}}(\omega_q) = \left[\bar{\sigma}_1^2(\omega_q), \bar{\sigma}_2^2(\omega_q), \dots, \bar{\sigma}_D^2(\omega_q)\right]^T$.

The measurement vectors $\bar{\mathbf{z}}(\omega_q)$, $q = 1, 2, \dots, Q$, can be combined to establish an appropriate multi-frequency linear model that permits DOA estimation within the sparse reconstruction framework. In the following sections, two cases of normalized source spectra are distinguished. In the first case, the normalized power of each source to be independent of frequency as in (3.15), whereas the normalized source powers are allowed to vary with frequency in the second case.

### 3.2.2. Sparsity-Based DOA Estimation Under Proportional Spectra

The angular region of interest into a finite set of $K$ ($K \gg D$) grid points, $\{\theta_{g_1}, \theta_{g_2}, \dots, \theta_{g_K}\}$, with $\theta_{g_1}$ and $\theta_{g_K}$ being the limits of the search space. Then, (3.22) can be rewritten as

$$\bar{\mathbf{z}}(\omega_q) = \widetilde{\mathbf{A}}^g(\omega_q)\bar{\mathbf{p}}^g(\omega_q) + \bar{\sigma}_n^2(\omega_q)\bar{\mathbf{i}}, \qquad (3.23)$$

where the columns of the $(2M + N - 1)^2 \times K$ matrix $\widetilde{\mathbf{A}}^g(\omega_q)$ are the steering vectors at $\omega_q$ corresponding to the defined angles in the grid. The vector $\bar{\mathbf{p}}^g(\omega_q)$ is a $D$-sparse vector whose support corresponds to the source directions with the nonzero values equal to the normalized source powers. Under proportional source spectra, the source vector $\bar{\mathbf{p}}(\omega_q)$ is no longer a function of $\omega_q$, i.e., $\bar{\mathbf{p}}(\omega_q) = \bar{\mathbf{p}} = [\bar{\sigma}_1^2, \bar{\sigma}_2^2, \ldots, \bar{\sigma}_D^2]^T$ for all $q$, which implies that vector $\bar{\mathbf{p}}^g(\omega_q) = \bar{\mathbf{p}}^g$ for all $q$. As such, the measurement vectors $\bar{\mathbf{z}}(\omega_q)$ at the $Q$ operating frequencies can be stacked to form a single $Q(2M + N - 1)^2 \times 1$ vector,

$$\bar{\mathbf{z}}^g = \widetilde{\mathbf{B}}^g \bar{\mathbf{p}}^g + \check{\mathbf{i}}^g, \tag{3.24}$$

where $\bar{\mathbf{z}}^g = \left[\bar{\mathbf{z}}(\omega_1)^T, \bar{\mathbf{z}}(\omega_2)^T, \ldots, \bar{\mathbf{z}}(\omega_Q)^T\right]^T$, $\check{\mathbf{i}}^g = \left[\bar{\sigma}_n^2(\omega_1)\check{\mathbf{i}}^T, \bar{\sigma}_n^2(\omega_2)\check{\mathbf{i}}^T, \ldots, \bar{\sigma}_n^2(\omega_Q)\check{\mathbf{i}}^T\right]^T$, and the dictionary $\widetilde{\mathbf{B}}^g = \left[\left[\widetilde{\mathbf{A}}^g(\omega_1)\right]^T, \left[\widetilde{\mathbf{A}}^g(\omega_2)\right]^T, \ldots, \left[\widetilde{\mathbf{A}}^g(\omega_Q)\right]^T\right]^T$. The measurement vector is equivalent to that of a virtual array, whose element positions are given by the combined difference coarrays at the $Q$ frequencies, i.e.,

$$S_g = \{\alpha_1 S_0, \alpha_2 S_0, \ldots, \alpha_Q S_0\}, \tag{3.25}$$

where $S_0$ is the set of difference coarray elements at $\omega_0$. It is noted that in the case of overlapping points in the $Q$ coarrays, an averaged value of the multiple measurements that correspond to the same coarray location can be used. This results in a reduction in the dimensionality of $\bar{\mathbf{z}}^g$. More specifically, the length of $\bar{\mathbf{z}}^g$ becomes equal to the total number of unique lags in the combined difference coarray, which is given by

$$S_g = \bigcup_{q=1}^{Q} \alpha_q S_0, \tag{3.26}$$

The dictionary matrix and the noise vector would be changed accordingly.

It should be noted that not all the physical sensors must operate at all $Q$ frequencies. Situations may arise due to cost and hardware restrictions that only a few sensors can accommodate a diverse set of frequencies. The overall difference coarray is still the union of coarrays at the individual frequencies. However, the difference coarray at each frequency may no longer be a scaled version of the difference coarray at the reference frequency.

Given the model in (3.24), DOA estimation proceeds in terms of sparse signal reconstruction by solving the following constrained minimization problem

$$\hat{\mathbf{p}}^g = \arg\min_{\overline{\mathbf{p}}^g} \|\overline{\mathbf{p}}^g\|_1 \text{ subject to } \left\|\overline{\mathbf{z}}^g - \widetilde{\mathbf{B}}^g\overline{\mathbf{p}}^g\right\|_2 < \epsilon \text{ and } \overline{\mathbf{p}}^g \succcurlyeq \mathbf{0}, \tag{3.27}$$

where $\epsilon$ is a user-specified bound which depends on the noise variance. The constraint $\overline{\mathbf{p}}^g \succcurlyeq \mathbf{0}$ forces the search space to be limited to nonnegative values [43]. This is due to the fact that the nonzero elements of $\overline{\mathbf{p}}^g$ correspond to the normalized source powers which are always positive. This constraint accelerates the convergence of the solution by reducing the search space. Various techniques can be used to solve the constrained minimization problem in (3.27), examples being Least absolute shrinkage and selection operator (Lasso), Orthogonal Matching Pursuit (OMP), and Compressive Sampling Matched Pursuit (CoSaMP) [44-46]. In this research, Lasso is used. Lasso solves an equivalent problem to (3.26),

$$\hat{\mathbf{p}}^g = \arg\min_{\overline{\mathbf{p}}^g} \left[\frac{1}{2}\left\|\overline{\mathbf{z}}^g - \widetilde{\mathbf{B}}^g\overline{\mathbf{p}}^g\right\|_2 + \lambda_t\|\overline{\mathbf{p}}^g\|_1\right] \text{ subject to } \overline{\mathbf{p}}^g \succcurlyeq \mathbf{0}, \tag{3.28}$$

where the $\ell_2-$ norm is the least squares cost function and the $\ell_1-$ norm encourages the sparsity constraint. The regularization parameter $\lambda_t$ is used to control the weight of the sparsity constraint in the overall cost function. Increasing $\lambda_t$ results in a sparser solution at the cost of an increased least squares error. Several methods have been proposed to estimate the regularization parameter, such as the discrepancy principle [27, 47] and cross validation [44].

The maximum number of resolvable sources using the proposed method depends on the number of unique lags in the combined difference coarray. According to [48], the sparsity based minimization problem in (3.28) is guaranteed to have a unique solution under the condition $L_{u,Q} \geq 2D$, where $L_{u,Q}$ is equal to the number of independent observations or the number of unique lags in the combined difference coarray. As a result, the maximum number of resolvable sources is equal to the number of unique positive lags in the combined coarray. At the reference frequency, the difference coarray extends from $-(2M-1)Nd_0$ to $(2M-1)Nd_0$, and it has a total of $(M-1)(N-1)$ holes, which means that the number of unique lags at each frequency is equal to $(3MN+M-N)$, and the highest number of possible unique positive lags is $(3MN+M-N-1)/2$. Therefore, the maximum number of resolvable sources at each frequency is $(3MN+M-N-1)/2$. Taking into account the overlap between the lags at the different employed frequencies, the maximum number of resolvable sources with the multi-frequency technique is bounded as follows

$$\frac{(3MN+M-N-1)}{2} < D \leq Q \frac{(3MN+M-N-1)}{2} - (Q-1). \tag{3.29}$$

The term $(Q-1)$ is subtracted from the upper bound due to the unavoidable overlap between the $Q$ difference coarrays for the zero lag.

### 3.2.3. Sparsity-Based DOA Estimation Under Non-Proportional Spectra

When the source powers vary with frequency, the single measurement vector model of (3.24) is no longer applicable. However, the $D$ sources have the same directions $[\theta_1, \theta_2, ..., \theta_D]$ regardless of their power distribution with frequency. As such, the vectors $\bar{\mathbf{p}}^g(\omega_q), q = 1, 2, ..., Q$, have a common support. That is, if a certain element in, e.g., $\bar{\mathbf{p}}^g(\omega_1)$ has a nonzero value, the corresponding elements in $\bar{\mathbf{p}}^g(\omega_q)$, $q = 2, ... Q$, should be also nonzero. The common structure property suggests the application of a group sparse reconstruction. Therefore, the following group sparsity-based DOA estimation approach is proposed for the non-proportional spectra case.

The received signal vectors $\bar{\mathbf{z}}(\omega_q)$ in (3.23) corresponding to the $Q$ frequencies are stacked to form a long vector

$$\bar{\mathbf{z}}^g = \tilde{\mathbf{C}}^g \breve{\mathbf{p}}^g + \bar{\mathbf{i}}^g, \tag{3.30}$$

where $\tilde{\mathbf{C}}^g = bdiag\{\tilde{\mathbf{A}}^g(\omega_1), \tilde{\mathbf{A}}^g(\omega_2), \dots, \tilde{\mathbf{A}}^g(\omega_Q)\}$ and $\breve{\mathbf{p}}^g = \left[[\bar{\mathbf{p}}^g(\omega_1)]^T, \dots, [\bar{\mathbf{p}}^g(\omega_Q)]^T\right]^T$. The vector $\mathbfcal{X}$ is a group sparse vector where each group consists of the source powers corresponding to a specific direction at all operating frequencies. The group sparse solution is obtained by minimizing the following mixed $\ell_1 - \ell_2$ norm cost function

$$\min\left\|\bar{\mathbf{z}}^g - \tilde{\mathbf{C}}^g \breve{\mathbf{p}}^g\right\|_2 + \beta_t \|\breve{\mathbf{p}}^g\|_{2,1}, \tag{3.31}$$

where

$$\|\breve{\mathbf{p}}^g\|_{2,1} = \sum_{i=0}^{K-1} \left\|\left[[\bar{\mathbf{p}}^g(\omega_1)]_i, \dots, [\bar{\mathbf{p}}^g(\omega_Q)]_i\right]\right\|_2. \tag{3.32}$$

This means that the variables belonging to the same group are combined using the $\ell_2 - \text{norm}$, and the $\ell_1 - \text{norm}$ is then used across the groups to enforce group sparsity. Different algorithms can be utilized to perform sparse reconstruction with grouped variables. These algorithms include group Lasso and Block Orthogonal Matching Pursuit (BOMP) [49, 50], among many others. In this paper, group Lasso is considered to perform DOA estimation in the case of sources with non-proportional spectra. Further, similar to the method discussed in Section 3.2.2, a constraint can be added to force the elements of the solution vector $\mathbfcal{X}$ to be nonnegative.

It is noted that this formulation results in a smaller number of achievable DOFs compared to the case where the sources have proportional spectra. The maximum number of resolvable sources is now limited by the number of observations or unique lags at each frequency [27]. This means that up to $(3MN + M - N - 1)$ sources can be resolved.

### 3.2.4. Numerical Results

Both cases of proportional and non-proportional source spectra are considered in this section. For all of the examples in this section, an extended co-prime array configuration with six physical elements is considered with $M$ and $N$ chosen to be 2 and 3, respectively. The six sensors' positions are given by $[0, 2d_0, 3d_0, 4d_0, 6d_0, 9d_0]$. The physical array and the corresponding coarray are shown in Fig. 2.9.

In the first example, sparse signal reconstruction is applied under single frequency operation to perform DOA estimation. Since the difference coarray has eight positive lags, sparse reconstruction can be applied to resolve up to eight sources. A total of eight BPSK sources, uniformly spaced between –60° and 60°, are considered. The number of snapshots used is 1,000. Spatially and temporally white Gaussian noise is added to the observations and the SNR is set to 10 dB for all sources. The search space is discretized uniformly between –90° and 90° with a 0.2° step size, and the regularization parameter $\lambda_t$, chosen empirically, is set to 0.7 in this example. The normalized spectrum obtained using sparse signal recovery is shown in Fig. 3.20. A small bias can be noticed in the estimates, and the RMSE, computed across the angles of arrival, is found to be 1.05° for this particular run.

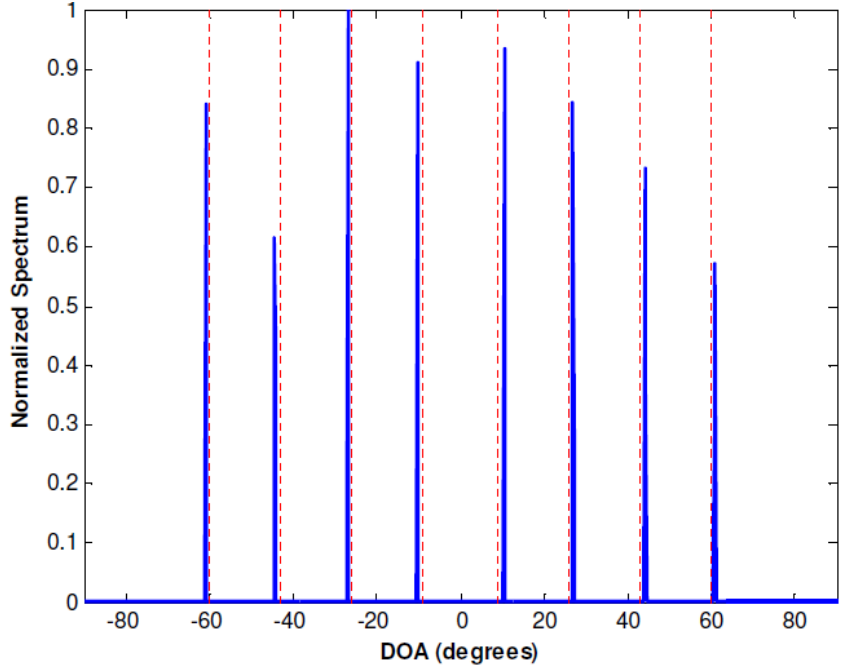

**Figure 3.20:** Single-frequency sparse reconstruction: $D = 8$ sources.

In the second example, sparse reconstruction is applied under dual-frequency operation. The physical co-prime array is now operated at both frequencies $\omega_0$ and $\omega_1 = 8/9\,\omega_0$. Sources with proportional spectra are assumed and, thus, the single measurement vector formulation of Section 3.2.2 can be used. The combined difference coarray is shown in Fig. 3.21. It has a total number of 33 unique lags, which makes it capable of resolving up to 16 sources, theoretically.

However, this number is not achievable because of the high mutual coherence of the dictionary. Since some of the virtual sensors in the combined coarray are closely separated, leading to highly correlated observations, deterioration in performance is observed if the number of sources is increased beyond eleven. Eleven BPSK sources with proportional spectra, uniformly spaced between −75° and 75°, are considered. The SNR is set to 10 dB for the sources at the two frequencies, and the total number of snapshots at each frequency is equal to 2,000. The regularization parameter $\lambda_t$ is set to 0.25 and the search space is divided into 181 bins of size 1°.

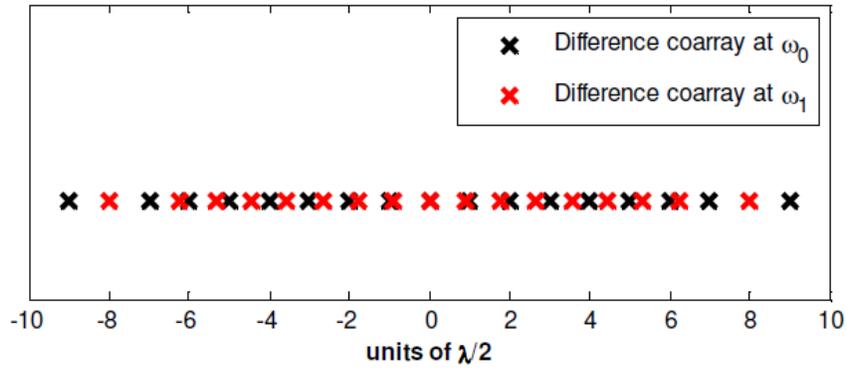

**Figure 3.21:** Dual frequency combined difference coarray, $\omega_1 = 8/9\omega_0$.

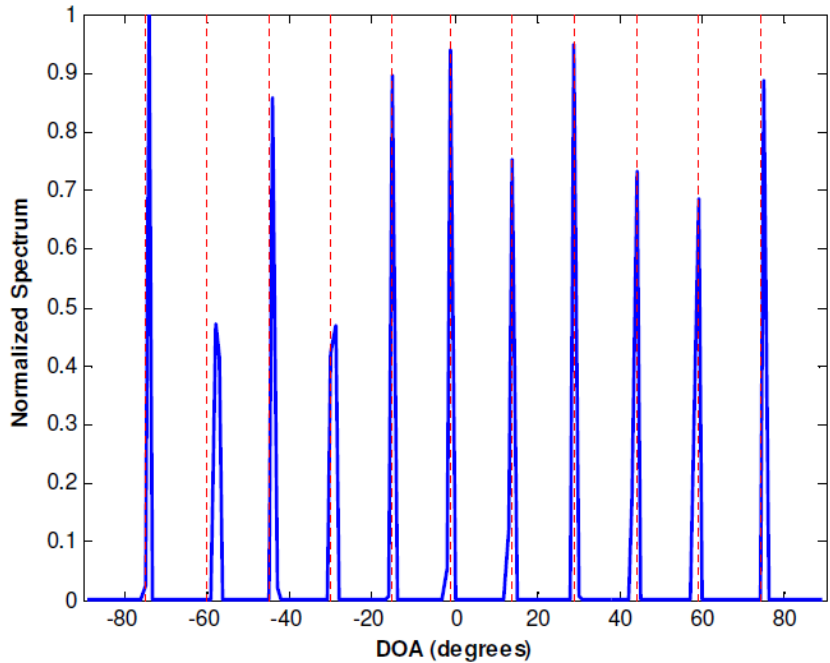

**Figure 3.22:** Dual-frequency sparse reconstruction, $D = 11$ sources.

Fig. 3.22 shows the normalized spectrum obtained using this method. It is evident that all the sources are correctly resolved. The RMSE in this example is equal to 0.84°.

A different choice of the two operational frequencies may reduce the mutual coherence, thereby permitting a larger number of sources to be estimated. For illustration, the second frequency is now set to $\omega_1 = 2\omega_0$. By choosing a frequency which is an integer multiple of $\omega_0$, the combined coarray positions are guaranteed to be integer multiples of $d_0$. As a result, the minimum separation between two consecutive coarray elements is equal to $d_0$. The combined difference coarray is shown in Fig. 3.23. The coarray has 13 unique positive lags, which means that the maximum number of resolvable sources is equal to 13. This is tested by considering 13 uniformly spaced sources between $-75°$ to $75°$. The SNR is again set to 10 dB and the number of snapshots is set to 2,000. The regularization parameter is again set to 0.25 and the search space is divided into 181 angle bins. Fig. 3.24 shows the normalized spectrum using the dual-frequency sparse reconstruction method. It is evident that all the sources are correctly estimated. The RMSE is found to be 0.26° in this case.

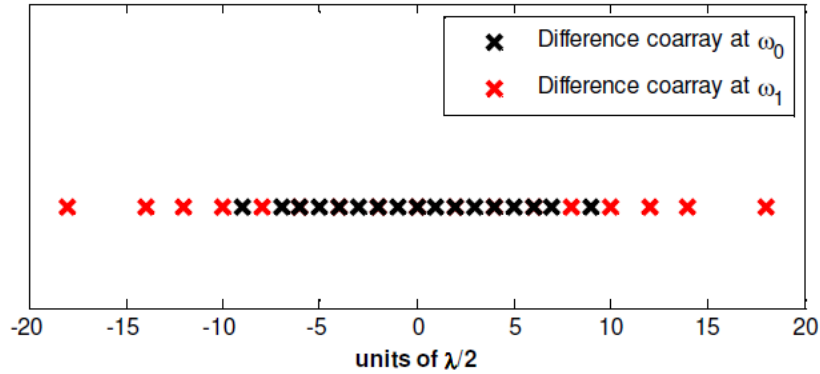

**Figure 3.23:** Dual frequency combined difference coarray, $\omega_1 = 2\omega_0$.

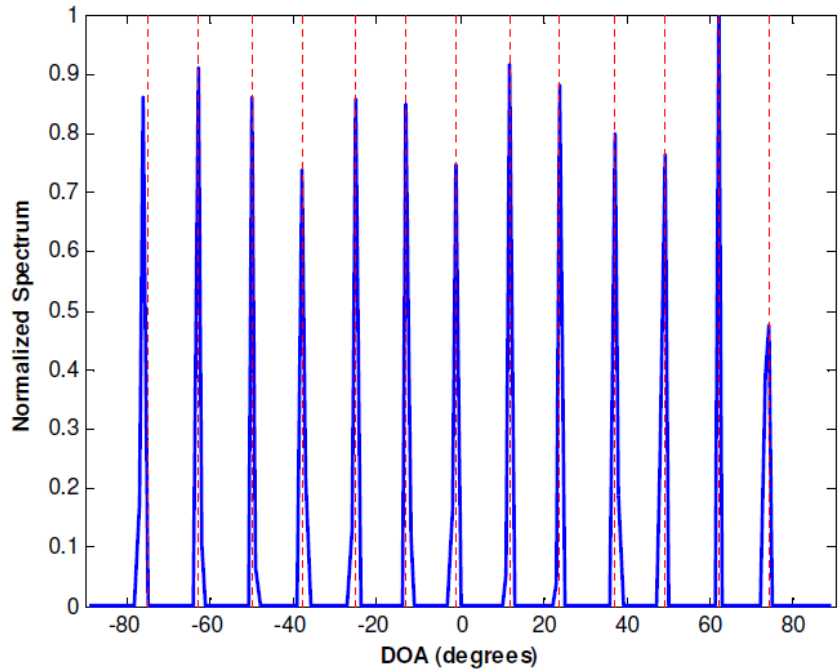

**Figure 3.24:** Dual-frequency sparse reconstruction, $D = 13$ sources.

In the following example, the entire array is operated at $\omega_0$, but only the elements at $[2d_0 \ 4d_0 \ 9d_0]$ also operate at the second frequency $\omega_1 = 2\omega_0$. The combined difference coarray is shown in Fig. 3.25, where the difference coarray at $\omega_0$ is shown in black, and the additional lags, obtained by operating the subarray at $\omega_1$, are shown in red. The overall difference coarray has 10 positive lags which implies that up to 10 sources can be resolved. This is tested by considering 10 uniformly spaced sources between –60° and 60°. The number of snapshots is set to 2,000 at each frequency, and the SNR is set to 10 dB. The regularization parameter is set to 0.7 in this example, and the search space is kept the same. Fig. 3.26 shows the normalized spectrum using the dual-frequency sparse reconstruction method. It can be noticed that all the sources are correctly estimated, and the corresponding RMSE is 1.09°.

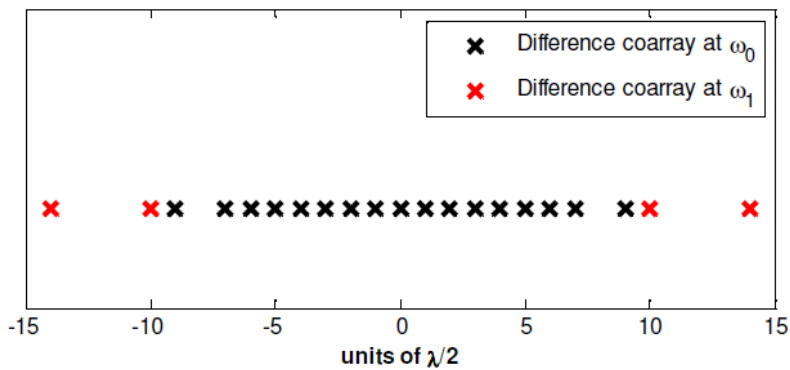

**Figure 3.25:** Dual frequency combined difference coarray, $\omega_1 = 2\omega_0$.

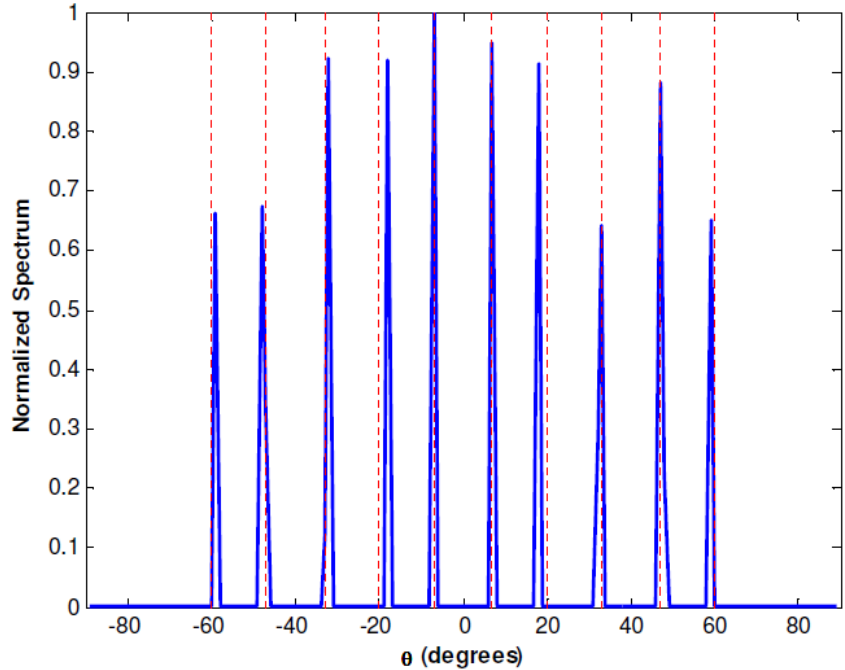

**Figure 3.26:** Dual-frequency sparse reconstruction, $D = 10$ sources.

The following example examines the case when the sources have non-proportional spectra. In this case, group sparse reconstruction is applied. The two operational frequencies are selected to be $\omega_0$ and $2\omega_0$. Eight sources with non-proportional spectra are considered. The SNR of all the sources at the first frequency is set to 10 dB. At the second frequency, the SNR of each source is a realization of a uniformly distributed random variable between 5 dB and 15 dB. This ensures that the sources have non-proportional spectra. The noise variance is set to unity at the two frequencies and a total of 2,000 snapshots are used. Fig. 3.27 shows the normalized spectrum obtained using the formulation in Section 3.2.2 which mistakenly assumes proportional source spectra. Consequently, this method is expected to fail as evident in the spectrum of Fig. 3.27. One of the sources is not resolved and several spurious peaks appear in the spectrum. The DOA estimation is next repeated using group sparse reconstruction which was discussed in Section 3.2.3. This method does not require the sources to have proportional spectra. The mean of the recovered spectra at the two employed frequencies is computed and shown in Fig. 3.28. It can be seen that group sparse reconstruction is successful in localizing the DOAs of all the sources. The RMSE is found to be 0.6° in this case.

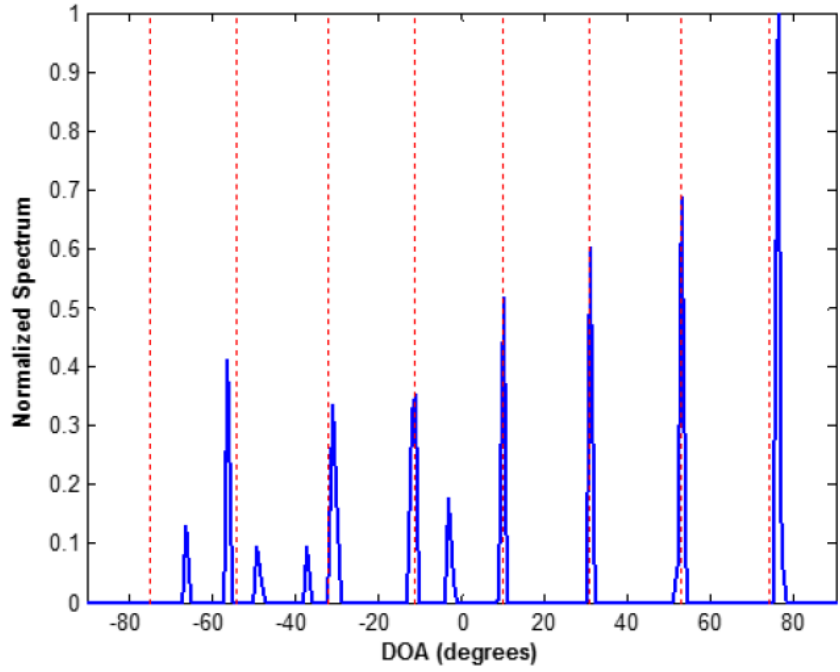

**Figure 3.27:** Dual-frequency sparse reconstruction, $D = 8$ sources with non-proportional spectra.

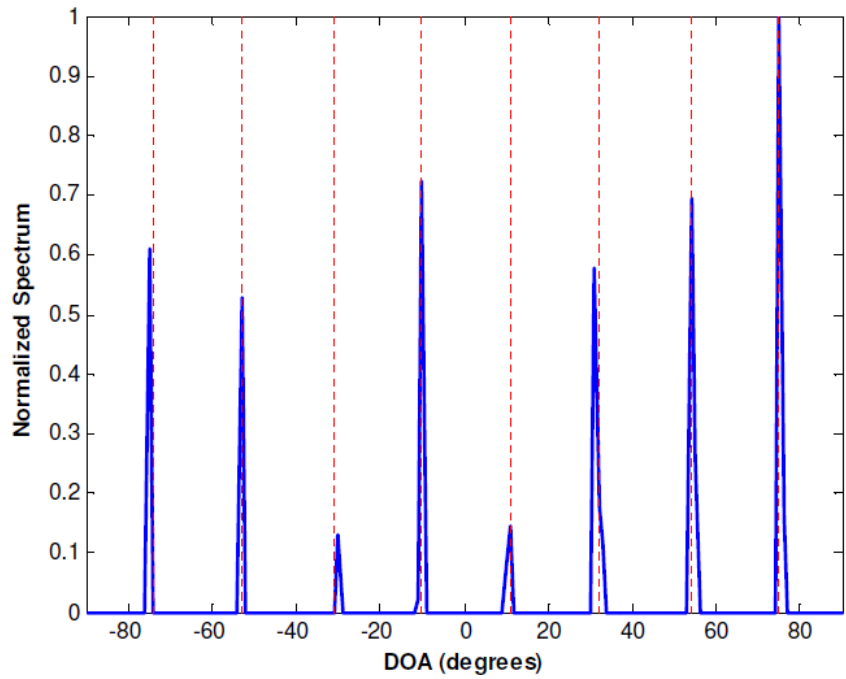

**Figure 3.28:** Dual-frequency group sparse reconstruction, $D = 8$ sources with non-proportional spectra.

The next example confirms the increase in the number of resolvable sources by using group sparse reconstruction compared to the single-frequency sparse reconstruction. As stated in the first example, the maximum number of resolvable sources using single-frequency sparse reconstruction is equal to the number of unique positive lags in the difference coarray, which is eight in this case. A total of 16 sources with non-proportional spectra is considered in the example. The sources are uniformly spaced between –75° and 75°. Twenty uniformly spaced frequencies between $\omega_0$ and $2\omega_0$ are employed. The SNR of each source at each frequency is chosen randomly between -5 dB and 5 dB, and the number of snapshots at each frequency is set to 1,000. Fig. 3.29 shows the normalized mean spectrum obtained using group sparse reconstruction. It can be seen that all the sources are correctly estimated, and the RMSE is equal to 0.35° in this case. Fig. 3.30 shows the normalized spectrum for the single-frequency sparse reconstruction case. This figure confirms that sparse reconstruction using a single frequency completely fails in estimating the sources. This is due to the fact that single-frequency sparse reconstruction can only resolve up to eight sources which is smaller than the total number of sources in this example.

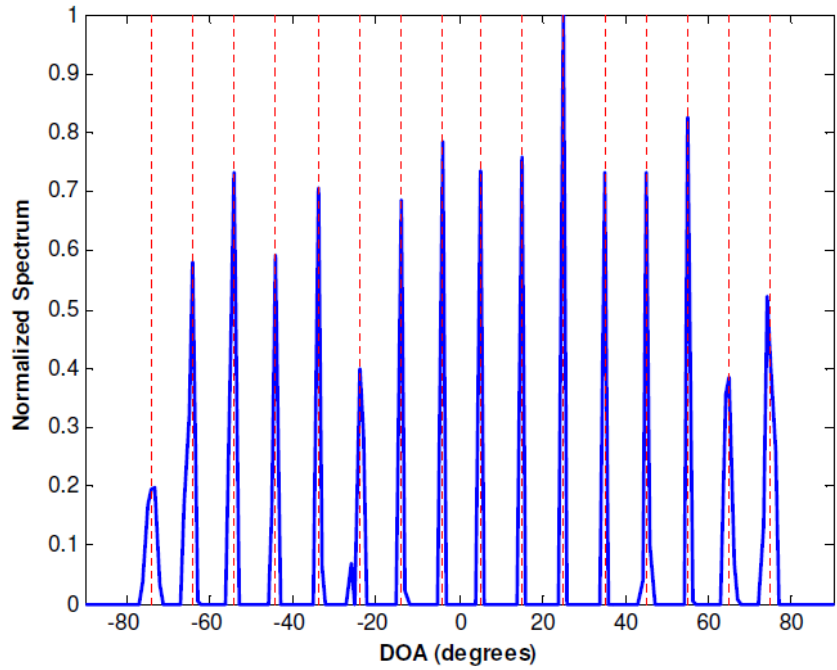

**Figure 3.29:** Multi-frequency group sparse reconstruction, $D = 16$ sources with non-proportional spectra.

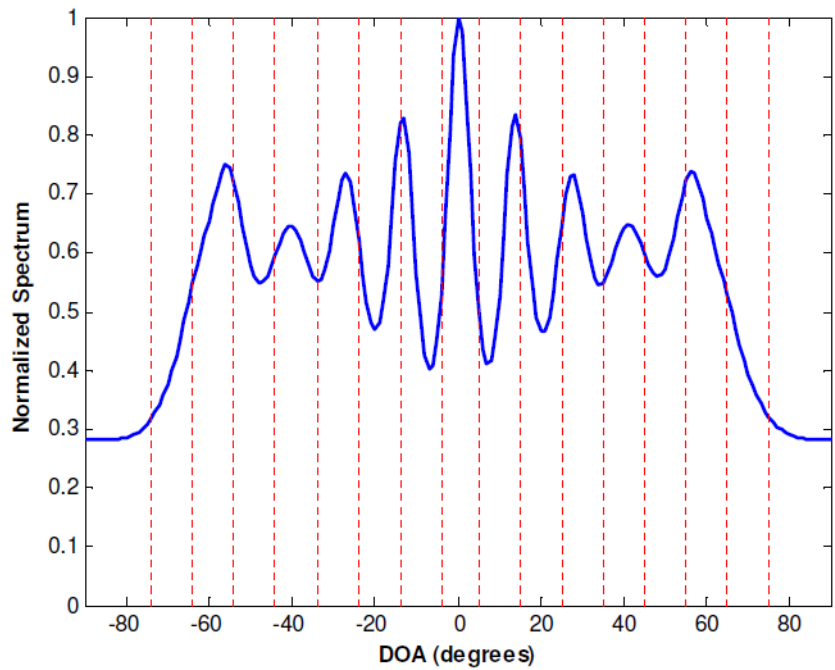

**Figure 3.30:** Single-frequency sparse reconstruction, $D = 16$ sources.

The final example examines the case where the source signals have overlapping spectra but do not share the same bandwidth. Group sparse reconstruction can still be used to perform DOA estimation. Thirty percent of the source powers at the employed frequencies in the previous example are randomly set to zero. The remaining parameters are kept the same. Fig. 3.31 shows the normalized spectrum using group sparse reconstruction. It is evident that all sources are correctly estimated. Some spurious peaks are present in the spectrum, and an increase in the estimates bias is obtained. The RMSE is found to be 0.61°.

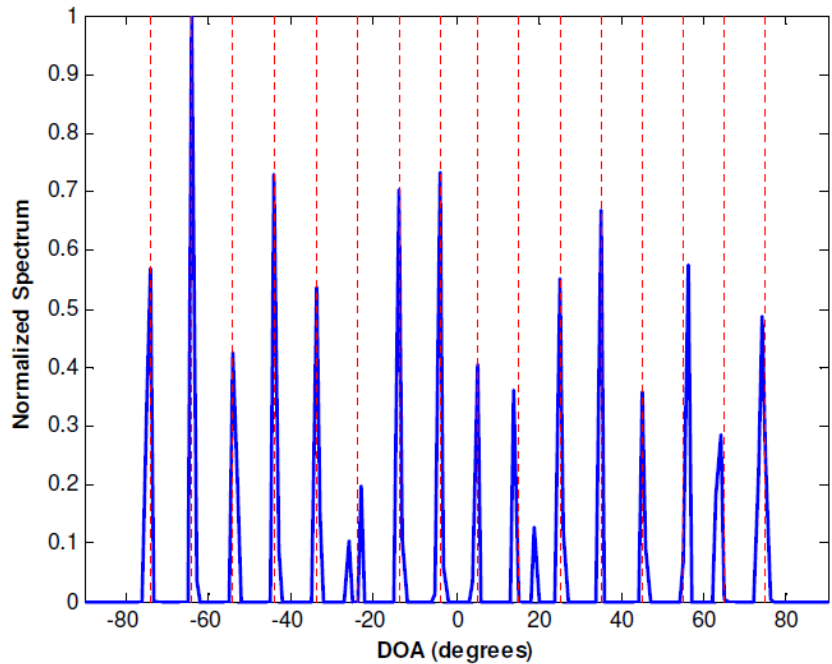

**Figure 3.31:** Dual-frequency group sparse reconstruction, $D = 8$ sources with non-proportional spectra.

### A.    *Comparison with High-Resolution Approach*

In order to compare the performance of sparse reconstruction and MUSIC based multi-frequency approaches, the following example is considered. The same extended co-prime array configuration with $M = 2$ and $N = 3$ is used. Two frequencies, $\omega_0$ and $\omega_1 = (8/9)\omega_0$, are employed; the latter can fill the holes in the corresponding difference coarray so that the multi-frequency MUSIC technique can be applied. Nine sources with directions uniformly spaced between –0.9 and 0.9 in the reduced angular coordinate $\sin(\theta)$ are used, which is the maximum number of sources that can be resolved using the multi-frequency MUSIC approach. Two separate cases are considered in this example. The first case assumes sources with proportional spectra, while the second considers sources with non-proportional spectra. For the latter, the source powers at $\omega_0$ are assumed to be identical and equal to unity, whereas the source powers associated with $\omega_1$ are assumed to independently follow a truncated Gaussian distribution with a mean of 5.5 and a variance of 2. Fig. 3.32 shows the RMSE, averaged over 1,000 Monte Carlo runs, as a function of the SNR for both cases. The SNR is assumed to be identical for all sources at $\omega_0$ and is varied between –10 dB and 10 dB with a 2.5 dB increment. It can be readily observed that the multi-frequency MUSIC approach outperforms the sparse reconstruction method for all SNR values when the sources have proportional spectra. In case of sources with non-proportional spectra, the multi-frequency MUSIC method outperforms the sparse reconstruction approach for low values of SNR, whereas both methods achieve similar performance at high SNR values. For both proportional and non-proportional spectra cases, the sparse reconstruction approach exhibits significantly degraded performance at low SNR values. This is expected since the accuracy of the sparse reconstruction methods suffers in high noise cases.

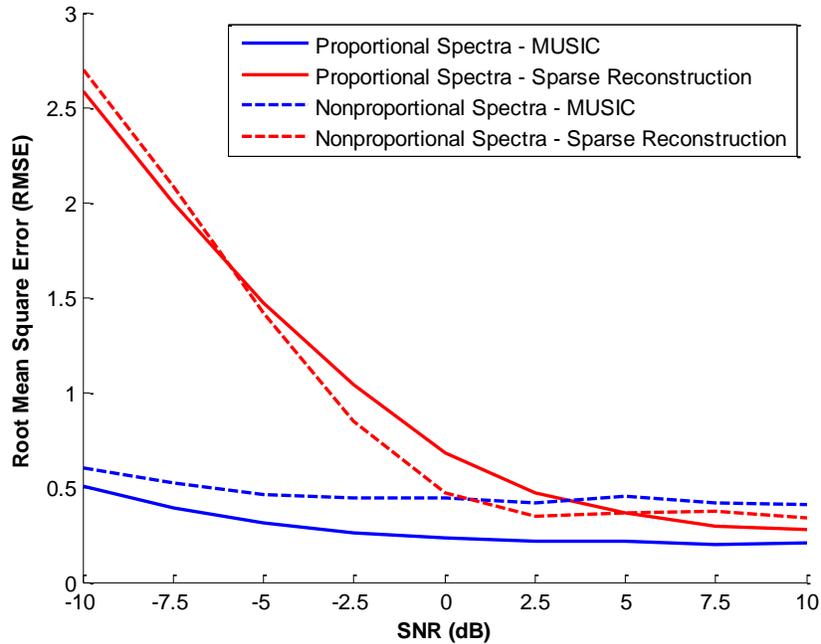

**Figure 3.32:** RMSE vs. SNR comparison between MUSIC and sparse reconstruction based multi-frequency approaches.

### 3.2.5.  Contributions

The main contributions of this research are listed below.

1)  Application of sparse reconstruction to multi-frequency co-prime arrays in order to make use of all generated DOFs.

2)  Providing a model when the sources have proportional spectra and determining a bound for the maximum number of resolvable sources.

3)  Providing a model when the sources have non-proportional spectra and determining a bound for the maximum number of resolvable sources.

### 3.3.    Concluding Remarks

In this chapter, multi-frequency operation was considered to perform DOA estimation with increased DOFs. First, a multi-frequency technique was presented for high-resolution DOA estimation using co-prime arrays. A virtual covariance matrix at the reference frequency was created using elements of the narrowband covariance matrices corresponding to the different employed frequencies. The virtual covariance matrix corresponds to a uniform linear array with a difference coarray of the same extent as that of the co-prime array, except that the coarray of the ULA is filled whereas that of the co-prime array has holes. Observations and insights were provided with regards to i) the maximum frequency separation required to fill all the holes in the difference coarray, ii) the lower bound on the number of sensors required to operate at more than one frequency, and iii) the performance under non-proportional source spectra case. These insights contribute towards better understanding the offerings and limitations of the proposed multi-frequency high-resolution approach. Supporting simulation examples were provided for DOA estimation of the proposed approach under both proportional and non-proportional spectra. The results demonstrated that the proposed approach can estimate DOAs with high accuracy for sources with proportional spectra, while for non-proportional spectra, the estimation error varies with the SNR as well as the values of $M$ and $N$. The effect of non-proportionality was shown to be not as significant at high SNR for higher values of $M$ and $N$ as for lower values. The same approach was also used to fill the missing elements in the difference coarray of minimum hole arrays.

Second, a sparse reconstruction method was presented for DOA estimation using multi-frequency non-uniform arrays. The proposed approach offers an enhancement in the degrees of freedom over the single-frequency operation. For sources with proportional spectra, all observations at the employed frequencies were combined to form a received signal vector at a larger virtual array, whose elements are given by the combination of the difference coarrays at the individual frequencies, thereby increasing the number of resolvable sources. In the case of sources with non-proportional spectra, the common support that is shared by the observations at the employed frequencies was exploited through group sparse reconstruction. Numerical examples demonstrated the superior performance of the proposed multi-frequency approach compared to its single-frequency counterpart.

# CHAPTER IV

# SPARSITY-BASED INTERPOLATION FOR DOA ESTIMATION USING NON-UNIFORM ARRAYS

In this chapter, sparsity-based interpolation is employed to fill the missing elements in the different coarray of non-uniform arrays. Several methods have recently been reported in the literature to alleviate the problem of missing elements in the difference coarray of a co-prime array and fully exploit the available DOFs [23, 33, 34, 51]. In [23], array motion was employed to collect measurements at the missing elements in the difference coarray. However, it requires data collection to be performed at precise locations. Any measurement errors can lead to performance degradation. In Chapter 2, the co-prime array was operated at multiple frequencies, and some of the measurements at the additional frequencies were used to fill in the missing elements in the difference coarray. Some of the drawbacks of this approach include the increased hardware complexity due to the multi-frequency operation, the imposed restrictions on the sources' power spectra, and the requirement of the sources to have a certain bandwidth [33, 34].

Sparsity-based imputation has been widely used in speech recognition to replace unreliable data samples that are corrupted by noise [52-55]. The difference between the proposed method and sparsity-based imputation is that the missing or unknown measurements in the coarray are not random and depend on the array geometry. In [56], a sparsity-based extrapolation technique was utilized to extend the aperture of a ULA beyond its physical extent.

The proposed method starts with the observations at the unique elements in the difference coarray and then applies sparsity-based interpolation to fill the missing elements. A combined measurements vector, consisting of the actual and interpolated measurements, is then formed to produce the effect of a difference coarray with no missing elements. MUSIC, in conjunction with spatial smoothing, is then applied to the combined measurement vector. The proposed technique is not only limited to filling the holes in the coarray. It can also be employed to extend the difference coarray aperture to beyond that achieved by the physical array. Extensive numerical simulations, which validate the performance enhancements of the proposed method, are also provided.

The remainder of the chapter is organized as follows. Section 4.1 discusses the proposed sparsity-based interpolation technique. The performance of the proposed technique is evaluated in Section 4.2, and Section 4.3 concludes the chapter by summarizing the contributions.

## 4.1. Sparsity-Based Interpolation

In this section, sparse reconstruction is employed to interpolate measurements to fill the holes in the difference coarray. After the missing elements are filled, MUSIC with spatial smoothing is then applied to a combined measurements vector.

The angular region of interest is discretized into $K$ grid points. A fully populated difference coarray with no missing elements is defined. The difference coarray is assumed to have contiguous elements between $-L_{full}d_0$ and $L_{full}d_0$. The corresponding array manifold matrix is denoted by $\overline{\mathbf{A}}_{full}$. $\overline{\mathbf{A}}_{full}$ has dimensions $(2L_{full} + 1) \times K$, and its $k$th column is the steering vector of the fully populated difference coarray corresponding to the $k$th grid point $\theta_k$. The $(i, k)$th element of $\overline{\mathbf{A}}_{full}$ can be expressed as

$$\left[\overline{\mathbf{A}}_{full}\right]_{i,k} = \exp\left(jk_0 x_{d,i} \sin \theta_k\right),\tag{4.1}$$

where $x_{d,i}$ is the location if the $i$th element in the fully populated difference coarray.

Similar to the procedure in Section 2.3.8, the vector $\mathbf{z}_u$ which holds the measurement at the unique difference coarray elements is first formed. Then, sparse reconstruction is applied to obtain an estimate of the source powers vector $\hat{\mathbf{p}}^g$. This is achieved by solving the constrained minimization in (2.33). After obtaining $\hat{\mathbf{p}}^g$, an estimate of the measurements at the fully populated difference coarray can then be computed using

$$\hat{\mathbf{z}}_e = \overline{\mathbf{A}}_{full}\hat{\mathbf{p}}^g. \tag{4.2}$$

A combined measurements vector $\mathbf{z}_e$ is then formed using the elements of $\mathbf{z}_u$ and $\hat{\mathbf{z}}_e$. The combination procedure is summarized as follows

$$\mathbf{z}_e\langle l\rangle = \begin{cases} \mathbf{z}_u\langle l\rangle, l \in S \\ \hat{\mathbf{z}}_e\langle l\rangle, l \notin S' \end{cases} \tag{4.3}$$

where $\mathbf{z}_e\langle l\rangle$ denotes the element of $\mathbf{z}_e$ corresponding to the measurement at lag $l$, and $S$ is the set of element positions of the difference coarray. If the $l$th lag in the fully populated difference coarray is present in the original difference coarray, the corresponding measurement is obtained from $\mathbf{z}_u$. Otherwise, the measurement is taken from $\hat{\mathbf{z}}_e$ since it corresponds to a missing element. In the final step, MUSIC with spatial smoothing is applied to the combined measurements vector $\mathbf{z}_e$.

It is essential to retain the available measurements at the original difference coarray in order to obtain a reliable performance. This is due to the fact that the original measurements contain information about the actual sources, some of which may not be accurately reconstructed or go undetected during the sparse reconstruction step. An example of this scenario would be when dealing with a weak source in close proximity to a strong one. The weak source might be masked by the strong source and goes undetected in the sparse reconstruction step.

## 4.2. Numerical Results

### 4.2.1. Strong Source Next to Weak Source

First, an extended co-prime array with $M = 3$ and $N = 5$ is considered. The array consists of two ULAs, with the first one having six elements with positions $[0, 5, 10, 15, 20, 25]d_0$ and the second having five elements with positions $[0, 3, 6, 9, 12]d_0$. The overall array comprises 11 elements with positions $[0, 3, 5, 6, 9, 10, 12, 15, 20, 25]d_0$. The corresponding difference coarray extends from $-25d_0$ to $25d_0$, and has contiguous elements between $-17d_0$ and $17d_0$. The physical array and the corresponding coarray are shown in Fig. 4.1.

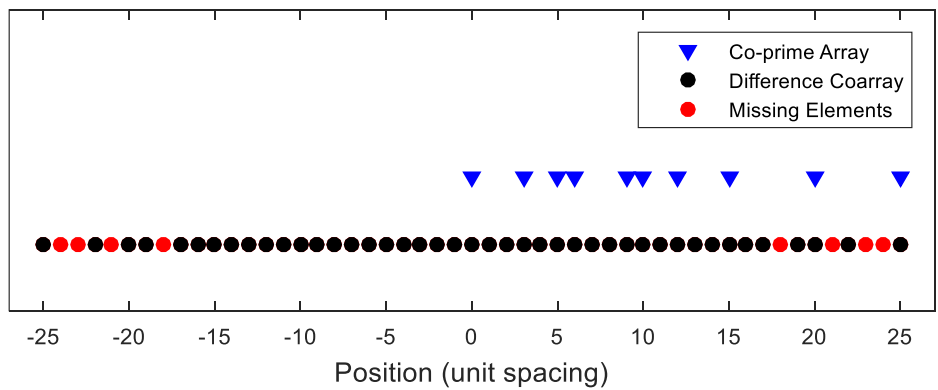

**Figure 4.1:** Extended co-prime array ($M = 3$, $N = 5$) array configuration and difference coarray.

Two sources with directions $[0, 0.03]$ in the reduced angular coordinate $\sin \theta$ are considered. The SNR of the first source is set to 20 dB and the SNR of the second source is set to 0 dB. This scenario simulates the case of two closely separated sources where the power of one source is much larger than that of the second one. The total number of snapshots is set to 500. MUSIC with spatial smoothing is first applied to the measurements at the contiguous part of the coarray, i.e., between $-17d_0$ and $17d_0$. Fig. 4.2(a) shows the estimated MUSIC spectrum. This figure shows that the weak source is completely missed. The reconstructed spectrum using sparse reconstruction is shown in Fig. 4.2(b). In this figure, the weak source is again missed. Finally, the proposed method is applied to the fully populated coarray which extends between $-25d_0$ and $25d_0$. In other words, the proposed technique is used to fill in the missing elements in the difference coarray. The obtained spectrum is shown in Fig. 4.2(c). Clearly, the two sources are correctly estimated. Even though the weak source is missed in the sparse reconstruction step, the original coarray measurements which are kept after interpolation contain information about it.

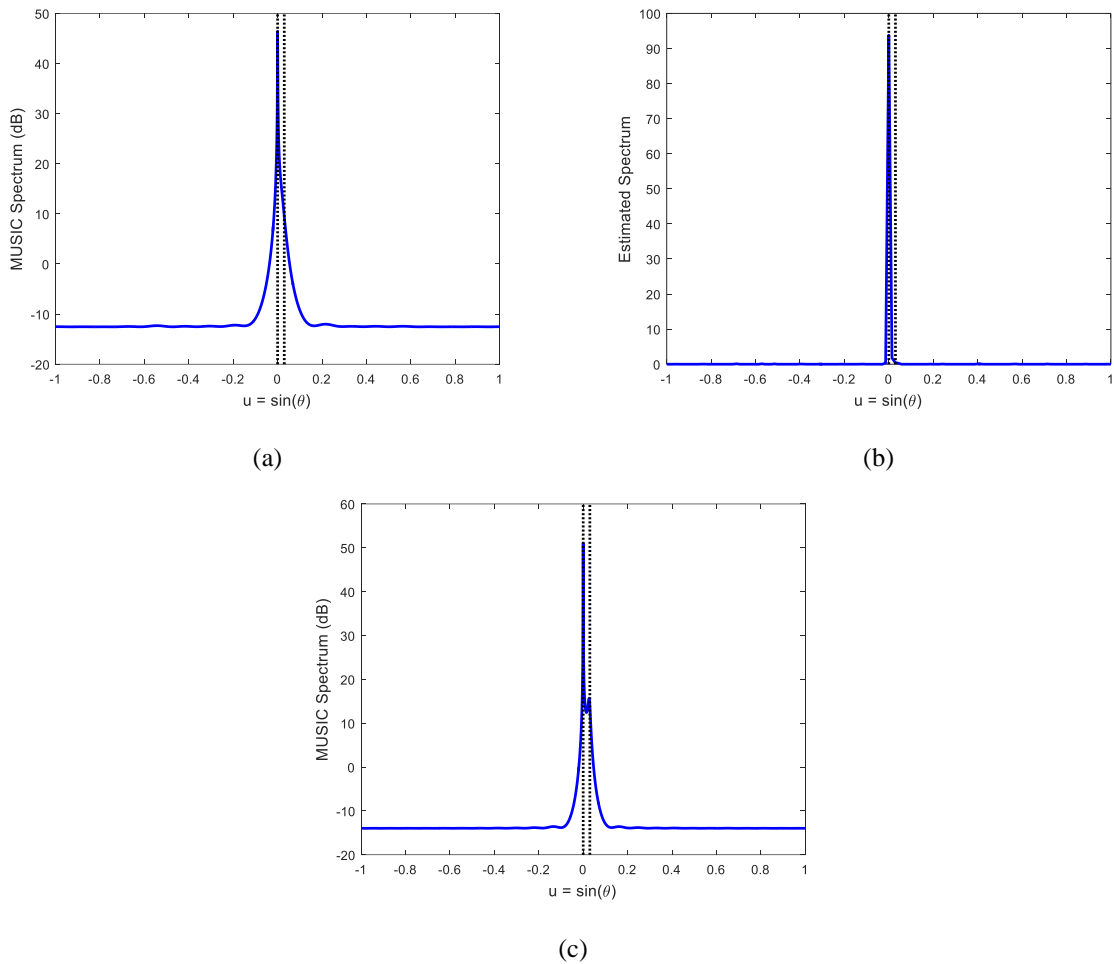

(a)

(b)

(c)

**Figure 4.2:** Extended co-prime array ($M = 3$, $N = 5$), $D = 2$, $\Delta$SNR = 20dB, $\Delta u = 0.03$ (a) MUSIC to contiguous part of coarray (b) Sparse reconstruction to unique coarray observations (c) MUSIC to fully populated coarray.

### 4.2.2. Multiple Source with Varying Powers

In the second example, the same co-prime array is used, but with a different number of sources. A total number of 17 sources, uniformly distributed between –0.85 and 0.75 in the reduced angular coordinate $\sin\theta$, are considered. The SNR for each source is randomly picked from a uniform distribution between –5 dB and 5 dB. The total number of snapshots is again set to 500. Fig. 4.3(a) shows the estimated spectrum when MUSIC with spatial smoothing is applied to the measurements at the contiguous part of the difference coarray. Clearly, some of the sources are completely missed and a considerable number of the remaining estimates are biased. The reconstructed spectrum using sparse reconstruction is shown in Fig. 4.3(b). It is evident that the reconstructed spectrum contains spurious peaks and one of these peaks is even larger than the power of an actual source. The proposed sparsity-based interpolation technique is then applied to generate the measurements at the fully populated difference coarray. MUSIC with spatial using is then applied the combined measurement vector and the estimated spectrum is depicted in Fig. 4.3(c). It is evident that all the sources are correctly estimated.

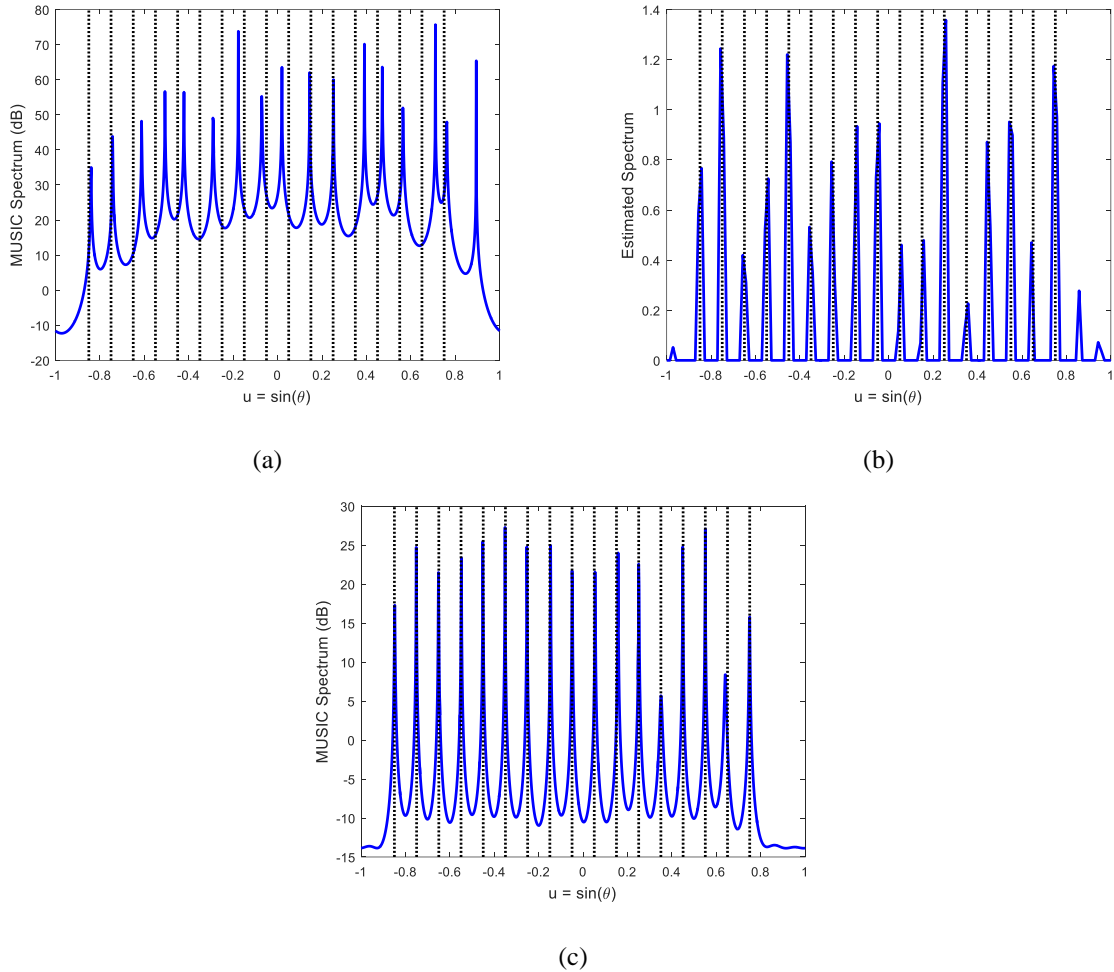

(a)    (b)

(c)

**Figure 4.3:** Extended co-prime array ($M = 3$, $N = 5$), $D = 17$ (a) MUSIC to contiguous part of coarray (b) Sparse reconstruction to unique coarray observations (c) MUSIC to fully populated coarray.

### 4.2.3. Performance Analysis

In the third example, the same 11-element co-prime array is used with varying source powers and varying source separations. Two sources are considered, with the direction of the first source fixed at $u_1 = \sin\theta_1 = 0$, and the SNR of the second source is fixed at 0 dB. The SNR of the first source is varied between 0 dB and 30 dB, and the direction of the second $u_2 = \sin\theta_2$ is varied between 0.03 and 0.20 to simulate different source separation scenarios. For each set of parameters, 100 Monte Carlo runs are used and the RMSE is computed. Fig. 4.4 compares the performance of MUSIC with spatial smoothing applied to the contiguous part of the coarray to that of the proposed method.

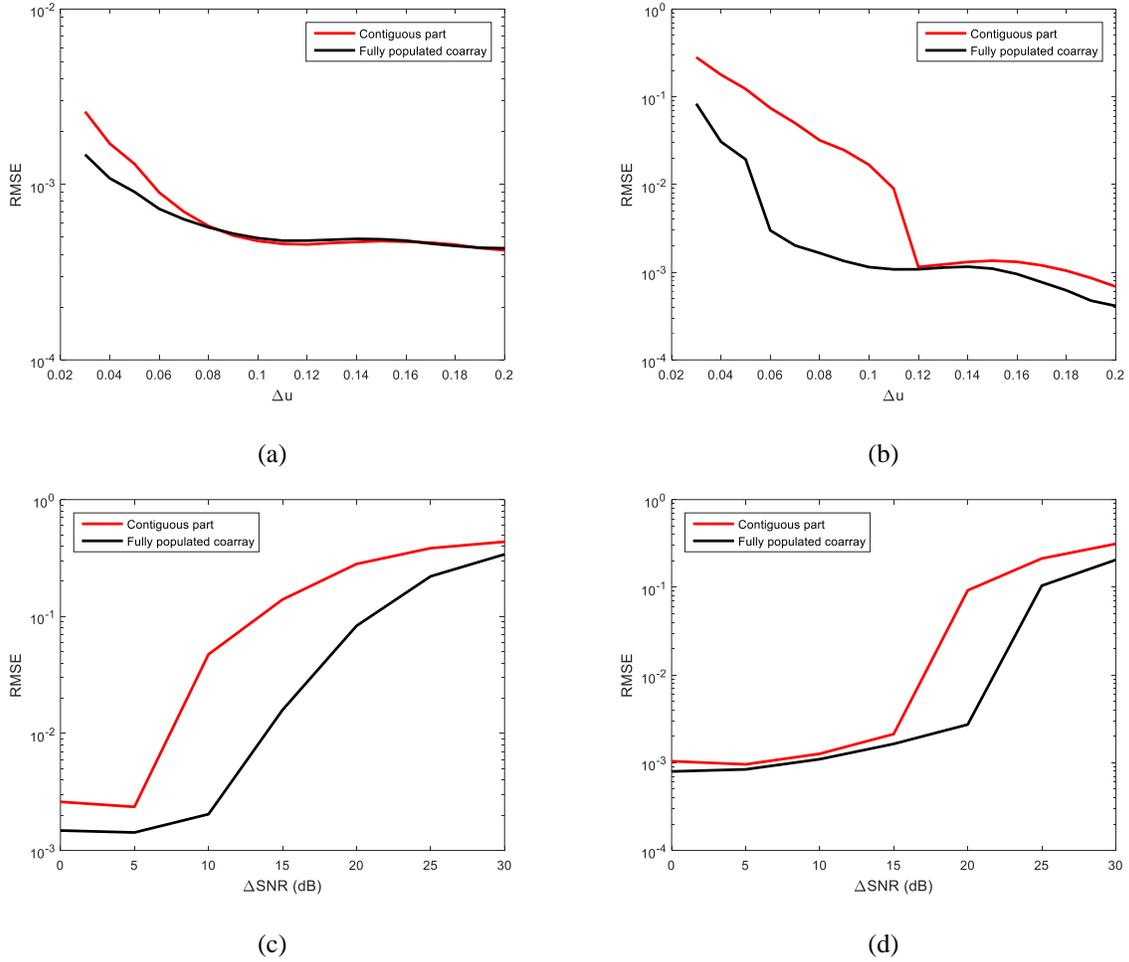

**Figure 4.4:** Extended co-prime array ($M = 3$, $N = 5$), $D = 2$ (a) $\Delta$SNR = 0dB: RMSE vs. $\Delta u$ (b) $\Delta$SNR = 20dB: RMSE vs. $\Delta u$ (c) $\Delta u = 0.03$: RMSE vs. $\Delta$SNR (d) $\Delta u = 0.05$: RMSE vs. $\Delta$SNR.

In Fig. 4.4(a) and Fig. 4.4(b), the difference between the source powers, $\Delta$SNR, is set to 0 dB and 20 dB, respectively, and the average RMSE is plotted as a function of the source separation. By examining these two figures, several conclusions can be reached. First, the estimation performance improves as the separation between the sources increases. Second, for sources with equal powers, the improvement of the proposed method is only noticeable for small source separations. Third, for sources with large SNR discrepancy, the performance enhancement is large, especially for small source separations. Figs. 4.4(c) and 4.4(d) show the average RMSE as a function of $\Delta$SNR for $\Delta u = 0.03$ and $\Delta u = 0.05$, respectively. It can be noticed that the performance deteriorates as ΔSNR increases. The proposed method shows an improvement over the MUSIC with spatial smoothing algorithm applied to the contiguous part of the coarray. The performance improvement is more noticeable for $\Delta = 0.03$. As the separation between the source powers increases, the performance improvement increases until a certain point is reached where the two methods begin to fail to resolve the two sources and the two plots begin to converge to the same value.

### 4.2.4. Co-Prime Arrays vs. Minimum Hole Arrays

In this section, the performance of a co-prime array is compared with the performance of a minimum hole array having the same number of physical sensors. It should be noted that for the same number of elements, multiple co-prime array or minimum hole array configurations can be found. Table 4.1 groups the different combinations for varying numbers of elements. The percentage of missing elements in the coarray aperture is also included in the table. By comparing the different combinations, it can be noticed that, for the same number of physical elements, co-prime arrays consistently result in a smaller percentage of holes compared to minimum hole arrays.

**Table 4.1:** Co-prime arrays vs. minimum hole arrays

| $N_A$ | CPA | | % of holes in coarray | MHA | % of holes in coarray |
|---|---|---|---|---|---|
| | $M$ | $N$ | | | |
| 6 | 2 | 3 | 10.52% | $[0,1,4,10,12,17]d_0$ $[0,1,4,10,15,17]\,d_0$ $[0,1,8,11,13,17]\,d_0$ $[0,1,8,12,14,17]\,d_0$ | 11.43% |
| 8 | 2 | 5 | 12.90% | $[0,1,4,9,15,22,32,34]\,d_0$ | 17.39% |
| 9 | 3 | 4 | 14.63% | $[0,1,5,12,25,27,35,41,44]\,d_0$ | 17.97% |
| 10 | 2 | 7 | 13.95% | $[0,1,6,10,23,26,34,41,53,55]\,d_0$ | 18.01% |
| | 3 | 5 | 15.68% | | |
| 12 | 2 | 9 | 14.54% | $[0,2,6,24,29,40,43,55,68,75,76,85]\,d_0$ | 22.22% |
| | 3 | 7 | 16.90% | | |
| | 4 | 5 | 16.90% | | |

First, a six-element co-prime array with $M = 2$ and $N = 3$ is considered. Fig. 3.15(a) shows the array and the corresponding difference coarray. For comparison, a six-element MHA with elements at $[0, 1, 4, 10, 12, 17]d_0$ is considered. The MHA and its coarray are depicted in Fig. 3.17(b). Since the difference coarray of the MHA has a larger contiguous part than that of the co-prime array, MUSIC with spatial smoothing, applied to the contiguous part, is expected to result in a better performance using the MHA. After interpolation, the fully populated coarray of the MHA is much larger than that of the co-prime array and an improved performance using the MHA is expected as well. Fig. 4.5 compares the performance of these arrays for different source separations and different source powers. Two sources are also considered in this example. The direction of the first source is fixed to $u = 0$, and the direction of the second source is varied. The SNR of the first source is varied between 0 dB and 30 dB, and the second source's SNR is set to 0 dB. For each set of parameters, the number of Monte Carlo runs is set to 100 and the average RMSE is calculated. The total number of snapshots is set to 1,000. By examining Fig. 4.5, it is clear that the MHA consistently outperforms the co-prime array even before applying the proposed method. In addition, applying the proposed technique results in improving the estimation performance of both arrays. For a fixed ΔSNR and a large source separation Δ$u$, the performance improvement is not noticeable.

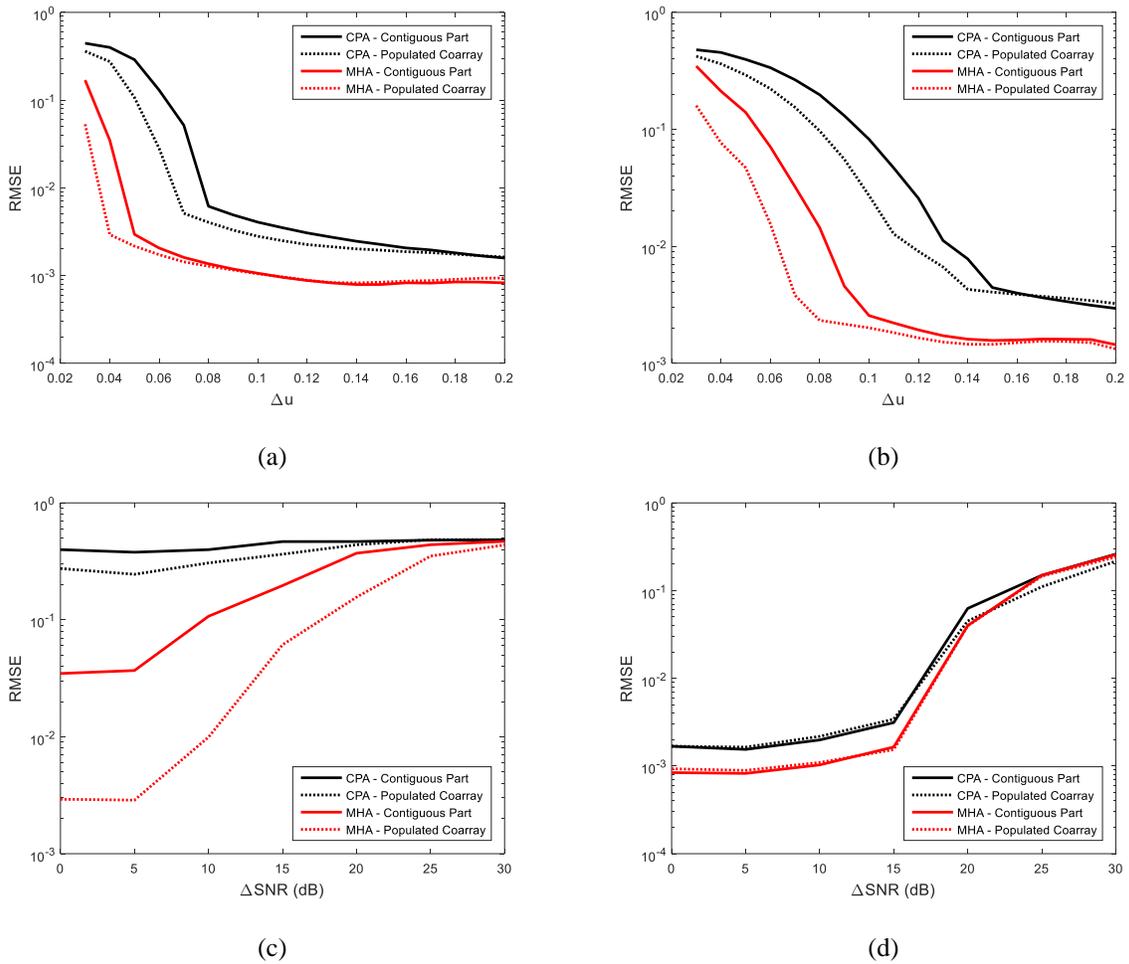

(a)

(b)

(c)

(d)

**Figure 4.5:** Extended co-prime array ($M = 2$, $N = 3$) vs. six-element MHA ([0, 1, 4, 10, 12, 17]d$_0$), $D = 2$ (a) ΔSNR = 0dB: RMSE vs. Δ$u$ (b) ΔSNR = 15dB: RMSE vs. Δ$u$ (c) Δ$u$ = 0.04: RMSE vs. ΔSNR (d) Δ$u$ = 0.19: RMSE vs. ΔSNR.

In the following example, the same six-element co-prime array is used; however, an MHA with a different configuration is considered. Fig. 3.17 shows the MHA configuration and the corresponding difference coarray. Fig. 4.6 shows the different RMSE plots. The same simulation parameters as in the previous example are used. One notable difference between the results of Fig. 4.5 and those of Fig. 4.6 is that, in Fig. 4.6, the co-prime array consistently outperforms the MHA before applying the proposed method. This is expected since the difference coarrays have the same contiguous part, and the difference coarray of the co-prime array has more redundancies than that of the MHA.

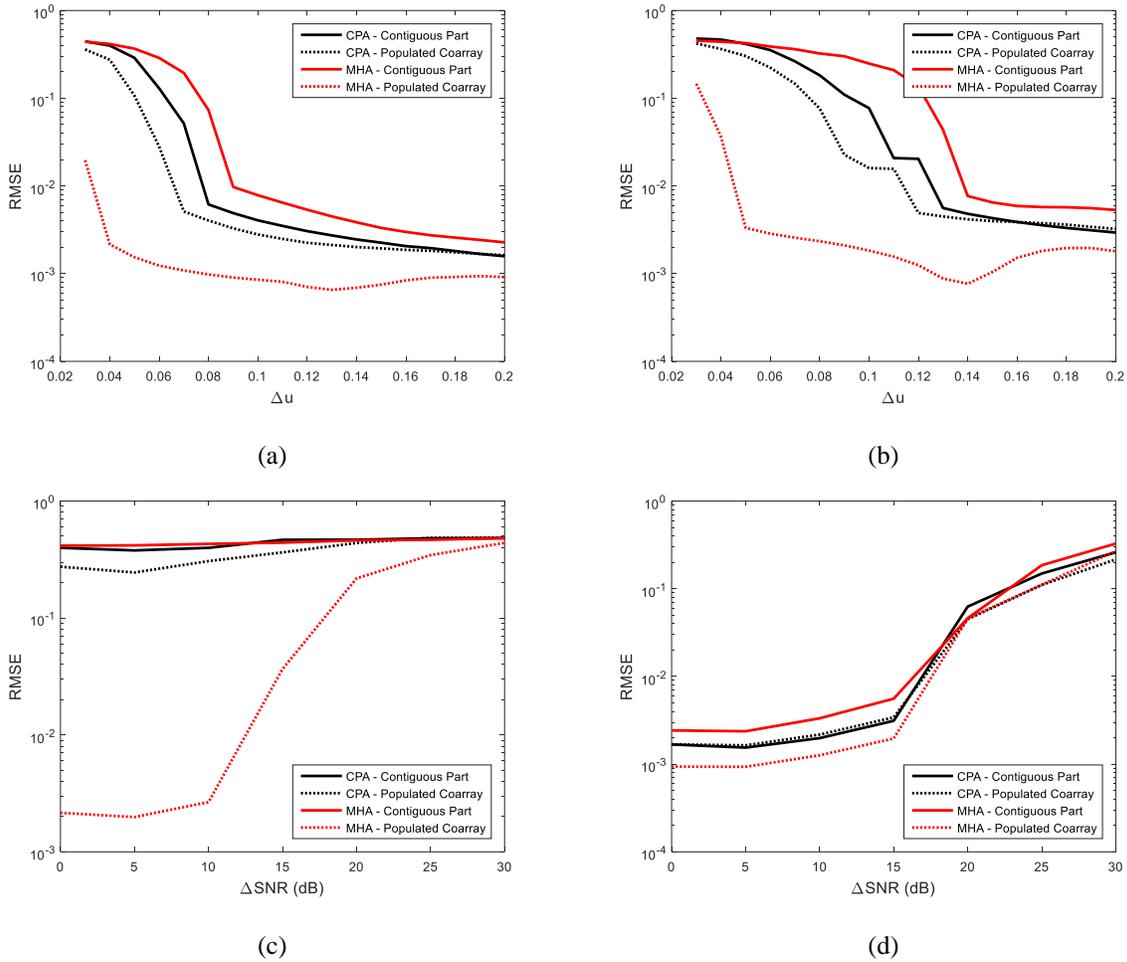

(a)

(b)

(c)

(d)

**Figure 4.6:** Extended co-prime array ($M = 2$, $N = 3$) vs. six-element MHA ([0, 1, 4, 10, 15, 17]$d_0$), $D = 2$ (a) $\Delta$SNR = 0dB: RMSE vs. $\Delta u$ (b) $\Delta$SNR = 15dB: RMSE vs. $\Delta u$ (c) $\Delta u = 0.04$: RMSE vs. $\Delta$SNR (d) $\Delta u = 0.19$: RMSE vs. $\Delta$SNR.

### 4.2.5. Sparsity-Based Interpolation vs. Multi-Frequency Approach

In this example, the performance of the proposed method is compared to the performance of the high-resolution multi-frequency method that was presented in Chapter 2.

For this comparison, the sources are assumed to have proportional spectra at the operating frequencies in order to meet the requirements of the multi-frequency approach. The extended co-prime array, shown in Fig. 4.1, is used for the comparison. The multi-frequency approach requires operating the array at four additional frequencies in order to fill all the holes in the coarray. Two sources are considered with varying separations and varying $\Delta$SNR. The number of snapshots is set to 1,000 for the proposed approach, and the total number of snapshots at each additional frequency is set to 1,000, resulting in 5,000 snapshots for the multi-frequency approach. The number of Monte Carlo runs for each set of parameters is fixed to 100. Fig. 4.7 shows the average RMSE plots for the two approaches. It can be noticed the multi-frequency approach slightly outperforms the proposed approach.

In order to make a fairer comparison, the total number of snapshots needs to be similar for both approaches. This number is first fixed to 5,000 snapshots and the results are shown in Fig. 4.8. For this scenario, the number of snapshots at each frequency is set to 1,000. It is evident that the proposed approach outperforms the multi-frequency approach. The same conclusion is drawn when each frequency employs 200 snapshots and the proposed approach employs 1,000 snapshots. The corresponding results are shown in Fig. 4.9.

### 4.2.6. Sparsity-Based Interpolation vs. Nuclear Norm Minimization

In this example, the performance of the proposed method is compared to the performance of the coarray interpolation via nuclear norm minimization technique [51]. The extended co-prime

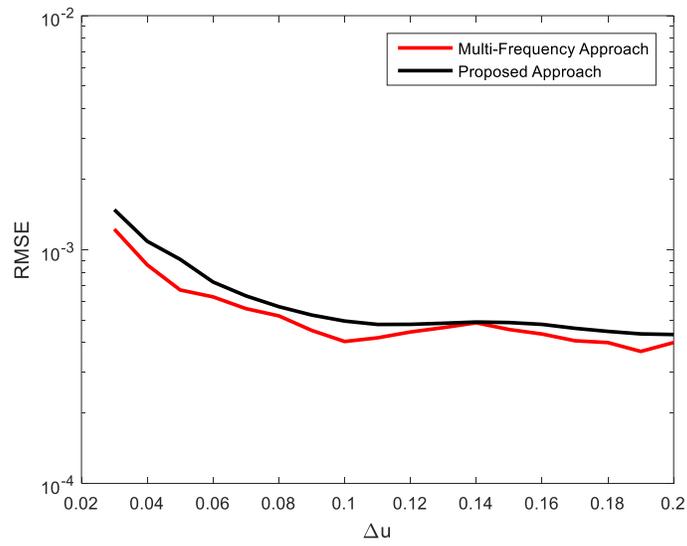

(a)

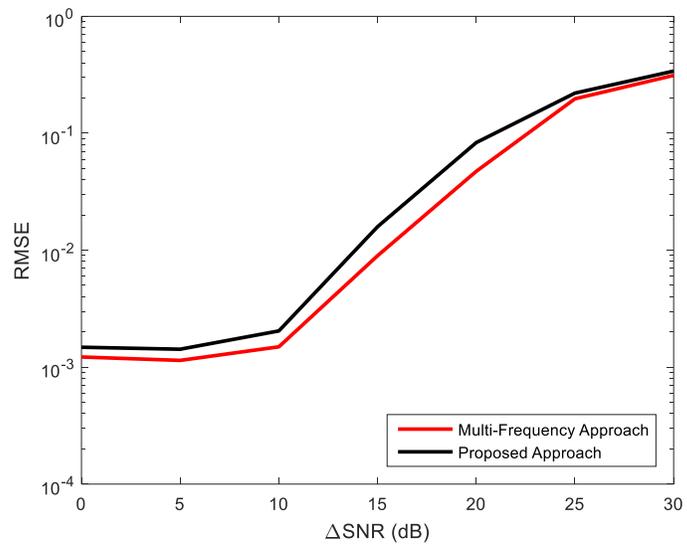

(b)

**Figure 4.7:** Proposed method vs. multi-frequency approach: extended co-prime array ($M = 3$, $N = 5$), $T_{SBE} = 1,000$, $T_{MF} = 5,000$ (a) $\Delta SNR = 0$dB: RMSE vs. $\Delta u$ (b) $\Delta u = 0.03$: RMSE vs. $\Delta SNR$.

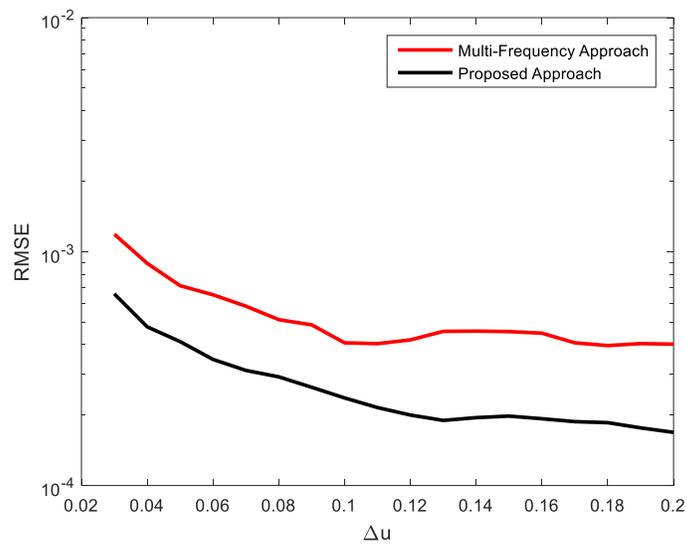

(a)

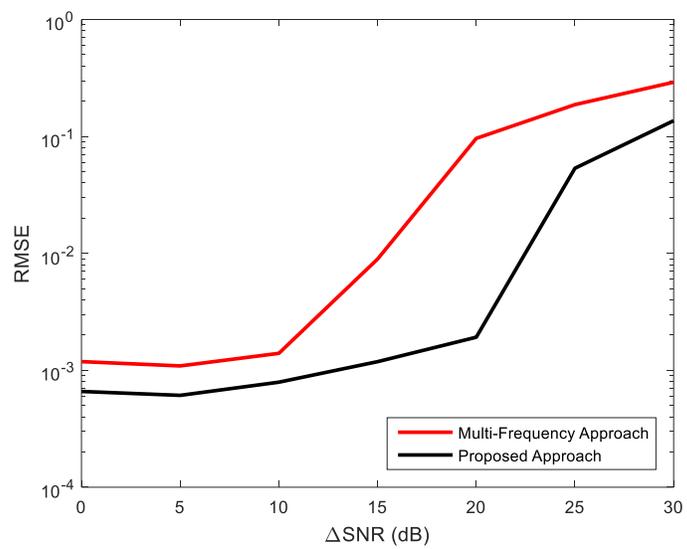

(b)

**Figure 4.8:** Proposed method vs. multi-frequency approach: extended co-prime array ($M = 3$, $N = 5$), $T_{SBE} = 5,000$, $T_{MF} = 5,000$ (a) $\Delta$SNR = 0dB: RMSE vs. $\Delta u$ (b) $\Delta u = 0.03$: RMSE vs. $\Delta$SNR.

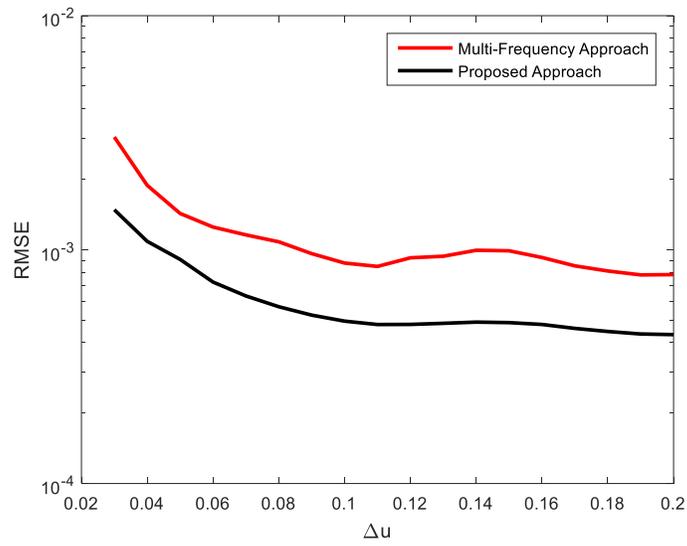

(a)

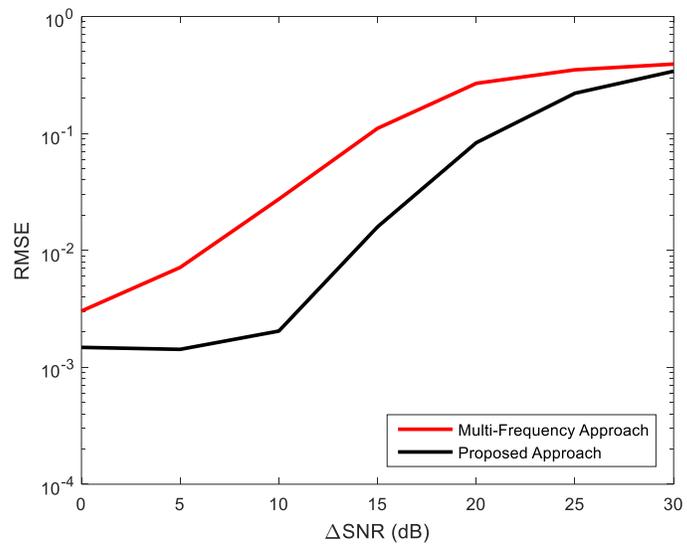

(b)

**Figure 4.9:** Proposed method vs. multi-frequency approach: extended co-prime array ($M = 3$, $N = 5$), $\text{T}_{\text{SBE}} = 1,000$, $\text{T}_{\text{MF}} = 1,000$ (a) $\Delta\text{SNR} = 0\text{dB}$: RMSE vs. $\Delta u$ (b) $\Delta u = 0.03$: RMSE vs. $\Delta\text{SNR}$.

array of Fig. 4.1 is again used for the comparison. Two sources with varying separations and varying ΔSNR are considered. The number of snapshots is set to 500 for both techniques. The number of Monte Carlo runs for each set of parameters is fixed to 100. Fig. 4.10 shows the average RMSE plots as a function ΔSNR for the two approaches. It is evident that both approaches provide a similar performance.

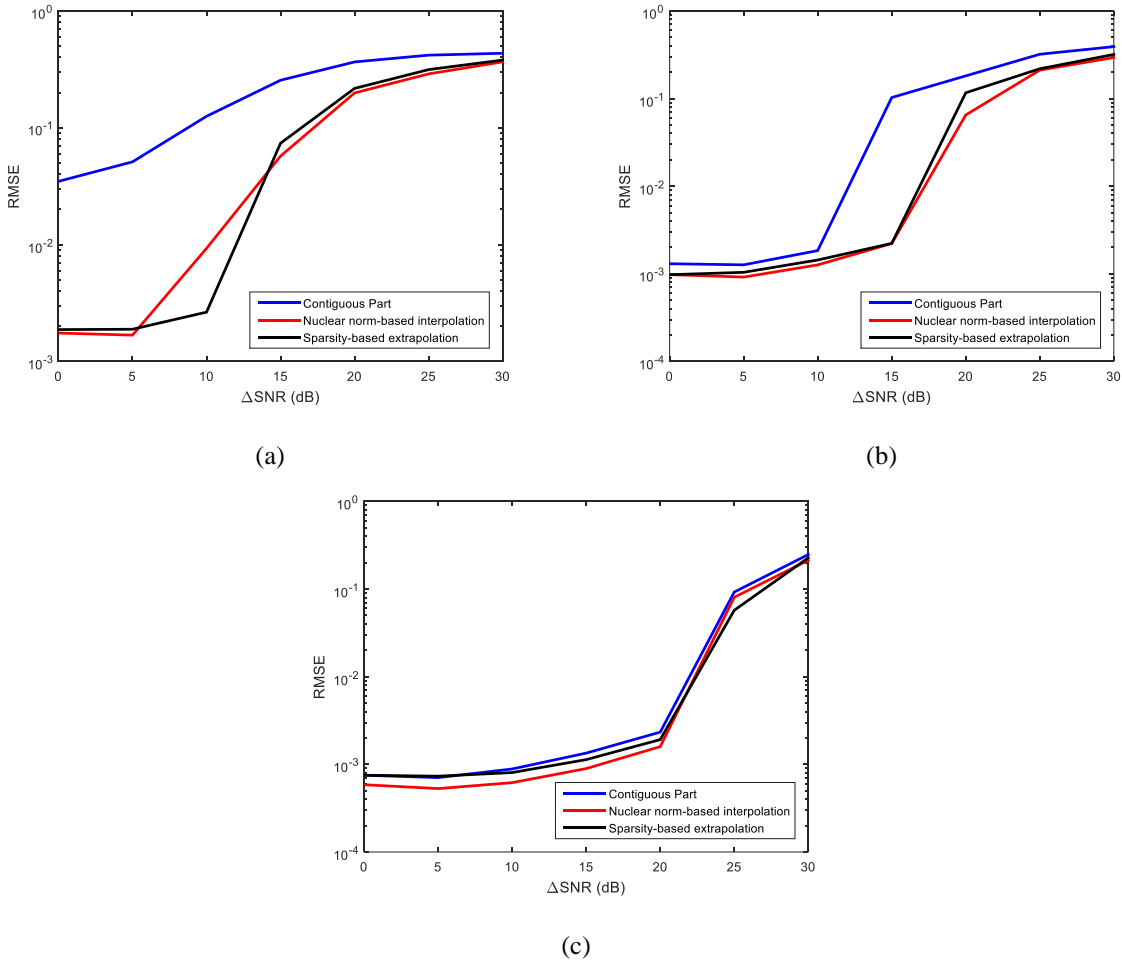

**Figure 4.10:** Proposed method vs. nuclear norm minimization approach: extended co-prime array ($M = 3$, $N = 5$), (a) $\Delta u = 0.03$: RMSE vs. ΔSNR (b) $\Delta u = 0.05$: RMSE vs. ΔSNR (c) $\Delta u = 0.15$: RMSE vs. ΔSNR.

### 4.2.7. Extended Coarray Aperture

In the previous examples, the proposed approach was used to fill the missing elements in the difference coarrays. However, the same approach can be employed to extrapolate measurements beyond the coarray aperture as shown in the following example. The six-element co-prime array, shown in Fig. 3.15(a), is used in this example. The different considered coarrays are shown in Fig. 4.11(a). The contiguous part refers to the contiguous part of the original coarray which extends between $-7d_0$ and $7d_0$. The populated coarray refers to the original coarray with the elements at $\pm 8d_0$ being filled using the proposed approach. The extended coarray refers to a fully populated coarray between $-14d_0$ and $14d_0$ which extends beyond the aperture of the original coarray. In this example, two sources with directions $[0, \ 0.14]$ in $\sin\theta$ and powers $[30, \ 0]$ dB are considered. The total number of snapshots is set to 500. The MUSIC spectrum applied to the different coarrays is shown in Fig. 4.11(b). In this scenario, only the extended coarray is able to resolve the sources successfully.

### 4.3. Concluding Remarks

In this chapter, a sparsity-based technique was proposed to interpolate missing coarray measurements and exploit the degrees-of-freedom offered by non-uniforms arrays in DOA estimation. Starting with the observations at the unique difference coarray locations, sparse reconstruction was used to interpolate measurements at the missing elements in the difference coarray. A combined measurements vector which comprises the original observations and the interpolated ones was then formed. MUSIC with spatial smoothing was then applied to the combined measurements vector. The proposed method was successfully applied to co-prime arrays as well as minimum hole arrays. Numerical simulations validated the proposed method and evaluated its performance under different scenarios.

### 4.3.1. Contributions

The following are the contributions of the research in this chapter.

1) Application of sparse reconstruction to interpolate observations at the missing elements in the difference coarray.

2) Investigating the effects of source separation and the dynamic range of source powers on the performance of the proposed method.

3) Comparison between the performance of co-prime arrays and minimum hole arrays using sparsity-based interpolation.

4) Comparison between the performance of the proposed method and the performance of previously developed methods under different scenarios.

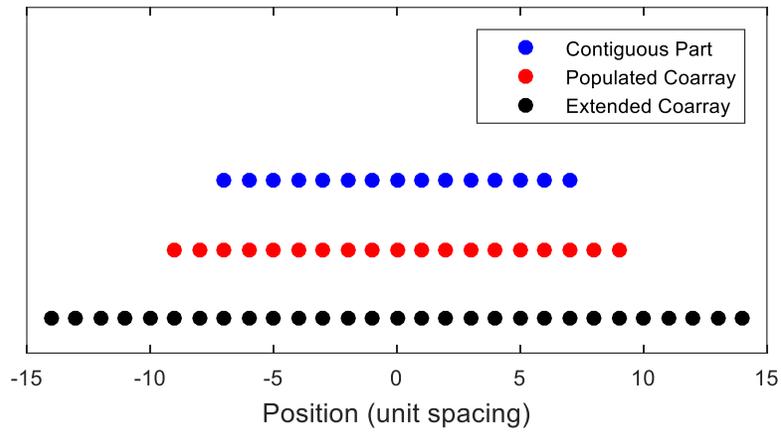

(a)

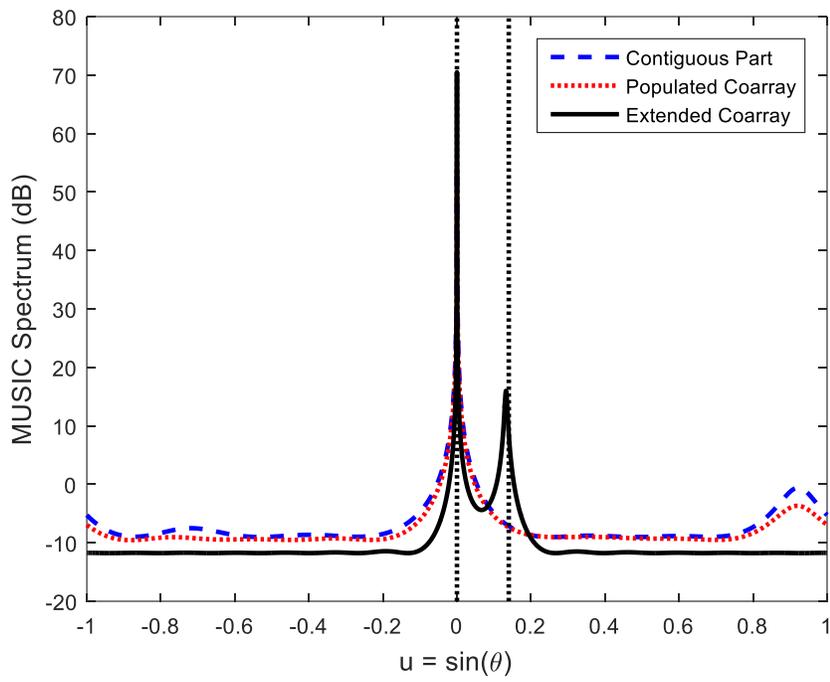

(b)

**Figure 4.11:** Extended co-prime array ($M = 2$, $N = 3$) (a) different populated coarray configurations (b) MUSIC to considered coarrays.

# CHAPTER V

# MUTUAL COUPLING EFFECT AND COMPENSATION IN NON-UNIFORM ARRAYS FOR DOA ESTIMATION

In this chapter, the effect of mutual coupling on DOA estimation using non-uniform arrays is investigated. The DOA estimation accuracy in the presence of mutual coupling is compared and contrasted for three different non-uniform array geometries, namely, MRAs, nested arrays, and co-prime arrays, and for two antenna types, namely dipole antennas and microstrip antennas. Through numerical simulations, it is demonstrated that the mutual coupling, if unaccounted for, can, in general, lead to performance degradation, with the MRA faring better against mutual coupling than the other two non-uniform structures for both antenna types. Two methods that can compensate for the detrimental effects of mutual coupling are also proposed, leading to accurate and reliable DOA estimation. Supporting numerical simulation results are provided which show the effectiveness of the proposed compensation methods.

The mutual coupling between the array elements can be captured in a matrix called the mutual coupling matrix (MCM). Two major trends exist in the literature for performing DOA estimation in the presence of mutual coupling. The first deals with the case of perfectly known or modeled MCM, wherein the DOA estimation procedure is modified to account for the coupling [57]. In the second trend, the MCM is assumed to be unknown or imprecisely known with a specific structure, and is jointly estimated along with the source directions.

Electromagnetic theory and numerical or analytical modeling techniques are typically employed to characterize the MCM [58-63]. The MCM depends on the self and mutual impedances between the array elements. One of the earliest methods that model the coupling matrix is the open-circuit method [58]. This method treats the array as a bilateral terminal network and relates the uncoupled voltages with the coupled voltages through a mutual impedance matrix. For dipole antennas, the elements in the mutual impedance matrix can be approximated by closed-form expressions [64]. An extension of the open-circuit method has been proposed in [59], where two types of mutual impedances are defined, namely, the transmission mutual impedance and the re-radiation mutual impedance. In [60], the receiving-mutual-impedance method (RMIM) is described for use in receive-only antenna arrays. As such, it provides a more accurate coupling model in DOA estimation applications. RMIM considers each antenna pair separately to compute the receiving mutual impedances. An enhancement of RMIM is presented in [61], which takes into account all the elements simultaneously in order to compute the receiving mutual impedances.

For a perfectly known or modeled MCM, DOA estimation algorithms can be modified to incorporate the coupling and compensate for it in order to achieve accurate source directions [57]. However, if the modeled MCM is not exact, the performance of the DOA estimation is degraded. Moreover, the MCM must be re-calibrated periodically to account for any changes in local conditions. For instance, the presence of a new scatterer in the vicinity of the antenna array changes the mutual coupling. Several methods have been proposed to circumvent these issues. These methods assume the coupling matrix to be unknown or imprecisely known and aim to jointly estimate the MCM along with the source DOAs [30-32]. Ref. [30] presents an iterative method to estimate the MCM, the DOAs, and the antenna gains, wherein the cost function is minimized with respect to one unknown quantity at a time while keeping the remaining two unknowns fixed. A maximum likelihood estimator for DOA estimation under unknown multipath and unknown mutual coupling has been proposed in [31]. Ref. [32] employs sparse reconstruction to perform DOA estimation in the presence of unknown mutual coupling. However, all of these aforementioned methods have been developed for ULAs and take advantage of the special structure of the corresponding MCMs. Although these methods can be modified and applied to non-uniform arrays, they fail to take advantage of the increased DOFs offered by non-uniform arrays for DOA estimation. An iterative method for DOA estimation using non-uniform arrays in the presence of mutual coupling was proposed in [65]. This method treats the non-uniform array as a subset of a ULA and, therefore, cannot take full advantage of the increased DOFs as well.

In this chapter, the mutual coupling effect in non-uniform arrays is addressed. First, the impact of coupling on the DOA estimation accuracy for different array geometries is examined. The performance is evaluated for different array sizes and for two antenna element types, namely, dipole antenna and microstrip antenna. The latter is becoming increasingly popular in radar and wireless communications due to its low profile, ease of fabrication, low cost, and compatibility with radio frequency (RF) circuit boards. A computational electromagnetics software package, FEKO [66], is used to model the antenna arrays, and the RMIM [60, 61] is used to compute the coupling matrices based on the obtained measurements. Through numerical simulations, it is shown that the MRA provides superior performance compared to the nested and co-prime geometries, irrespective of the antenna type. Second, two compensation methods that allow accurate DOA estimation using non-uniform arrays in the presence of mutual coupling are proposed. The first method assumes partial knowledge of the mutual coupling and employs an iterative approach to update the perturbed MCM and DOAs. Sparse signal reconstruction is used to find the source directions for a given coupling matrix, and a global optimization algorithm called covariance matrix adaptation evolution strategy (CMA-ES) [67] is used to update the MCM while keeping the DOAs fixed. The second method assumes unknown coupling and simultaneously estimates the MCM, the source powers, and sources directions by minimizing a cost function using CMA-ES. Finally, the effectiveness of the proposed methods is evaluated through numerical examples.

The remainder of this chapter is organized as follows. The signal model in the presence of mutual coupling is presented in Section 5.1. In Section 5.2, DOA estimation performance of different non-uniform array geometries is evaluated and compared for the case of uncompensated mutual coupling. Section 5.3 discusses the two proposed compensation methods that allow accurate DOA estimation under mutual coupling and provides supporting numerical results. Section 5.4 concludes the chapter by summarizing its contributions.

## 5.1.    Signal Model

Thus far, mutual coupling has been ignored in the signal model. However, in practical antenna arrays, coupling between the antenna elements is a real issue and thus needs to be taken into account. The signal model in (2.6) can be modified to incorporate mutual coupling as

$$\mathbf{x}(t) = \mathbf{CAs}(t) + \mathbf{n}(t), \tag{5.1}$$

where $\mathbf{C}$ is the $N_A \times N_A$ mutual coupling matrix. It should be noted that the coupling-free model, discussed in (2.6), is a particular case of (5.1) corresponding to $\mathbf{C}$ being an identity matrix. The covariance matrix of the measurements in (5.1) is given by

$$\mathbf{R}_{xx} = E\{\mathbf{x}(t)\mathbf{x}^H(t)\} = \mathbf{CAR}_{ss}\mathbf{A}^H\mathbf{C}^H + \sigma_n^2\mathbf{I}. \tag{5.2}$$

Proceeding with the vectorization and spatial smoothing, followed by DOA estimation without compensating for the MCM, is likely to degrade performance, owing to the mismatch between the assumed model (2.6) and the actual measurements (5.1). The severity of performance degradation, however, is a function of the array configuration and the choice of antennas, as shown in the following section.

## 5.2. Mutual Coupling Impact on DOA Estimation

The performance degradations due to mutual coupling effect is quantified in terms of DOA estimation accuracy for three different non-uniform linear array configurations, namely, the minimum redundancy, nested, and co-prime geometries. For comparison, the performance of a ULA in the presence of mutual coupling is also provided.

### 5.2.1. Mutual Coupling Matrix Modeling and Measurement

The mutual coupling matrix for each considered array configuration is modeled using the RMIM [60]. Two conditions must be satisfied in order to render the application of this method feasible [60, 61]. First, the array should be in the receiving mode. Second, the antenna elements should be terminated with a known load impedance $Z_L$. Assuming these conditions have been fulfilled, the received voltage across the terminal load of a particular antenna can be expressed as a superposition of two external excitations

$$\mathrm{v}_i = Z_L \mathrm{i}_i = \mathrm{w}_i + \tilde{\mathrm{v}}_i, i = 1, 2, \dots, N_A \tag{5.3}$$

where $\mathrm{v}_i$ is the terminal load voltage of the $i$th antenna, $\mathrm{i}_i$ is the current induced in the $i$th antenna, $\mathrm{w}_i$ is the voltage due to the external sources, and $\tilde{\mathrm{v}}_i$ is the voltage due to the mutual coupling from the other elements in the array. The coupled voltage $\tilde{\mathrm{v}}_i$ is given by

$$\tilde{\mathrm{v}}_i = Z_{i,1}\mathrm{i}_1 + Z_{i,2}\mathrm{i}_2 + \cdots + Z_{i,i-1}\mathrm{i}_{i-1} + Z_{i,i+1}\mathrm{i}_{i+1} + \cdots + Z_{i,N_A}\mathrm{i}_{N_A}, \tag{5.4}$$

where $Z_{i,j}$ is the receiving mutual impedance between the $i$th and $j$th elements. Substituting (5.3) in (5.4) and rearranging, the uncoupled voltages $w_i$, $i = 1,2,\ldots,N_A$ can be stacked in a vector $\mathbf{w}$ as

$$\mathbf{w} = \mathbf{Z}\mathbf{v} = \begin{bmatrix} 1 & -Z_{1,2}/Z_L & \cdots & -Z_{1,N_A}/Z_L \\ -Z_{2,1}/Z_L & 1 & \cdots & -Z_{2,N_A}/Z_L \\ \vdots & \vdots & \ddots & \vdots \\ -Z_{N_A,1}/Z_L & -Z_{N_A,2}/Z_L & \cdots & 1 \end{bmatrix} \begin{bmatrix} v_1 \\ v_2 \\ \vdots \\ v_{N_A} \end{bmatrix}, \tag{5.5}$$

where $\mathbf{Z}$ is the mutual impedance matrix.

In order to determine the elements of $\mathbf{Z}$, $R$ plane waves with different DOAs $\{\theta_1, \ldots, \theta_R\}$ are individually used to excite the array, and the corresponding received voltages, $v_m^{(r)}, m = 1, \ldots, N_A, r = 1, \ldots, R$, are recorded. $v_m^{(r)}$ denotes the received voltage at the $m$th array element when the $r$th plane wave is impinging on the array. The same set of plane waves is also used to excite each array element in isolation in order to measure the uncoupled voltages $w_m^{(r)}$. Given $v_m^{(r)}$ and $w_m^{(r)}$ for all $r$, the following system of linear equations is solved for each antenna element in order to compute the corresponding mutual impedance values.

$$\begin{bmatrix} v_m^{(1)} - w_m^{(1)} \\ v_m^{(2)} - w_m^{(2)} \\ \vdots \\ v_m^{(R)} - w_m^{(R)} \end{bmatrix} = \begin{bmatrix} v_1^{(1)} & \cdots & v_{m-1}^{(1)} & v_{m+1}^{(1)} & \cdots & v_{N_A}^{(1)} \\ v_1^{(2)} & \cdots & v_{m-1}^{(2)} & v_{m+1}^{(2)} & \cdots & v_{N_A}^{(2)} \\ \vdots & \ddots & \vdots & \vdots & \ddots & \vdots \\ v_1^{(R)} & \cdots & v_{m-1}^{(R)} & v_{m+1}^{(R)} & \cdots & v_{N_A}^{(R)} \end{bmatrix} \begin{bmatrix} Z_{m,1} \\ \vdots \\ Z_{m,m-1} \\ Z_{m,m+1} \\ \vdots \\ Z_{m,N_A} \end{bmatrix}. \tag{5.6}$$

In order to compute the mutual impedance between each element and the remaining elements in the array, the number of planes waves $R$ should be greater than or equal to $(N_A - 1)$ [60]. Once the matrix $\mathbf{Z}$ has been determined, the MCM is computed as its inverse, i.e., $\mathbf{C} = \mathbf{Z}^{-1}$.

## 5.2.2. Performance Comparisons

In this section, the effect of mutual coupling on the DOA estimation performance is investigated for the different array configurations. Two different antenna types, namely, a dipole antenna and a rectangular microstrip or the so-called patch antenna, are considered as array elements. Each antenna is designed for operation at 3 GHz. The dipoles are chosen as half-wavelength at 3 GHz. Each rectangular patch element has dimensions $L_p = 31.18$ mm and $W_p = 46.64$ mm, where $L_p$ and $W_p$ correspond to the resonant length and radiating edge of the patch, respectively. The patch antenna is printed on a 2.87 mm lossless FR4 substrate with dielectric constant of 2.2, as shown in Fig. 5.1(a). The ground plane is assumed to be infinite. The patch antenna is modeled using FEKO and the corresponding gain pattern is shown in Fig. 5.1(b). This antenna is directive with a maximum gain at $\theta = 0°$ and nulls at $\pm 90°$. The patch elements in the array are positioned with their resonant edges facing each other, as shown in Fig. 5.1(c) which depicts a six-element uniform linear patch array with an inter-element spacing of half-wavelength at 3 GHz.

For each array geometry, the number of elements, $N_A$, is varied from four to ten with a step size of two. The element positions of the corresponding MRA configurations are provided in Table 5.1, while those for nested and co-prime geometries are presented in Tables 5.2 and 5.3, respectively. It should be noted that, in case of MRAs, more than one array structure is available for $N_A > 4$. The configuration which has the least number of element pairs separated by half-wavelength is chosen. For co-prime arrays, for each $N_A$, the configuration with $M = N_A/2, N = M + 1$ is chosen. This choice was shown to have operational advantages in [18] and [34]. Further, for nested arrays, the configurations with $N_1 = N_2 = N_A/2$ are employed; this choice maximizes the DOFs for a given number of antennas [4]. FEKO is used to model the various microstrip and dipole array configurations and measure the required voltages for the RMIM. The

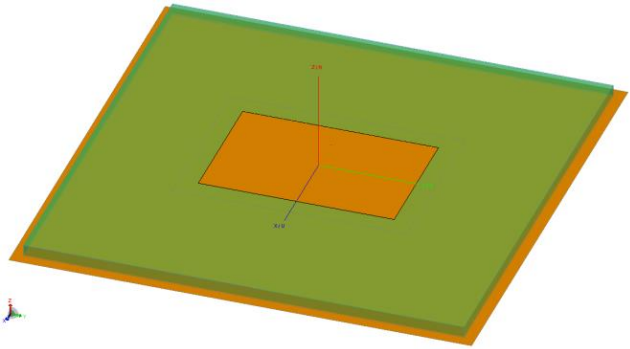

(a)

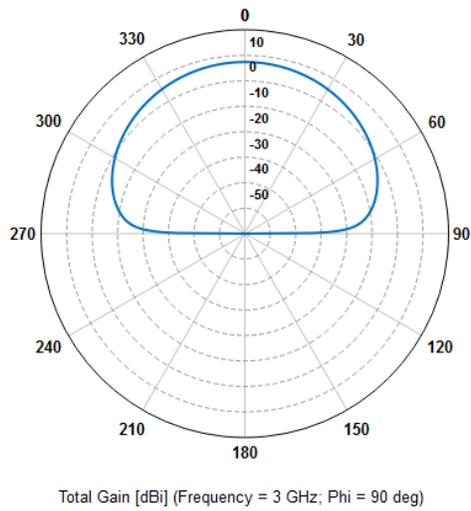

Total Gain [dBi] (Frequency = 3 GHz; Phi = 90 deg)

(b)

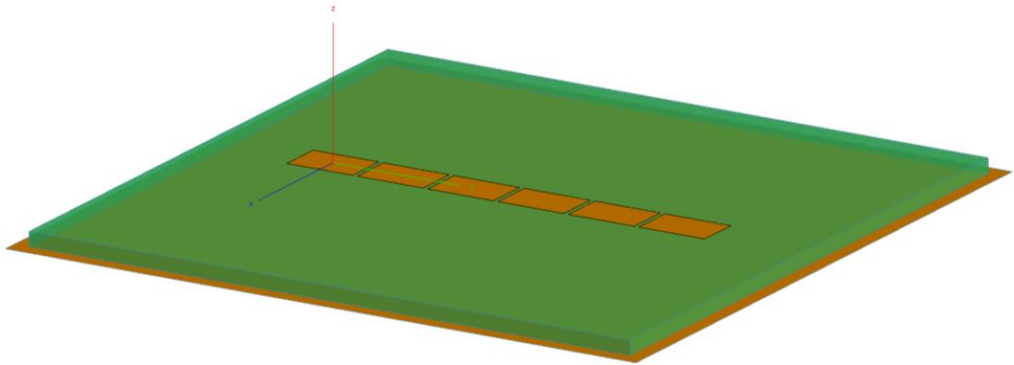

(c)

**Figure 5.1:** (a) Patch antenna, (b) gain pattern (dBi) of a single element in isolation, (c) six-element uniform linear patch array.

**Table 5.1:** Minimum redundancy array configurations

| $N_A$ | Positions |
|---|---|
| 4 | $[0, 1, 4, 6]d_0$ |
| 6 | $[0, 1, 6, 9, 11, 13]d_0$ |
| 8 | $[0, 1, 4, 10, 16, 18, 21, 23]d_0$ |
| 10 | $[0, 1, 3, 6, 13, 20, 27, 31, 35, 36]d_0$ |

**Table 5.2:** Nested array configurations

| $N_A$ | Positions |
|---|---|
| 4 | $[0, 1, 2, 5]d_0$ |
| 6 | $[0, 1, 2, 3, 7, 11]d_0$ |
| 8 | $[0, 1, 2, 3, 4, 9, 14, 19]d_0$ |
| 10 | $[0, 1, 2, 3, 4, 5, 11, 17, 23, 29]d_0$ |

**Table 5.3:** Co-prime array configurations

| $N_A$ | M | N | Positions |
|---|---|---|---|
| 4 | 2 | 3 | $[0, 2, 3, 4]d_0$ |
| 6 | 3 | 4 | $[0, 3, 4, 6, 8, 9]d_0$ |
| 8 | 4 | 5 | $[0, 4, 5, 8, 10, 12, 15, 16]d_0$ |
| 10 | 5 | 6 | $[0, 5, 6, 10, 12, 15, 18, 20, 24, 25]d_0$ |

corresponding mutual impedance and mutual coupling matrices are then computed for the different array geometries with varying number of elements. In the RMIM, the number of plane waves $R$ is set to 16 for all array configurations. The directions of the plane waves are uniformly distributed between $-74°$ and $76°$.

For each combination of array configuration, antenna type, and total number of elements, 1,000 Monte Carlo runs are performed with two sources at a fixed separation in the reduced angular coordinate, $u = \sin\theta$. That is, for each run, the first source direction $u_1$ is randomly chosen to lie between –0.95 and 0.95 and the second source direction $u_2$ is selected so that $\Delta u = |u_1 - u_2|$ is kept constant. Two source separations, $\Delta u = 0.1$ and $\Delta u = 0.2$, are considered. The model in (5.1) is used to generate the array measurements, with the SNR set to 0 dB. The total number of snapshots per run is chosen as 10,000. This high number is selected to remove the influence of i) varying coarray redundancy of different array configurations, and ii) small sample size for correlation matrix estimation as a sample average. Spatial smoothing method is applied in conjunction with MUSIC to estimate the DOAs without compensating for the MCM. It should be noted that, in case of co-prime configurations, the DOA estimation only exploits the contiguous part of the coarray. The estimation accuracy is evaluated in terms of the average RMSE which is given by

$$\overline{RMSE} = \frac{1}{D} \sum_{d=1}^{D} \sqrt{\frac{1}{N_{MC}} \sum_{n=1}^{N_{MC}} (\hat{u}_{d,n} - u_d)^2}, \tag{5.7}$$

where $N_{MC}$ is the total number of Monte Carlo runs and $\hat{u}_{d,n}$ is the estimate of the $d$th source at the $n$th Monte Carlo run.

### A.    *Dipole Arrays*

Fig. 5.2(a) depicts the average RMSE as a function of the number of elements for all considered geometries when $\Delta u = 0.1$, while the RMSE for $\Delta u = 0.2$ is plotted in Fig. 5.2(c). For reference, the corresponding RMSE plots in the coupling-free scenario are shown in Fig. 5.2(b) and Fig. 5.2(d), respectively. By comparing Fig. 5.2(a) to Fig. 5.2(b), and Fig. 5.2(c) to Fig. 5.2(d), it can be noticed that the results for the coupling-free scenario exhibit much smaller RMSE values than those in the presence of mutual coupling, thereby confirming the effect of mutual coupling on the DOA estimation performance. By examining Fig. 5.2(a) and Fig. 5.2(c),

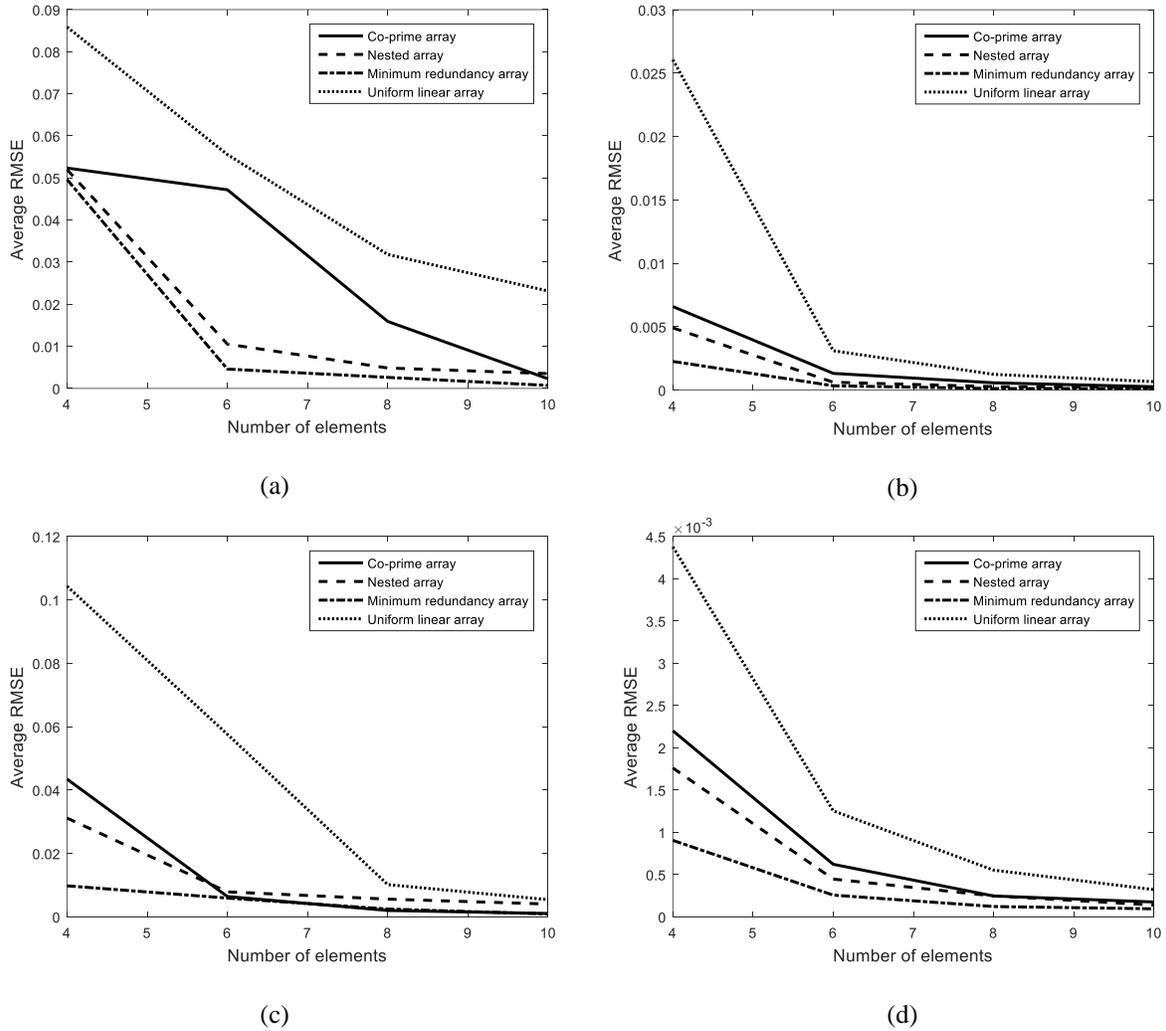

**Figure 5.2:** Dipole arrays: Average RMSE for different array geometries and different number of elements at SNR = 0dB, (a) $\Delta u = 0.1$ in the presence of mutual, (b) $\Delta u = 0.1$ in the absence of mutual coupling, (c) $\Delta u = 0.2$ in the presence of mutual coupling, (d) $\Delta u = 0.2$ in the absence of mutual coupling.

several observations can be made. First, the estimation error decreases as the array size increases for all configurations. Since mutual coupling depends on the distance between the array elements, larger arrays provide a much sparser MCM as compared to smaller arrays, thereby reducing the overall effect on performance. Second, irrespective of the number of elements, the ULA provides the worst performance while the MRA achieves the best performance for both source separations. This is expected because i) the ULA has the highest number of element pairs that are half-wavelength apart, ii) all considered MRAs have a reduced number of pairs of antennas separated by half-wavelength, and iii) the MRAs provide both the largest array size for a given number of antennas and largest filled coarray aperture, leading to better resolution capability. Finally, for $\Delta u = 0.2$, the co-prime array provides better performance than the nested array for $N_A = 6$, 8, and 10, as seen in Fig. 5.2(c). This is expected since the nested array has a greater number of element pairs separated by half-wavelength. For $N_A = 4$, however, the nested array outperforms the co-prime array. This can be explained by examining the two corresponding array structures in Tables 5.2 and 5.3. Both arrays have three contiguous elements at half-wavelength spacing, while the fourth element is closer to its nearest neighbor in the co-prime array as compared to the nested array. In the case of $\Delta u = 0.1$, the roles are reversed for $N_A = 6$ and $N_A = 8$, where the nested array outperforms the co-prime array. This is primarily due to the difference in the corresponding resolution capabilities. As mentioned earlier, since the difference coarray corresponding to a co-prime array has holes, a reduced coarray aperture is employed for spatial smoothing based DOA estimation. Even though the coupling effect is larger in nested arrays, its effect on the DOA estimation performance is outweighed by the resolution capability when the sources are closely separated.

### B.    *Microstrip Arrays*

The Monte Carlo experiments that were performed for dipole arrays are repeated for the microstrip arrays. Fig. 5.3 shows the obtained average RMSE plots for the different array configurations and different source separations. By comparing the corresponding plots in Fig. 5.2 and Fig. 5.3, an increase in the average RMSE is observed when using microstrip arrays. This can be attributed to the proximity of the edges of the consecutive elements. Further, similar to the case of dipole arrays, the microstrip MRA provides the smallest estimation error, while the microstrip ULA has the largest error for both source separations. In addition, the estimation error decreases with increasing number of array elements for all configurations. However, unlike the case of dipole arrays, nested arrays outperform the co-prime arrays for both source separations when microstrip antennas are employed. This performance difference between the co-prime and nested arrays for the two antenna types is due to the fact that mutual coupling in microstrip arrays comprises not only the edge coupling but also the coupling due to the presence of surface waves in the substrate. Since the aperture of co-prime arrays is smaller for the same number of antennas, the surface wave coupling influences the performance of co-prime arrays more than that of nested arrays.

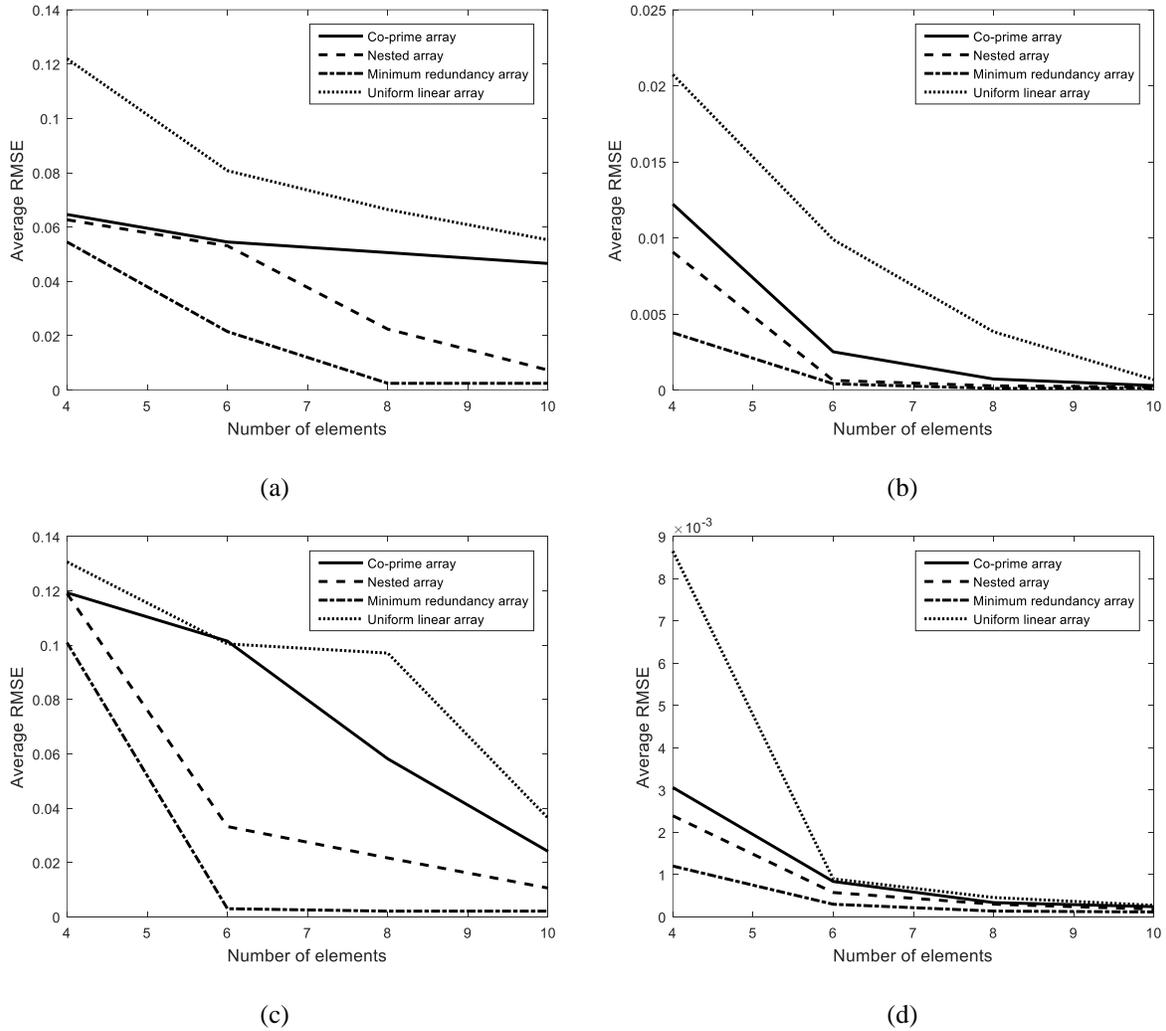

(a)

(b)

(c)

(d)

**Figure 5.3:** Microstrip arrays: Average RMSE for different array geometries and different number of elements at SNR = 0dB, (a) $\Delta u = 0.1$ in the presence of mutual, (b) $\Delta u = 0.1$ in the absence of mutual coupling, (c) $\Delta u = 0.2$ in the presence of mutual coupling, (d) $\Delta u = 0.2$ in the absence of mutual coupling.

To summarize, mutual coupling affects the DOA estimation performance. The degree of performance degradation depends on the array configuration, the number of elements and their types, the source directions, and the source separations.

## 5.3.    Mutual Coupling Compensation

The MCM modeling provides a characterization of the mutual coupling, which can be utilized to account for the coupling in DOA estimation methods. However, in practice, the model can suffer from inaccuracies and, as such, requires frequent re-calibration in order to account for any changes in local conditions. Maintaining an exact MCM model can be cumbersome, if not impossible, in many practical applications. In this section, two compensation methods are proposed for accurate DOA estimation under unknown or imperfectly known MCMs. The first method treats the modeling imperfections as perturbations in the MCM and employs an iterative approach to estimate the source directions and the perturbed MCM. The second method performs joint estimation of the MCM and the source directions simultaneously.

### 5.3.1.   Iterative Approach

Imperfections in the coupling matrix are modeled as arising from perturbations in the mutual impedance matrix, i.e., $\mathbf{Z}_{actual} = \mathbf{Z}_{model} + \Delta\mathbf{Z}$, where $\mathbf{Z}_{actual}$ is the actual mutual impedance matrix, $\mathbf{Z}_{model}$ is the initial modeled mutual impedance matrix, and $\Delta\mathbf{Z}$ is the perturbation matrix. The sources are assumed to be sparse in angle. The angular region of interest is discretized into a finite set of $K$ grid points, where $K \gg D$, with $D$ being the number of sources. Substituting $(\mathbf{Z}_{model} + \Delta\mathbf{Z})^{-1}$ for $\mathbf{C}$ in (5.2) and vectorizing $\mathbf{R}_{xx}$ yields

$$\begin{aligned}
\text{vec}\{\mathbf{R}_{xx}\} &= \text{vec}\{\mathbf{C}\mathbf{A}\mathbf{R}_{ss}\mathbf{A}^H\mathbf{C}^H + \sigma_n^2\mathbf{I}\} \\
&= \{[(\mathbf{Z}_{model} + \Delta\mathbf{Z})^{-1}\mathbf{A}]^* \otimes [(\mathbf{Z}_{model} + \Delta\mathbf{Z})^{-1}\mathbf{A}]\}\text{vec}\{\mathbf{R}_{ss}\} + \sigma_n^2\bar{\mathbf{i}},
\end{aligned} \tag{5.8}$$

In order to solve for the unknowns, namely the perturbations $\Delta\mathbf{Z}$, source directions and powers, and noise variance, a nested optimization problem can be posed as

$$\begin{aligned}
\min_{\Delta\mathbf{Z}} \min_{diag(\bar{\mathbf{R}}_{ss}),\,\sigma_n^2} \big\| \text{vec}\{\hat{\mathbf{R}}_{xx}\} \\
- \{[(\mathbf{Z}_{model} + \Delta\mathbf{Z})^{-1}\bar{\mathbf{A}}]^* \otimes [(\mathbf{Z}_{model} + \Delta\mathbf{Z})^{-1}\bar{\mathbf{A}}]\}\text{vec}\{\bar{\mathbf{R}}_{ss}\} \\
- \sigma_n^2\bar{\mathbf{i}} \big\|_2 + \lambda \|diag(\bar{\mathbf{R}}_{ss})\|_1
\end{aligned} \tag{5.9}$$

where $\hat{\mathbf{R}}_{xx}$ is the covariance matrix obtained as a sample average, $\bar{\mathbf{A}}$ is the $N_A \times K$ array manifold matrix corresponding to the grid of potential directions, $\bar{\mathbf{R}}_{ss}$ is the covariance matrix of the potential sources, and $\lambda$ is the regularization parameter. The elements on the main diagonal of $\bar{\mathbf{R}}_{ss}$ are the powers of the potential sources. The $D$ nonzero diagonal elements correspond to the powers of the actual sources.

The inner optimization in (5.9) over $diag(\bar{\mathbf{R}}_{ss})$ and $\sigma_n^2$ is convex and can be solved using sparse reconstruction techniques with the constraint that the unknowns are nonnegative. The outer minimization over $\Delta\mathbf{Z}$ is non-convex and can be solved by general nonlinear optimization methods. The nested optimization in (5.9) is solved iteratively until the maximum number of iterations is reached or until the cost function stagnates.

In this chapter, the outer optimization problem in (5.9) is solved using CMA-ES [67], which is a nature-based global optimization algorithm. Nature-based optimization algorithms try to emulate natural phenomena, such as swarm intelligence and the Darwinian model of natural evolution, in order to find optimal solutions. These algorithms can deal with highly nonlinear cost functions, which require simultaneous optimization of a large number of parameters. Nature-based optimization algorithms include many categories, such as Genetic Algorithms (GAs) [68], Particle Swarm Optimization (PSO) [69], Evolutionary Programming (EP) [70], and Evolution Strategies (ES) [71]. CMA-ES has been shown to outperform other evolutionary algorithms in many complex electromagnetic problems [72].

CMA-ES is a self-adaptive evolution strategy which requires no parameter tuning. Fig. 5.4 shows the block diagram of the main operation of CMA-ES. The algorithm starts by initializing the parameters to their default values. It then samples a new generation of potential solutions from a multivariate Gaussian distribution using

$$\mathbf{y}_i^{(g+1)} \sim N\left(\mathbf{m}^{(g)},\ \left(\sigma_c^{(g)}\right)^2 \mathbf{C}_c^{(g)}\right), \tag{5.10}$$

where $\mathbf{y}_i^{(g+1)}$ consists of the parameters of the $i$th potential solution at the $(g + 1)$th generation, $\mathbf{m}^{(g)}$ is the mean parameter vector of the best performing members of the previous generation, $\sigma_c^{(g)}$ is the step size, and $\mathbf{C}_c^{(g)}$ is the covariance matrix of the parameters. The parameters of the multivariate Gaussian distribution are then updated sequentially using the best performing members of the generation [67]. The performance of the members is measured by their fitness value or score on the outer optimization in (5.9). This process is then repeated until a termination criterion is met. This criterion can be, for instance, a target fitness value or a maximum number of generations.

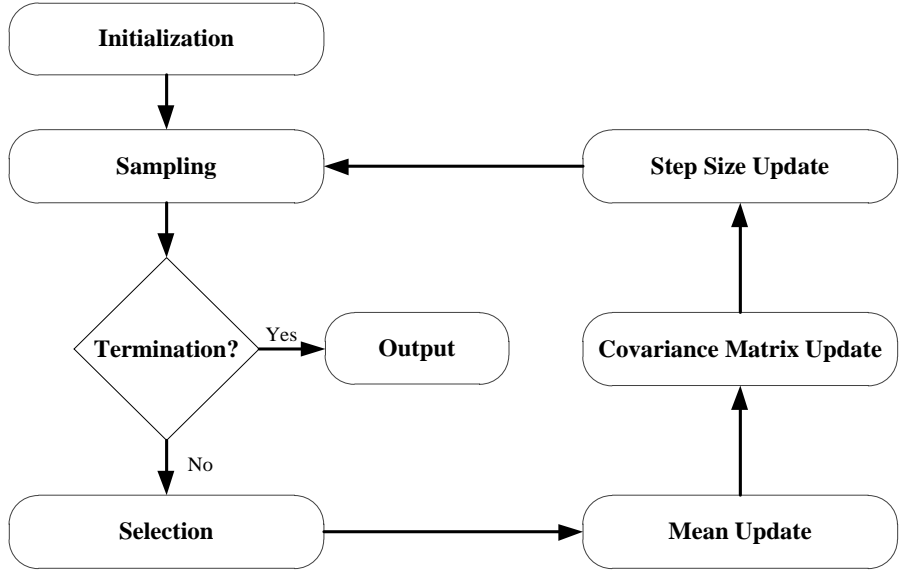

**Figure 5.4:** CMA-ES block diagram.

### 5.3.2. Simultaneous Approach

In this approach, the sources directions and the MCM are simultaneously estimated rather than in an iterative fashion [73]. Starting with the covariance matrix in (5.2), the joint DOA and MCM estimation is achieved by solving

$$\min_{\boldsymbol{\theta}, diag(\mathbf{R}_{ss}), \mathbf{z}, \sigma_n^2} \left\| \widehat{\mathbf{R}}_{xx} - \mathbf{CAR}_{ss} \mathbf{A}^H \mathbf{C}^H - \sigma_n^2 \mathbf{I} \right\|_F^2, \tag{5.11}$$

where $\|\cdot\|_F$ is the Frobenius norm, $\boldsymbol{\theta} = [\theta_1, \theta_2, \dots, \theta_D]^T$ contains the source DOAs, $diag(\mathbf{R}_{ss})$ consists of the source powers, and $\mathbf{z}$ holds the unique elements in $\mathbf{C}^{-1}$. The total number of unknowns is $(2D + 1 + 2|\mathbf{z}|)$, where $|\mathbf{z}|$ is the number of unique elements in the mutual impedance matrix. The multiplier 2 in front of $|\mathbf{z}|$ is due to the entries of $\mathbf{z}$ being complex valued. A mixed-parameter variation of CMA-ES is used to solve (5.11), since the sources DOAs are picked from a predetermined grid while the remaining unknowns are assumed to be continuous parameters [72, 74].

It should be noted that the perturbed mutual impedance matrix model can be employed in (5.11), with the minimization carried out with respect to $\boldsymbol{\theta}, diag(\mathbf{R}_{ss}), \sigma_n^2$, and $\Delta\mathbf{Z}$.

### 5.3.3. Supporting Results

In the first example, a dipole array with a six-element nested configuration is considered. The elements positions are given by $[1, 2, 3, 4, 8, 12]d_0$. The corresponding difference coarray extends from $-11d_0$ to $11d_0$ and is filled with no holes. The length of the dipoles is set to half-wavelength. The corresponding MCM is modeled using the RMIM and the signal model in (5.1) is used to generate the array measurements. The coupling matrix is then assumed to be unknown and is jointly estimated along with the DOAs using the simultaneous method. A total of 11 sources are considered. The sources are uniformly spaced between –0.85 and 0.8 in the reduced angular coordinate $u = \sin\theta$. Spatially and temporally white Gaussian noise is added to the observations, and the SNR is set to 10 dB. The total number of snapshots is fixed to 1,000. Mixed-parameter CMA-ES is used to minimize the cost function in (5.11), where the DOAs are assumed to fall on a grid with 1° step size and the remaining parameters are assumed to be continuous. The search space for the unknown mutual impedance matrix is restricted to be within 10% of the actual values. For the CMA-ES algorithm, the population size and the number of generations are each set to 1,000. Fig. 5.5 shows the estimated spectrum. Clearly, the proposed method is successfully able to compensate for the mutual coupling and estimate the correct source directions. The same array configuration is then used with a smaller number of sources ($D = 5$). The source directions are given by $[-58°, -26°, -1°, 23°, 53°]$. Fig. 5.6(a) shows the success rate as a function of the number of snapshots. For each snapshot value, a total number of 100 Monte Carlo runs are used, and the SNR of all sources is fixed to 10 dB. A solution is deemed successful if each DOA estimate is within $2°$ of the actual one. Fig. 5.6(b) shows the average RMSE values corresponding to the successful solutions. Figs. 5.6(c) and 5.6(d) show the success rate and the average RMSE as a function of SNR, respectively. In this scenario, the number of snapshots is fixed to 1,000 and 50 Monte Carlo runs are used for each SNR value. From these figures, it is evident that the performance of the proposed method improves with the increasing number of snapshots and the increasing SNR. In addition, the number of snapshots has a larger effect than the SNR on the performance. This is expected since this method depends on a good estimate of the covariance matrix using the sample average. It should be noted that the performance can be further improved by increasing the population size and the number of generations of the CMA-ES algorithm.

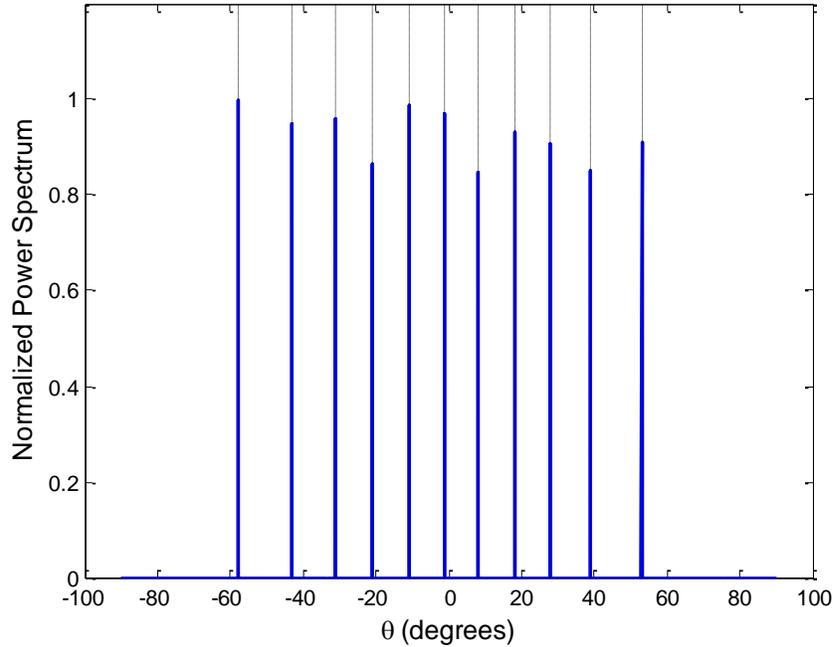

**Figure 5.5:** Simultaneous approach estimated spectrum: Six-element nested-array, $D = 11$.

In the second example, the iterative method is used to estimate the actual MCM along with the DOAs. A six-element microstrip array with an MRA configuration is used. The elements positions are given by $[0, 1, 6, 9, 11, 13]d_0$, and each microstrip element is similar to the one modeled in Section 5.2.2. The MCM is modeled using the RMIM and the data measurements are generated using the model in 5.1. The MCM is then perturbed to emulate the effect of changes in local conditions. The perturbations are drawn from uniform distributions that assume values between $-25\%$ and $25\%$ of the actual values. Eight sources, uniformly spaced between $u = -0.7$ and $u = 0.6$, are considered. The SNR of all sources is set to 10 dB and the number of snapshots is equal to 1,000. Fig. 5.7(a) shows the estimated spectrum using MUSIC with spatial smoothing without accounting for mutual coupling. Clearly, the estimation performance is severely degraded since the mutual coupling is not accounted for. Fig. 5.7(b) depicts the initial estimated spectrum, while Figs. 5.7(c) and 5.7(d) show the estimated spectra after the first and tenth iterations, respectively. The initial estimated spectrum of Fig. 5.7(b) is based on solving the inner optimization problem in (5.9) with the perturbations set to zero. The initial spectrum completely misses one source, provides biased estimates for some sources, and exhibits spurious peaks. The estimated spectrum after the first iteration in Fig. 5.7(c) finds all the sources, but exhibits some spurious peaks. The performance of the method improves with increasing number of iterations, and after ten iterations, all the sources are correctly estimated, as shown in Fig. 5.7(d). In order to validate the convergence of the proposed method, 100 Monte Carlo runs are performed. In each run, a new perturbation of the MCM is generated and the minimization is performed over 20 iterations. Fig. 5.8 shows the success rate as a function of the iteration number. In this example, a run is considered if each DOA estimate is within 1° of the actual value. It can be noticed that the success rate improves with the increasing number of iterations and reaches 100% after 20 iterations. It is to be noted that a faster convergence can be reached for smaller perturbations of the MCM.

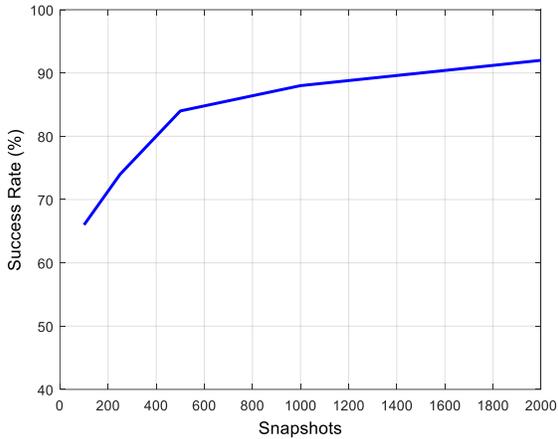
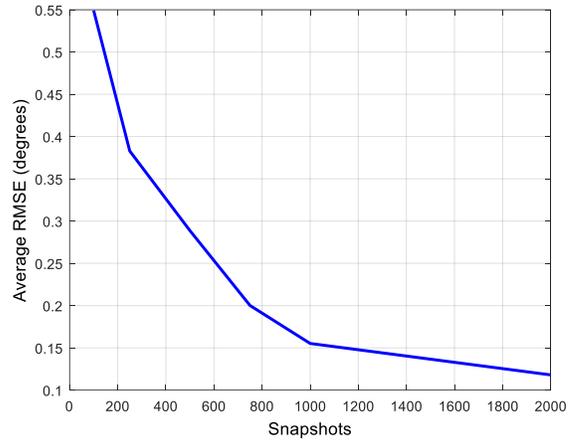

(a)            (b)

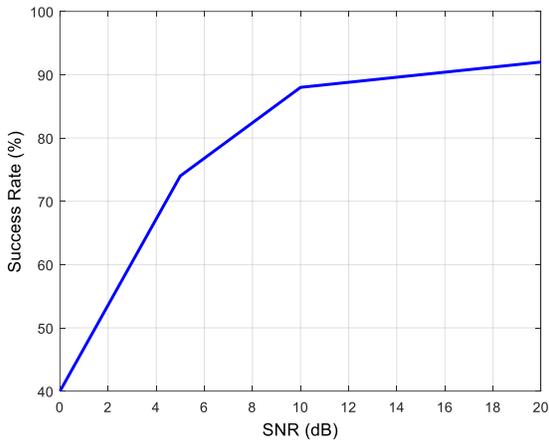
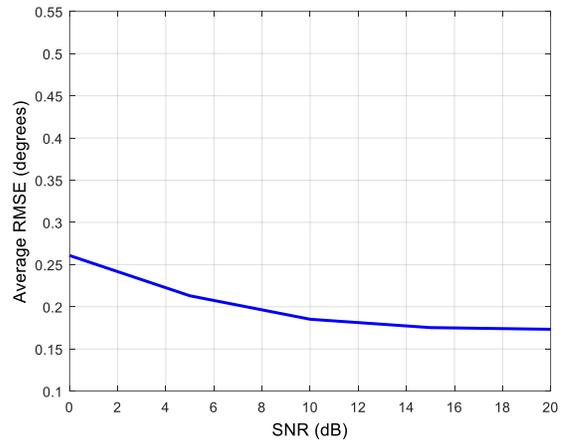

(c)            (d)

**Figure 5.6:** Simultaneous approach: Six-element nested-array, $D = 5$ (a) SNR = 10dB, success rate vs. snapshots, (b) SNR = 10dB, average RMSE vs. snapshots (c) $T = 1000$ snapshots, success rate vs. SNR (d) $T = 1000$ snapshots, average RMSE vs. SNR.

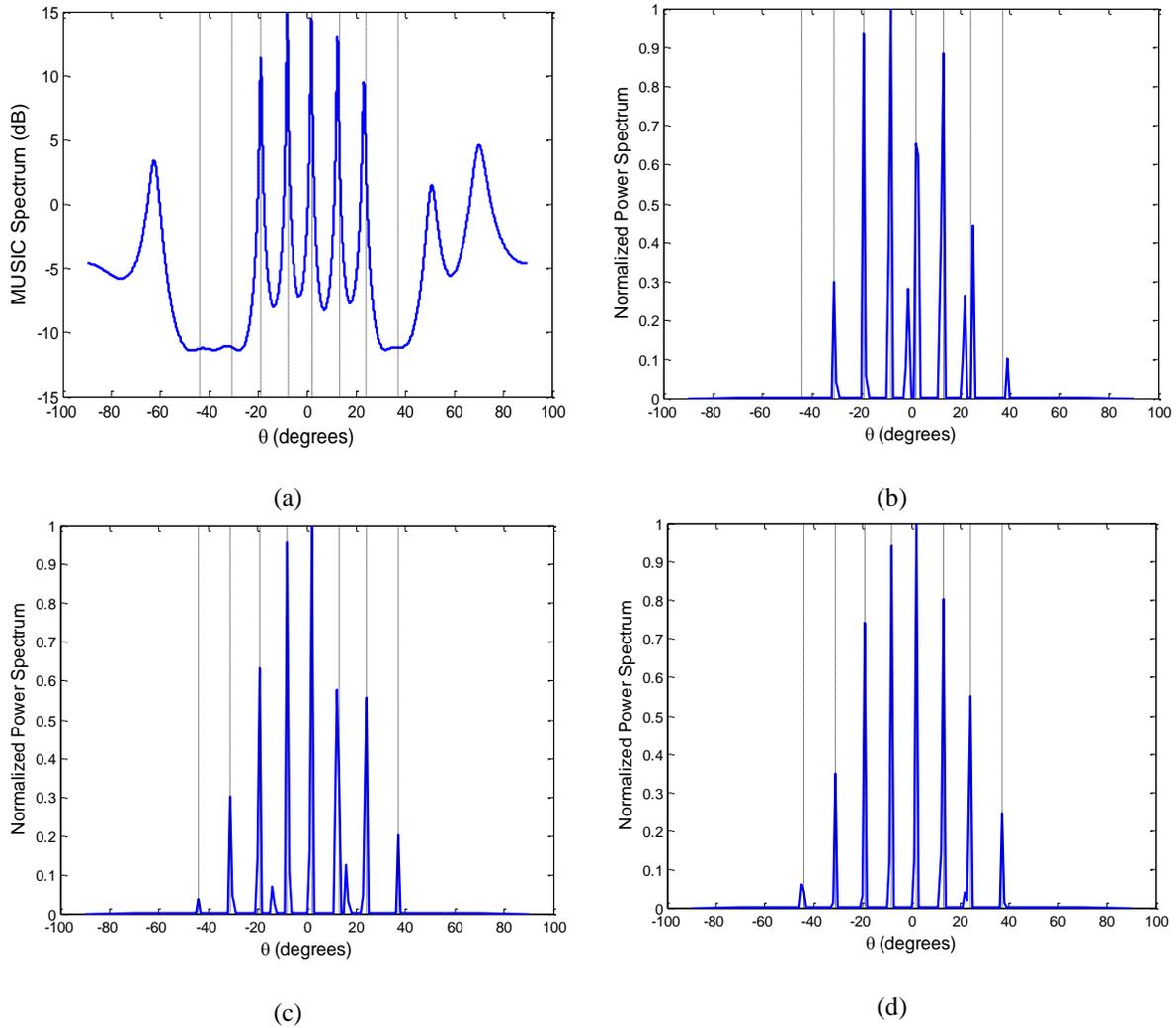

**Figure 5.7:** Six-element microstrip array (a) MUSIC with spatial smoothing without accounting for mutual coupling, Iterative method: (b) initial estimated spectrum, (c) estimated spectrum after first iteration, (d) estimated spectrum after tenth iteration.

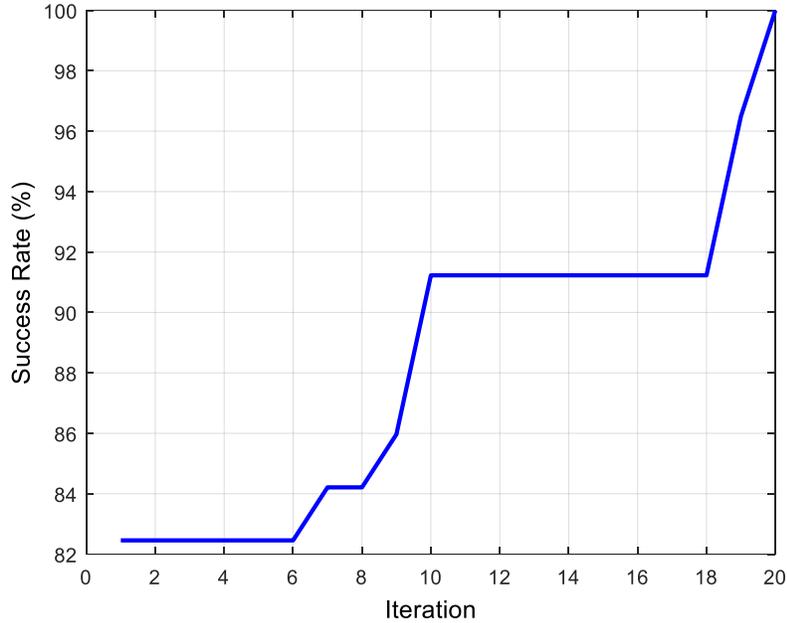

**Figure 5.8:** Six-element microstrip array: Success rate vs. iteration number.

In the final example, a six-element extended co-prime dipole array with $M = 2$ and $N = 3$ is considered. The simultaneous compensation method is used to jointly estimate the MCM and the DOAs. A total of seven uniformly spaced sources between $u = -0.8$ and $u = 0.8$ are considered. The SNR is fixed to 10 dB for all sources and the number of snapshots is set to 1,000 for each run. Mixed-parameter CMA-ES with 1,000 population size and 1,000 generations is used to minimize the cost function. A total of 100 Monte Carlo runs are performed to assess the ability of the proposed method to provide a unique solution. Fig. 5.9 shows the estimated spectrum of one of the successful runs. The DOA estimates of all 100 runs are superimposed in Fig. 5.10(a). It is evident that some of the runs result in wrong or biased estimates. The success rate as a function of the maximum bias of all estimates is plotted in Fig. 5.10(b). For instance, 76 percent of the runs result in a solution that has each estimated DOA within 2° of the actual value. The success rate can be improved by increasing the number of generations used in CMA-ES. This is validated by increasing the number of generations to 5,000 and introducing restarts after each 1,000 generations. Fig. 5.11 shows the corresponding results. It is observed that runs in excess of 60 percent result in unbiased estimates, while all 100 runs produce solutions having each source estimate within 2° of the actual value.

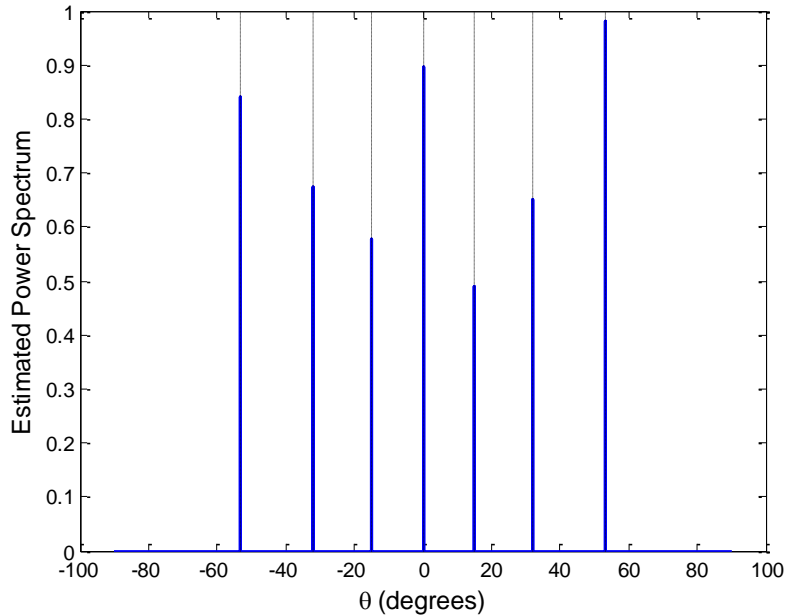

**Figure 5.9:** Estimated spectrum: Extended co-prime array with $M = 2$, $N = 3$, $D = 7$.

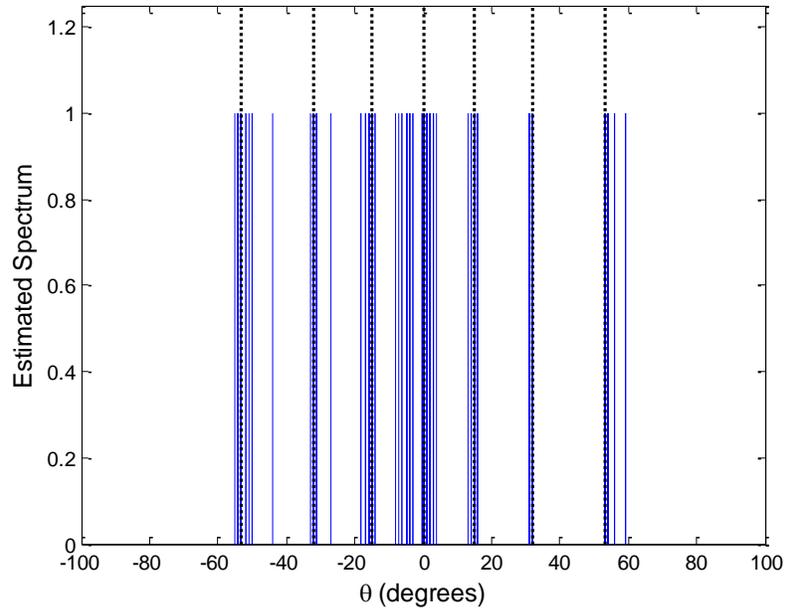

(a)

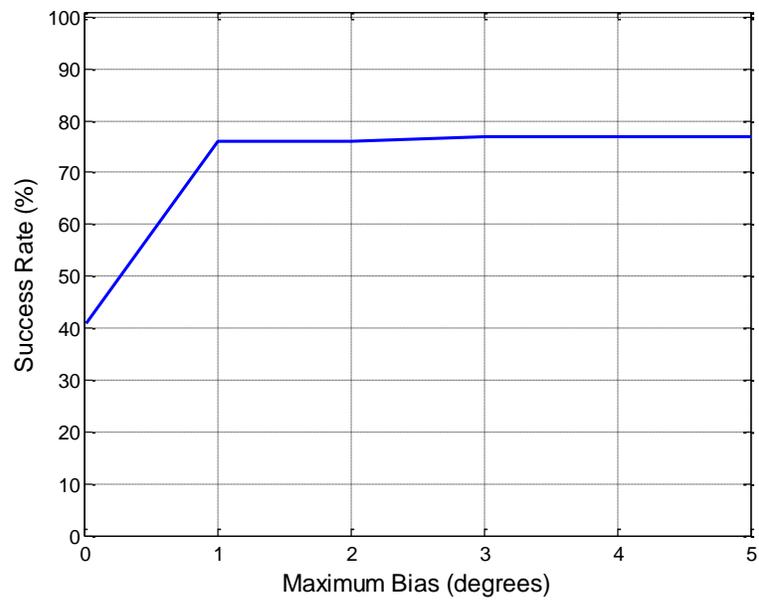

(b)

**Figure 5.10:** CMA-ES population size: 1000 with no restarts, (a) Estimated DOAs of 100 the Monte Carlo runs, (b) Success rate of the obtained solutions.

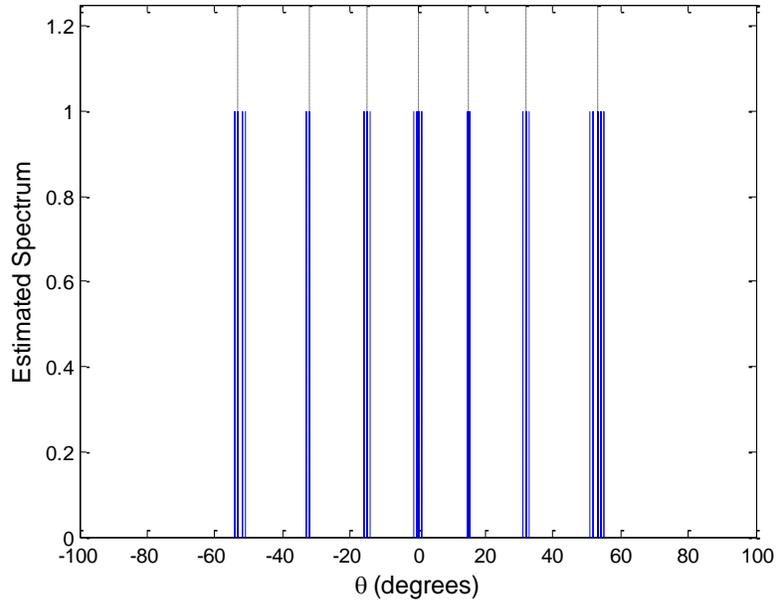

(a)

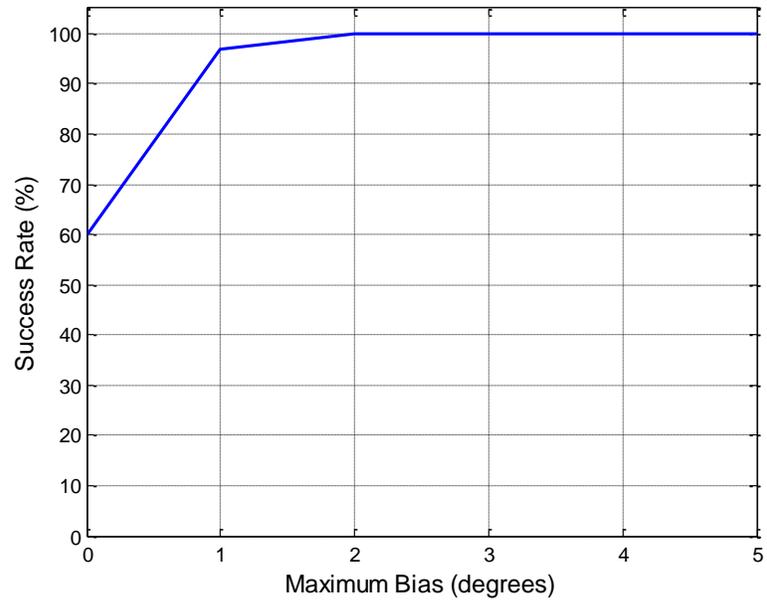

(b)

**Figure 5.11:** CMA-ES population size: 5000 with five restarts, (a) Estimated DOAs of 100 the Monte Carlo runs, (b) Success rate of the obtained solutions.

## 5.4. Mutual Coupling in Multi-Frequency Operation

Investigating mutual coupling under multi-frequency operation requires careful treatment. Assuming that the same physical array, designed for operation at $\omega_0$, is operated at a higher frequency $\omega_1$ ($\omega_1 > \omega_0$), the mutual coupling properties of the array change at $\omega_1$. Since $\omega_1$ is larger than $\omega_0$, the wavelength $\lambda_1$ at $\omega_1$ is smaller than $\lambda_0$. As a result, the unit spacing $d_0$, which is usually set to $\lambda_0/2$, is larger than $d_1 = \lambda_1/2$. This means that the electrical separations between the elements at $\omega_1$ are larger than those at $\omega_0$. However, this does not often lead to a decreased coupling effect at the larger frequency. This is due to the fact that mutual coupling not only depends on the separations between the elements, but also on the physical properties of the antenna elements. Taking a dipole array as an example, the coupling effect depends on the separations between the elements, the lengths of the dipoles, and their radii. At the larger frequency, the electrical length of the dipole becomes larger which might lead to an increased coupling. For illustration, a four-element dipole ULA is considered. The separation between consecutive elements is set to $\lambda_0/2$ and the length of each dipole is also set to $\lambda_0/2$. The antenna is operated at three frequencies: $\omega_1 = \omega_0$, $\omega_2 = 2\omega_0$, and $\omega_3 = 3\omega_0$. The magnitudes of the MCM elements at the three considered frequencies, modeled using the open-circuit method [58], are as follows

$$|C(\omega_1)| = |C(\omega_0)| = \begin{bmatrix} 0.962 & 0.198 & 0.092 & 0.061 \\ 0.198 & 0.941 & 0.190 & 0.092 \\ 0.092 & 0.190 & 0.941 & 0.198 \\ 0.061 & 0.092 & 0.198 & 0.962 \end{bmatrix}, \tag{5.12}$$

$$|C(\omega_2)| = \begin{bmatrix} 0.941 & 0.260 & 0.121 & 0.078 \\ 0.260 & 0.913 & 0.244 & 0.121 \\ 0.121 & 0.244 & 0.913 & 0.260 \\ 0.078 & 0.121 & 0.260 & 0.941 \end{bmatrix}, \tag{5.13}$$

$$|C(\omega_3)| = \begin{bmatrix} 1.001 & 0.025 & 0.014 & 0.016 \\ 0.025 & 1.001 & 0.025 & 0.014 \\ 0.014 & 0.025 & 1.001 & 0.025 \\ 0.016 & 0.014 & 0.025 & 1.001 \end{bmatrix}. \tag{5.14}$$

The off-diagonal terms in $C(\omega_2)$ have a larger magnitude than those in $C(\omega_1)$ which leads to a larger mutual coupling effect. The opposite scenario happens at $\omega_3 = 3\omega_0$, where the off-diagonal terms are much smaller. This leads to the conclusion that in order to examine mutual coupling in multi-frequency operation, the array configuration as well as the element properties need to be considered. This also requires the antenna gains at the multiple frequencies to be accounted for in the signal model. For calibration or compensation, all of these elements need to be taken into account while performing DOA estimation using multi-frequency arrays.

## 5.5. Concluding Remarks

In this chapter, the impact of mutual coupling on DOA estimation performance using non-uniform arrays was investigated. Direction finding accuracy was compared for three different non-uniform array configurations and two antenna element types. The MRA configuration was shown to provide superior estimation performance compared to nested and co-prime array configurations. Further, choice of dipole antennas as array elements fared better in terms of RMSE over microstrip antennas; the latter suffer from additional coupling arising from surface waves in the substrate. Additionally, two mutual coupling compensation methods were proposed for non-uniform arrays. The first method is iterative in nature and assumes imprecisely known MCM. The second method simultaneously estimates the coupling matrix and the DOAs and is better suited to scenarios where no prior knowledge of the MCM is available. Numerical examples were used to demonstrate the effectiveness of the proposed compensation methods.

### 5.5.1. Contributions

The following are the contributions of the research in this chapter.

1) Comparing the effect of mutual coupling on DOA estimation using different array configurations and different antenna element types.

2) Proposing two compensation methods which allow DOA estimation using non-uniform arrays in the presence of mutual coupling.

3) Validating the convergence of the proposed methods.

# CHAPTER VI

## SPARSITY-BASED DIRECTION FINDING OF COHERENT AND UNCORRELATED TARGETS USING ACTIVE NON-UNIFORM ARRAYS

In this chapter, direction finding of a mixture of coherent and uncorrelated targets is performed using sparse reconstruction and active non-uniform arrays. The data measurements from multiple transmit and receive elements can be considered as observations from the sum coarray corresponding to the physical transmit/receive arrays. The vectorized covariance matrix of the sum coarray observations emulates the received data at a virtual array whose elements are given by the difference coarray of the sum coarray (DCSC). Sparse reconstruction is used to fully exploit the significantly enhanced degrees-of-freedom offered by the DCSC for DOA estimation. Simulated data from multiple-input multiple-output (MIMO) minimum redundancy arrays and transmit/receive co-prime arrays are used for performance evaluation of the proposed sparsity-based active sensing approach.

The problem of DOA estimation becomes challenging in the presence of coherent sources or a mixture of coherent and uncorrelated sources, which often arise in the presence of multipath propagation. Traditional subspace-based DOA estimation techniques, such as MUSIC, can no longer be directly applied due to the rank deficiency of the noise-free covariance matrix. Spatial smoothing can be used to restore the rank of the covariance matrix [16]. However, it can only be applied to specific array structures and always results in reducing the DOFs that are available for DOA estimation.

Sparse reconstruction techniques have also been applied for DOA estimation of coherent sources [27, 75, 76]. In [27], an $\ell_1 - $ SVD method was proposed to perform sparsity-based DOA estimation. In this method, SVD is employed to reduce the dimensionality of the signal model, followed by a mixed $\ell_{1,2} -$ norm minimization, which assumes group sparsity across the time snapshots. The number of resolvable sources in $\ell_1 -$SVD is limited by the number of sensors in the array. Joint $\ell_0$ approximation, which is a related method to $\ell_1 - $SVD, was proposed in [75]. This method uses a mixed $\ell_{0,2} -$ norm minimization, instead of $\ell_{1,2}$, in order to enforce sparsity in the reconstructed DOAs. Another sparsity-based method for DOA estimation of more correlated sources than sensors was presented in [76]. This method adopts a dynamic array configuration, wherein different sets of elements of a ULA are activated in different time slots, and uses sparse reconstruction to estimate the vectorized form of the source covariance matrix to resolve the sources.

All of the aforementioned schemes employ passive or receive-only arrays for DOA estimation. An active or transmit/receive sensing method was proposed in [77] for direction finding in a coherent environment. This method generalizes the spatial smoothing decorrelation technique to encompass active arrays, where the transmitters illuminate the field of view, and the receivers detect the reflections from the targets. The recorded data emulates measurements at the corresponding sum coarray. Using the coarray equivalence principle, the sum coarray measurements can be considered as originating from a virtual transmit/receive array, which, compared to the physical transmit/receive array, provides a different tradeoff between the number of resolvable targets and the maximum number of mutually coherent targets that can be resolved. The number of resolvable targets for this active sensing scheme is limited by the number of receivers in the virtual transmit/receive array. In [78], a sparse reconstruction scheme for DOA estimation in co-located MIMO radar was proposed. The received data is arranged in a vector which emulates measurements at the sum coarray, and either $\ell_1 - $SVD or a reweighted minimization is applied to reconstruct the signal. For this method, the number of resolvable targets is limited by the number of sum coarray elements.

In this chapter, direction finding of a mixture of coherent and uncorrelated targets is performed by using the covariance matrix of the data vector that emulates measurements at the sum coarray of active non-uniform arrays. In so doing, the number of DOFs is significantly increased, owing to the fact that the vectorized covariance matrix of the sum coarray observations can be thought of as a single measurement at a virtual array whose elements are given by the difference coarray of the sum coarray. The DCSC has a much higher number of elements compared to the sum coarray itself [79]. Sparse reconstruction is employed to fully exploit the enhanced DOFs by estimating the vectorized form of the source covariance matrix, which is linearly related to the vectorized data covariance matrix of the sum coarray observations. Two different non-uniform array geometries are considered for performance evaluation using simulated data. The first configuration is the MIMO MRA, which maximizes the number of contiguous elements in the DCSC [79], whereas the second is the transmit/receive co-prime arrays. Simulation results clearly demonstrate the superior performance of the proposed scheme over existing methods in terms of the number of resolvable targets for a given number of transmitters/receivers.

The remainder of the chapter is organized as follows. In Section 6.1, the signal model for active sensing is reviewed. The proposed sparsity-based DOA estimation approach is presented in Section 6.2. The maximum number of resolvable targets is discussed in Section 6.3. The performance of the proposed method is evaluated in Section 6.4 through numerical simulations, and Section 6.5 concludes the chapter.

## 6.1. Signal Model

An $M_t$-element linear transmit array and an $N_r$-element linear receive array are considered. The two arrays may or may not share common elements. These arrays are assumed to be co-located so that a target in the far-field appears to have the same direction at all transmitters and receivers. Fig. 6.1 shows a general transmit/receive configuration. The scene is illuminated by multiple sequential narrowband transmissions of center frequency $\omega_0$ from the different transmitters. This group of transmissions, one from each transmitter, is referred to as a single "snapshot". The field of view is assumed to consist of $D$ point targets with directions $[\theta_1, \theta_2, \ldots, \theta_D]$, where $\theta$ is the angle relative to broadside of the transmit or receive array. The target distribution consists of both uncorrelated and coherent targets. Then, the output of the receive array can be expressed as an $M_t N_r \times 1$ vector [80, 81]

$$\mathbf{x}(t) = \sum_{d=1}^{D} \mathbf{a}_t(\theta_d) \otimes \mathbf{a}_r(\theta_d) s_d(t) + \mathbf{n}(t), \tag{6.1}$$

where $s_d(t)$ is the reflection coefficient of the $d$th target at snapshot $t$, and $\mathbf{a}_t(\theta_d)$ and $\mathbf{a}_r(\theta_d)$ are the steering vectors of the transmit and receive arrays corresponding to the direction of the $d$th target, respectively. The $m$th element of $\mathbf{a}_t(\theta_d)$ is given by $\exp(jk_0 t_m \sin \theta_d)$ where $t_m$ is the location of the $m$th transmitter and $k_0$ is the wavenumber at frequency $\omega_0$, and the $n$th element of $\mathbf{a}_r(\theta_d)$ is given by $\exp(jk_0 r_n \sin \theta_d)$ where $r_n$ is the location of the $n$th receiver. The vector $\mathbf{n}(t)$ in (6.1) is the $M_t N_r \times 1$ noise vector. The noise is assumed to be independent and identically distributed following a complex Gaussian distribution.

The term $\mathbf{a}_t(\theta_d) \otimes \mathbf{a}_r(\theta_d)$ in (6.1) is equivalent to the steering vector of a virtual receive-only array, whose elements are given by the sum coarray of the transmit and receive arrays. The sum coarray elements were defined in (2.1). Assuming that the number of unique elements in the sum coarray is $L_{SC}$, a new $L_{SC} \times 1$ received data vector can be formed from (6.1) as

$$\mathbf{x}_{sum}(t) = \mathbf{A}_{sum}\mathbf{s}(t) + \mathbf{n}_{sum}(t),$$ (6.2)

where $\mathbf{A}_{sum} = [\mathbf{a}_{sum}(\theta_1), \mathbf{a}_{sum}(\theta_2), ..., \mathbf{a}_{sum}(\theta_D)]$ is the $L_{SC} \times D$ array manifold matrix corresponding to the sum coarray with $\mathbf{a}_{sum}(\theta_d)$ being the steering vector of the sum coarray corresponding to direction $\theta_d$, $\mathbf{s}(t) = [s_1(t), s_2(t), ..., s_D(t)]^T$, and $\mathbf{n}_{sum}(t)$ is the noise vector at the sum coarray. It should be noted that if two or more transmit/receive element pairs contribute to the same sum coarray point, one of the corresponding measurements could be used in $\mathbf{x}_{sum}(t)$. The $\ell_1 - \text{SVD}$ method can be applied to the sum coarray data vector $\mathbf{x}_{sum}(t)$ for sparsity-based DOA estimation [78]. However, the maximum number of resolvable targets in this case is limited to the number of unique elements in the sum coarray [27].

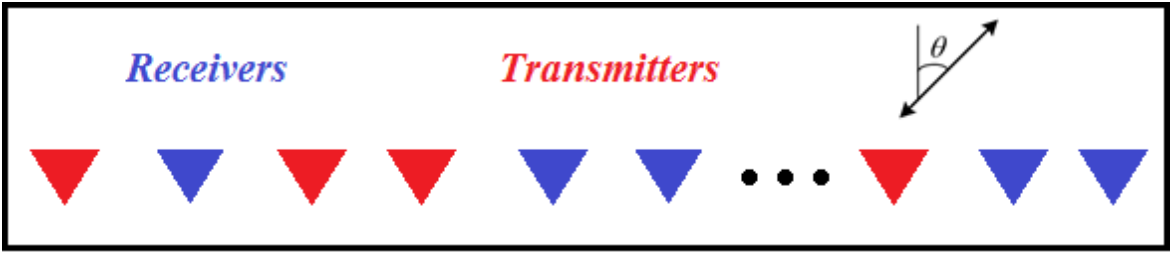

**Figure 6.1:** General transmit/receive configuration.

## 6.2. Proposed Direction Finding Approach

The $L_{sum} \times L_{sum}$ covariance matrix of the sum coarray data can be expressed as

$$\mathbf{R}_{sum} = E\{\mathbf{x}_{sum}(t)\mathbf{x}_{sum}^H(t)\} = \mathbf{A}_{sum}\mathbf{R}_{ss}\mathbf{A}_{sum}^H + \sigma_n^2\mathbf{I},$$ (6.3)

where $E\{\cdot\}$ is the expectation operator, $\sigma_n^2$ is the noise variance, and $\mathbf{I}_{sum}$ is an $L_{SC} \times L_{SC}$ identity matrix. $\mathbf{R}_{ss}$ is the $D \times D$ source correlation matrix, which contains the powers of the reflections from the targets on its main diagonal and the cross-correlations between the targets in the off-diagonal terms. In practice, the covariance matrix is estimated by a sample average over multiple snapshots.

In order to perform DOA estimation of the coherent and uncorrelated targets, $\mathbf{R}_{ss}$ is estimated from $\mathbf{R}_{sum}$ using sparse reconstruction. The angular region of interest is discretized into a finite set of $K \gg D$ grid points, $\{\theta_{g_1}, \theta_{g_2}, \dots, \theta_{g_K}\}$, with $\theta_{g_1}$ and $\theta_{g_K}$ being the limits of the search space. The targets are assumed to be located on the grid. As previously mentioned, several methods can be used to modify the model in order to deal with off-grid targets. The $L_{SC} \times K$ array manifold matrix whose columns are the steering vectors corresponding to the defined angles in the grid is denoted by $\widetilde{\mathbf{A}}_{sum}$, and the $K \times K$ target covariance matrix which holds the auto- and cross-correlations between the potential targets at the defined angles is denoted by $\widetilde{\mathbf{R}}_{ss}$. Equation (6.3) can then be rewritten as

$$\mathbf{R}_{sum} = \widetilde{\mathbf{A}}_{sum}\widetilde{\mathbf{R}}_{ss}\widetilde{\mathbf{A}}_{sum}^{H} + \sigma_n^2 \mathbf{I}_{sum}, \qquad (6.4)$$

Since $K \gg D$, $\widetilde{\mathbf{R}}_{ss}$ is a sparse matrix. Sparse reconstruction can then be applied to estimate $\widetilde{\mathbf{R}}_{ss}$, and, consequently, resolve the targets. The nonzero terms on the main diagonal of $\widetilde{\mathbf{R}}_{ss}$ correspond to the powers of the target reflections present in the field of view, and the nonzero off-diagonal terms correspond to the correlations between the coherent targets. As a result, the target directions can be obtained by identifying the nonzero terms on the main diagonal.

The covariance matrix $\mathbf{R}_{sum}$ is vectorized by stacking its columns to form a tall vector, which emulates a single snapshot at a virtual array whose elements are given by the DCSC of the transmit and receive arrays. With the sum coarray containing $L_{SC}$ unique elements at positions $x_\ell, \ell = 0, \dots, L_{SC} - 1$, the DCSC elements are given by the set $\Omega = \{x_{\ell_1} - x_{\ell_2}, \ell_1 = 0, \dots, L_{SC} - $ 1 and $\ell_2 = 0, \ldots, L_{SC} - 1$}. It can be readily shown that the $L_{SC}^2 \times 1$ vectorized form of the noise-free term of $\mathbf{R}_{sum}$ can be expressed as [76, 82],

$$vec(\widetilde{\mathbf{A}}_{sum}\widetilde{\mathbf{R}}_{ss}\widetilde{\mathbf{A}}_{sum}^H) = (\widetilde{\mathbf{A}}_{sum}^* \otimes \widetilde{\mathbf{A}}_{sum})vec(\widetilde{\mathbf{R}}_{ss}),  \tag{6.5}$$

Given the model in (6.5), the constrained optimization problem for reconstructing the $K^2 \times 1$ $vec(\widetilde{\mathbf{R}}_{ss})$ can be expressed as

$$\widehat{\mathbf{R}}_{ss} = \arg\min_{\widetilde{\mathbf{R}}_{ss}} \left\| vec\left(\mathbf{R}_{sum} - (\widetilde{\mathbf{A}}_{sum}^* \otimes \widetilde{\mathbf{A}}_{sum})vec(\widetilde{\mathbf{R}}_{ss})\right) \right\|_2 + \lambda \left\| vec(\widetilde{\mathbf{R}}_{ss}) \right\|_1.  \tag{6.6}$$

A constraint on the main-diagonal terms of $\widetilde{\mathbf{R}}_{ss}$ to be nonnegative can be added to reflect the fact that the nonzero terms represent powers which are always positive.

## 6.3.    Maximum Number of Resolvable Targets

The maximum number of resolvable targets using the proposed method depends on the number of unique lags in the DCSC and the number of coherent targets. Each pair of coherent targets corresponds to two nonzero off-diagonal terms in $\widetilde{\mathbf{R}}_{ss}$, and each target contributes a nonzero term on the main diagonal. Due to conjugate symmetry in $\widetilde{\mathbf{R}}_{ss}$, only the lower triangle matrix has to be estimated. This implies that, instead of $K^2$ terms, only $K(K+1)/2$ elements of $\widetilde{\mathbf{R}}_{ss}$ need to be estimated. According to [48], the sparsity based minimization problem in (6.6) is guaranteed to have a unique solution under the condition $L_{DCSC} \geq 2L_{nz}$, where $L_{DCSC}$ is equal to the number of independent observations or the number of unique elements in the DCSC, and $L_{nz}$ is the number of nonzero terms in the lower triangle of $\widetilde{\mathbf{R}}_{ss}$, which can be expressed as $L_{nz} = D + C$, where $C$ is the number of pairs of coherent targets.

The number of unique lags $L_{DCSC}$ in the DCSC is a function of the transmit and receive array geometries. For a given number of transmitters and receivers, active array configurations specifically designed to be optimal in the sense that the number of unique elements in the DCSC is maximized, would yield the highest number of resolvable sources. MIMO MRAs are one such type of arrays which are designed under the constraint that the DCSC has no holes [79]. However, the use of such optimal array configurations is not mandatory, and the proposed technique can be applied to other non-uniform arrays, such as co-prime arrays. Table 6.1 summarizes the number of unique elements in the sum coarray and the DCSC of three different implementations (Configurations A, B, and C) of a co-prime array comprising a $(2M - 1)$ element ULA with $N\lambda_0/2$ inter-element spacing and a second ULA having $N$ elements spaced by $M\lambda_0/2$; $M$ and $N$ are co-prime integers, and $\lambda_0$ is the wavelength at the frequency $\omega_0$. Configuration A uses the first ULA to transmit and the second ULA to receive. Configuration B employs the first ULA for transmission and both ULAs for reception. Configuration C uses the entire co-prime array to transmit and receive. These implementations provide different tradeoffs between cost, hardware complexity, and the maximum number of unique elements in the DCSC. It can be observed from Table 6.1 that the advantage of the proposed method over the $\ell_1 - \text{SVD}$ method applied directly to the sum coarray of the co-prime arrays is more evident for higher values of $M$ and $N$. For large $M$ and $N$ values, a three-fold increase in the DOFs occurs for configurations B and C.

**Table 6.1:** Unique elements in sum coarray and difference coarray of co-prime array

|  | $L_{SC}$ | $L_{DCSC}$ |
|---|---|---|
| Configuration A | $(2M - 1)N$ | $(5M - 3)N - M$ |
| Configuration B | $(2M - 1)(N + 1)$ | $(7M - 5)N + M$ |
| Configuration C | $(2M)(N + 1) - 1$ | $(7M - 3)N + M$ |

## 6.4.    Numerical Results

In this section, DOA estimation results for the proposed sparse reconstruction technique using non-uniform active arrays are presented, and a comparison with the $\ell_1 - \text{SVD}$ method is also provided. Both MIMO MRAs and co-prime arrays are considered. The RMSE with respect to the directions is used to compare the two methods.

In the first example, a four-element MIMO MRA, which consists of two receivers positioned at $[0, 7d_0]$ and three transmitters positioned at $[0, d_0, 3d_0]$, is considered. Fig. 6.2 shows the corresponding sum coarray and the DCSC. The sum coarray consists of six elements positioned at $[0, 1, 3, 7, 8, 10]d_0$, whereas the DCSC consists of 21 consecutive virtual elements and its aperture extends from $-10d_0$ to $10d_0$. As such, $\ell_1 - \text{SVD}$ applied to the sum coarray measurements can estimate up to six sources, whereas the proposed method can estimate up to ten nonzero elements in the lower triangle of the source covariance matrix. This is tested by first considering six targets from directions $[-60°, -20°, -15°, 10°, 30°, 40°]$, with the reflections from the first three targets being mutually coherent. The total number of snapshots is set to 500.

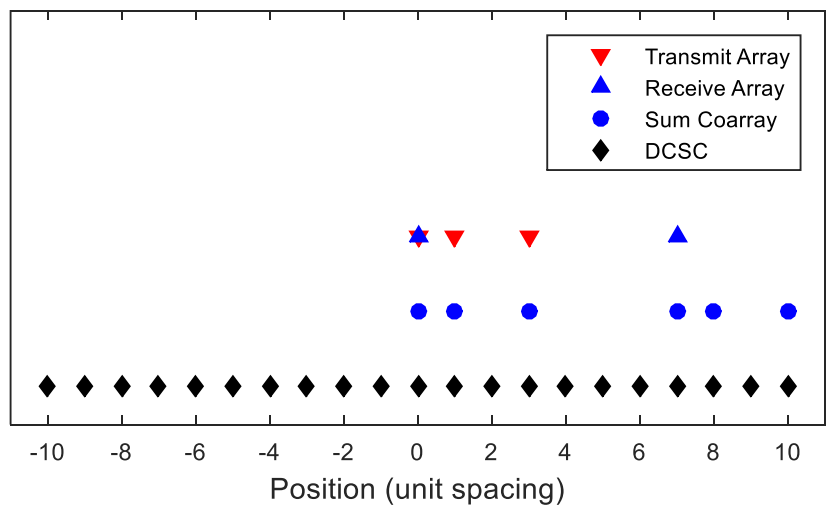

**Figure 6.2:** MIMO MRA, sum coarray, and DCSA.

Spatially and temporally white Gaussian noise is added to the observations, and the SNR for the six targets is set to $[10, 0, 5, 0, 10, 0]$ dB. The search space is discretized uniformly from $-90°$ and $90°$ with $1°$ increment, and the regularization parameter $\lambda$ is set empirically to 0.5 for the proposed method. The normalized spectrum obtained using $\ell_1 - \text{SVD}$ and averaged across the snapshots is shown in Fig. 6.3(a). Fig. 6.3(b) depicts the normalized values on the main diagonal of the estimated source covariance matrix using the proposed approach. It can be observed that the proposed method has correctly estimated the target directions. However, $\ell_1 - \text{SVD}$ misses two targets with low SNR, and produces biased estimates for the remaining targets. The RMSE is 0° for the proposed method.

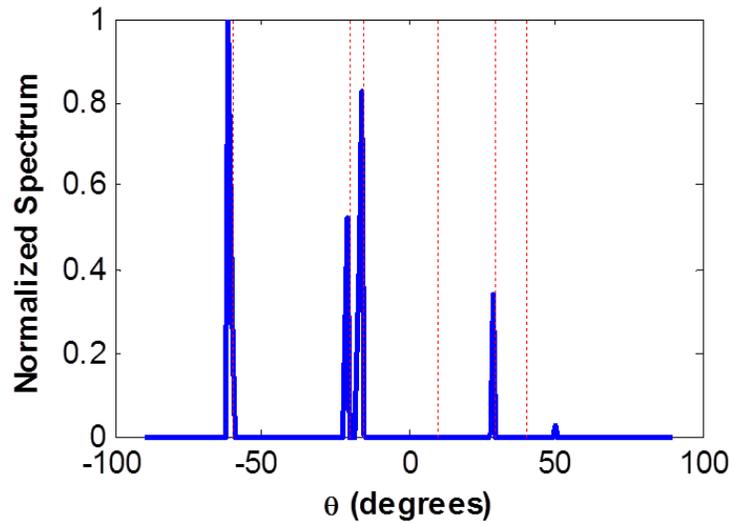

(a)

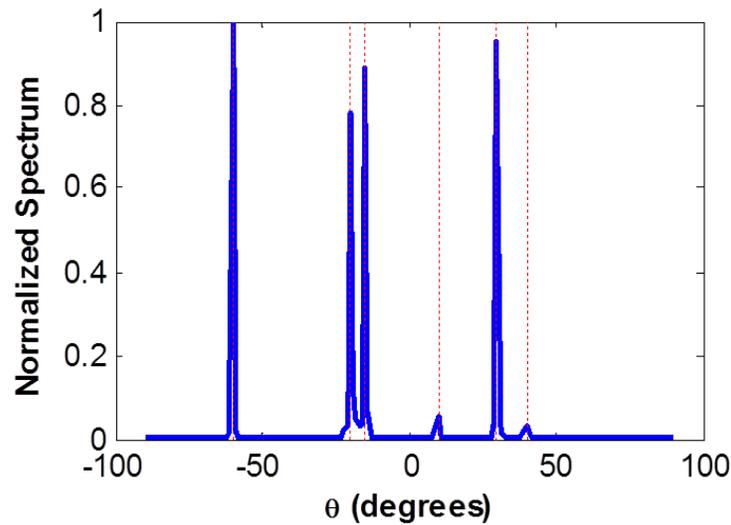

(b)

**Figure 6.3:** MIMO MRA, six targets (3 mutually coherent), (a) $\ell_1 - \text{SVD}$, (b) Proposed method.

Next, the same MIMO MRA is used, but the number of targets is increased to seven with the first three being mutually coherent. The targets are positioned at $[-55°, -40°, -15°, 5°, 20°, 45°, 65°]$. A 10 dB SNR is used for all the targets. The regularization parameter $\lambda$ is set to 0.3. Figs. 6.4(a) and 6.4(b) show the estimated spectra using $\ell_1 - \text{SVD}$ and the proposed method, respectively. Clearly, $\ell_1 - \text{SVD}$ fails to estimate the targets since the total number of targets exceeds the number of sum coarray elements. The proposed method, on the other hand, is successful since the number of nonzero elements in the lower triangle is equal to ten. The corresponding RMSE is 0.24°. The number of targets is then increased to ten, which is equal to the maximum number of nonzero elements in the lower triangle of the covariance matrix that can be estimated using the proposed method. The target directions are uniformly spaced between –50° and 50°. The reflections from all the targets are assumed to be uncorrelated in this example, and the other simulation parameters are kept the same as before. Fig. 6.5(a) shows the estimated spectrum using $\ell_1 - \text{SVD}$, which fails to estimate the target directions because the number of targets is larger than the number of sum coarray elements. The estimated spectrum using the proposed approach is shown in Fig. 6.5(b). As expected, this method correctly estimates all the DOAs, and the RMSE is equal to 0.2° in this example.

Next, a co-prime array with $M = 3$ and $N = 4$ is considered. The first ULA consists of five physical sensors with positions $[4, 8, 12, 16, 20]d_0$, and the second ULA consists of four sensors positioned at $[0, 3, 6, 9]d_0$. Configuration B is considered, which implies that the first ULA is used to transmit and both ULAs are used to receive. Fig. 6.6 shows the transmit array, the receive array, the corresponding sum coarray, and the DCSC. The sum coarray consists of 25 elements, and the DCSC consists of 67 elements. A total number of 30 targets, uniformly spaced between −0.95 and 0.95 in the reduced angular coordinate $\sin(\theta)$, is considered with three targets being mutually coherent. The rest of the simulation parameters are the same as in the previous examples. Figs. 6.7(a) and 6.7(b) show the estimated spectra using $\ell_1-$SVD and the proposed method, respectively. It is evident that $\ell_1-$SVD fails to estimate the target directions, since the number of targets exceeds the number of sum coarray elements. The proposed method correctly estimates the DOAs since the number of nonzero elements in the lower triangle of the source covariance matrix in this case is $L_{nz} = D + C = 30 + 3 = 33$, and the number of unique elements in the DCSC is $L_{DCSC} = 67$ which is greater than $2L_{nz}$. The corresponding RMSE is $0.03°$.

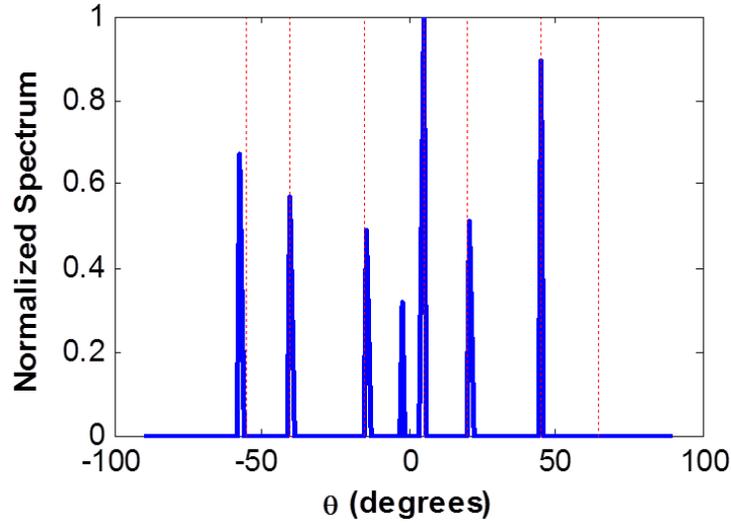

(a)

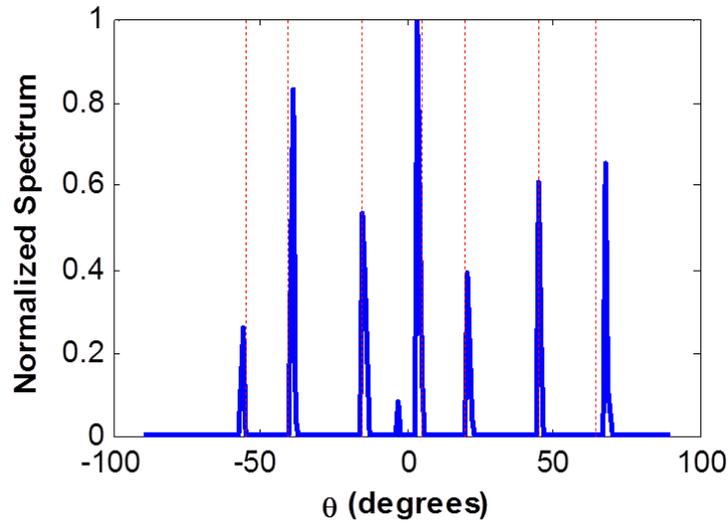

(b)

**Figure 6.4:** MIMO MRA, seven targets (3 mutually coherent), (a) $\ell_{1-}$SVD, (b) Proposed method.

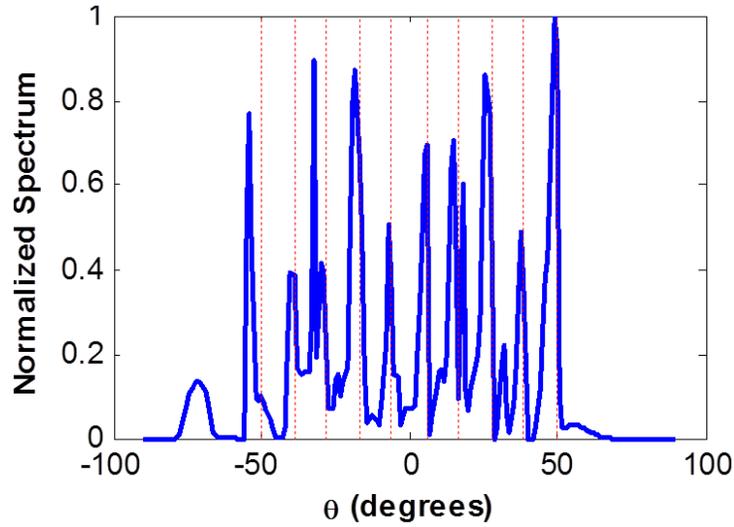

(a)

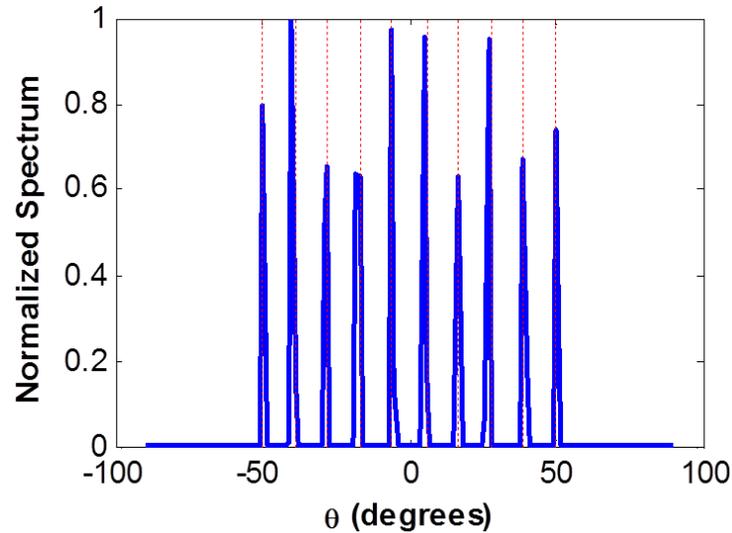

(b)

**Figure 6.5:** MIMO MRA, ten uncorrelated targets, (a) $\ell_1-$SVD, (b) Proposed method.

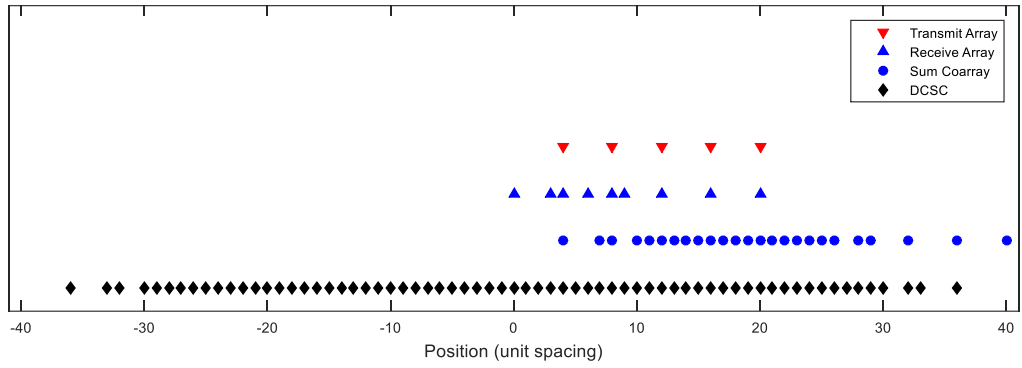

**Figure 6.6:** Co-prime array ($M = 3$, $N = 4$) with Configuration B, sum coarray, and DCSA.

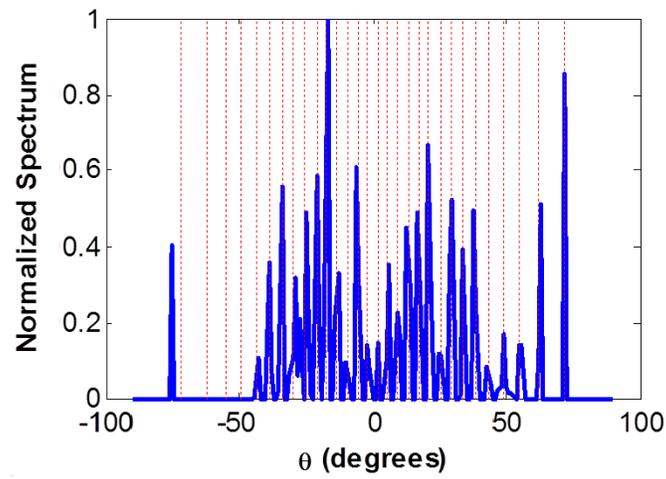

(a)

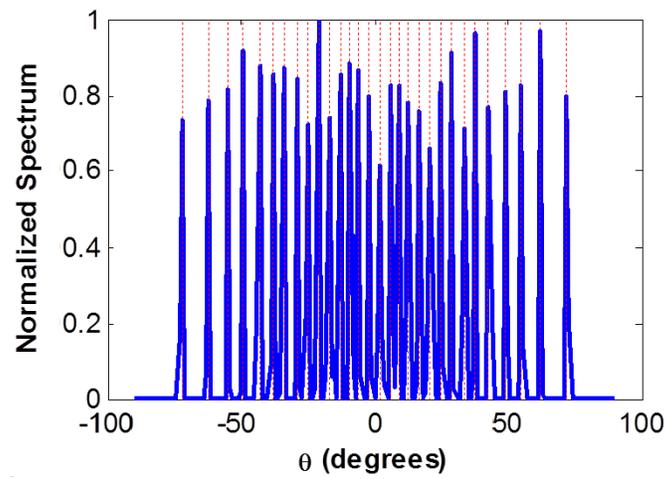

(b)

**Figure 6.7:** Co-prime array, 30 targets (3 mutually coherent), (a) $\ell_{1\_}$SVD, (b) proposed method.

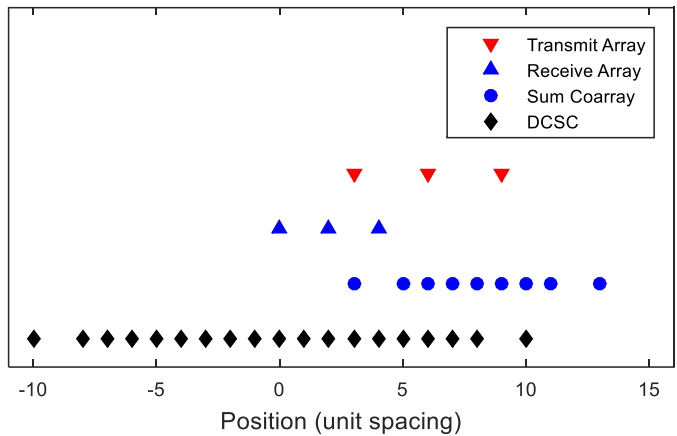

**Figure 6.8:** Co-prime array ($M = 2$, $N = 3$) with Configuration A, sum coarray, and DCSA.

In the final example, the proposed method is tested for two closely separated targets with varying reflection powers. First, a co-prime array with $M = 2$ and $N = 3$ is considered. The first ULA consists of three elements with positions $[3, 6, 9]d_0$ and the second ULA consists of three elements positioned at $[0, 2, 4]d_0$. Fig. 6.8, shows the transmit and receive arrays, the sum coarray, and the DCSC. Two targets are considered. The power of the reflection from the first target is fixed to 20 dB and the power of the second target is varied between 0 dB and 20 dB with 5 dB increments. The direction of the target is fixed to $-1°$ and the direction of the second target is varied. The total number of snapshots is set to 500. Fig. 6.9 shows the reconstructed spectrums when the second target is positioned at $1°$. Figs. 6.10 and 6.11 show the same set of results when the second target is positioned at $5°$ and $10°$, respectively. Several conclusions can be made by examining these figures. First, when the source separation is small, i.e., $\Delta\theta = 2°$ or $\Delta\theta = 6°$, the directions of the two targets cannot be estimated correctly. The target with the small SNR is completely missed. Second, the two targets can be resolved when $\Delta SNR = 0\ dB$

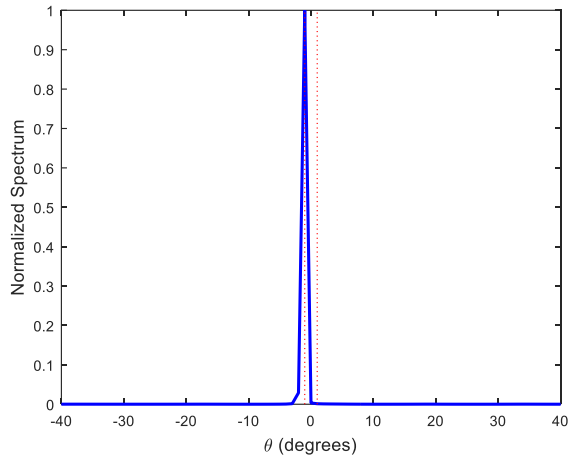

(a)

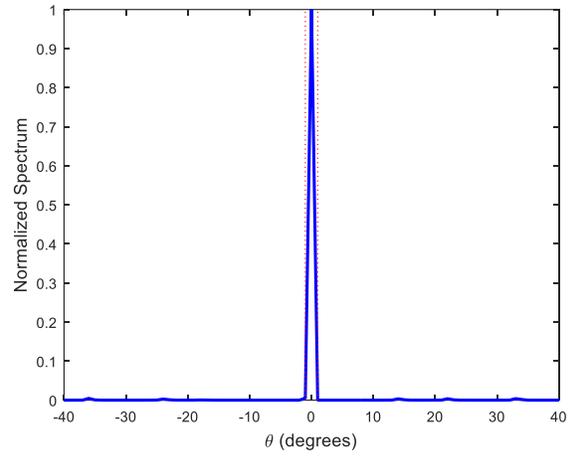

(b)

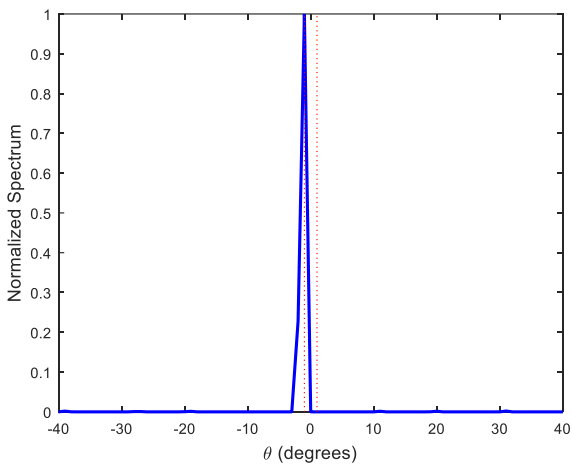

(c)

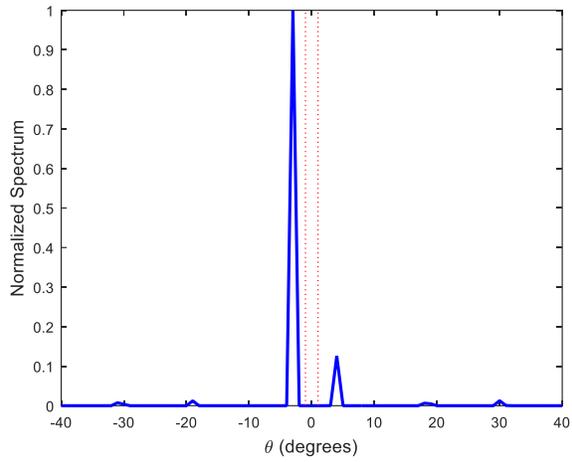

(d)

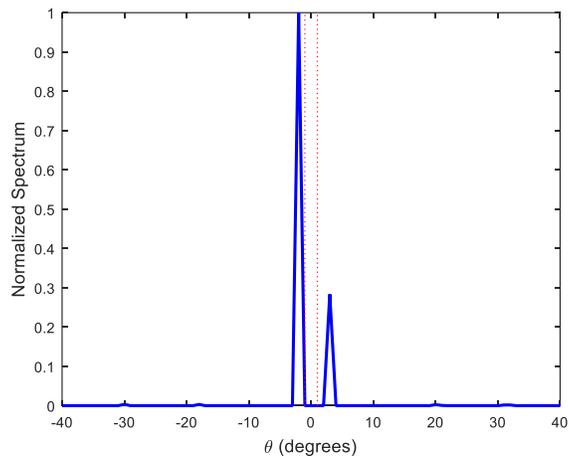

(e)

**Figure 6.9:** Co-prime array ($M = 2$, $N = 3$) with Configuration A, $D = 2$, $\Delta\theta = 2°$, (a) $\Delta$SNR = 20dB (b) $\Delta$SNR = 15dB (c) $\Delta$SNR = 10dB (d) $\Delta$SNR = 5dB (e) $\Delta$SNR = 0dB.

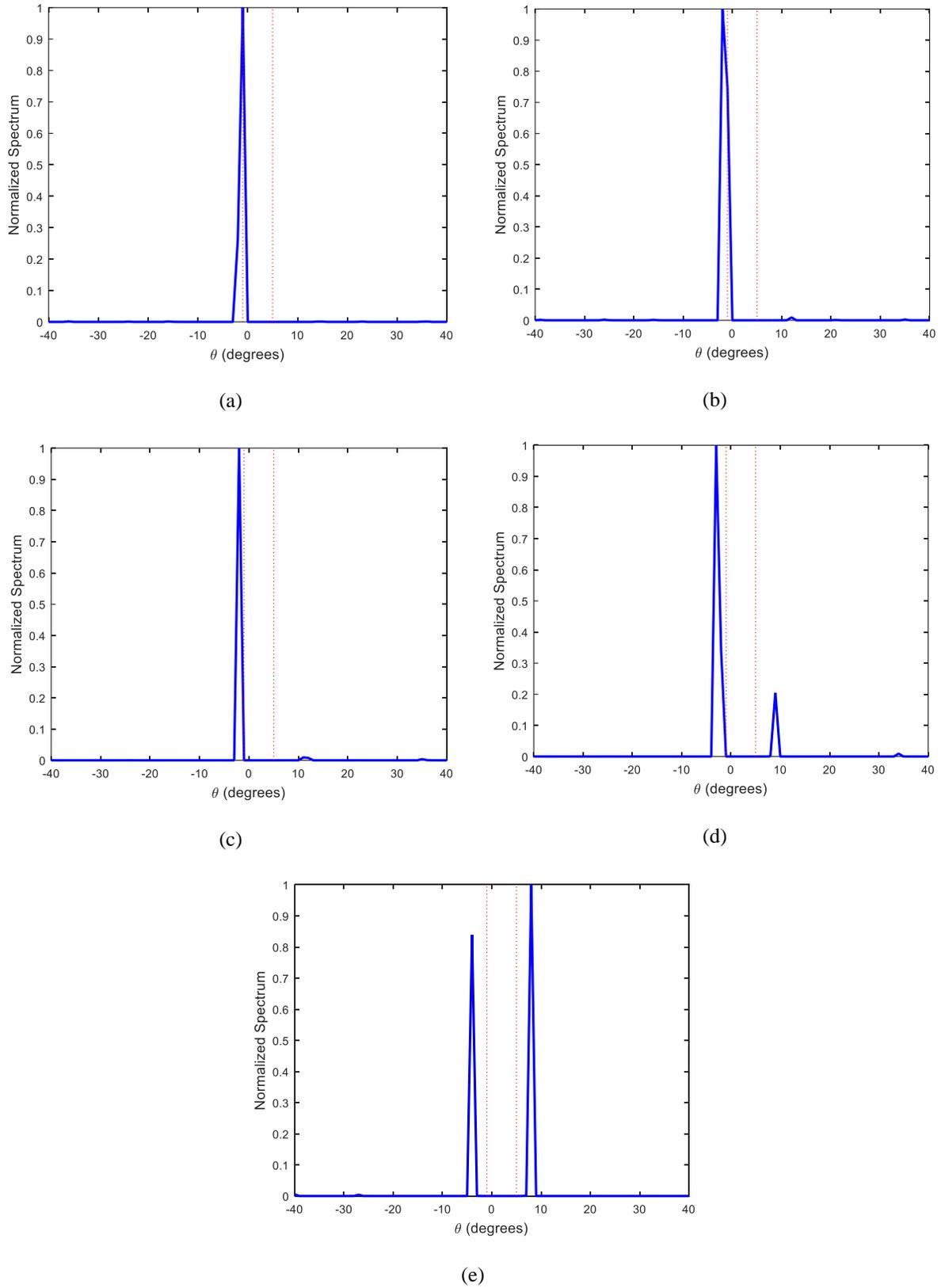

**Figure 6.10:** Co-prime array ($M = 2$, $N = 3$) with Configuration A, $D = 2$, $\Delta\theta = 6°$, (a) $\Delta$SNR = 20dB (b) $\Delta$SNR = 15dB (c) $\Delta$SNR = 10dB (d) $\Delta$SNR = 5dB (e) $\Delta$SNR = 0dB.

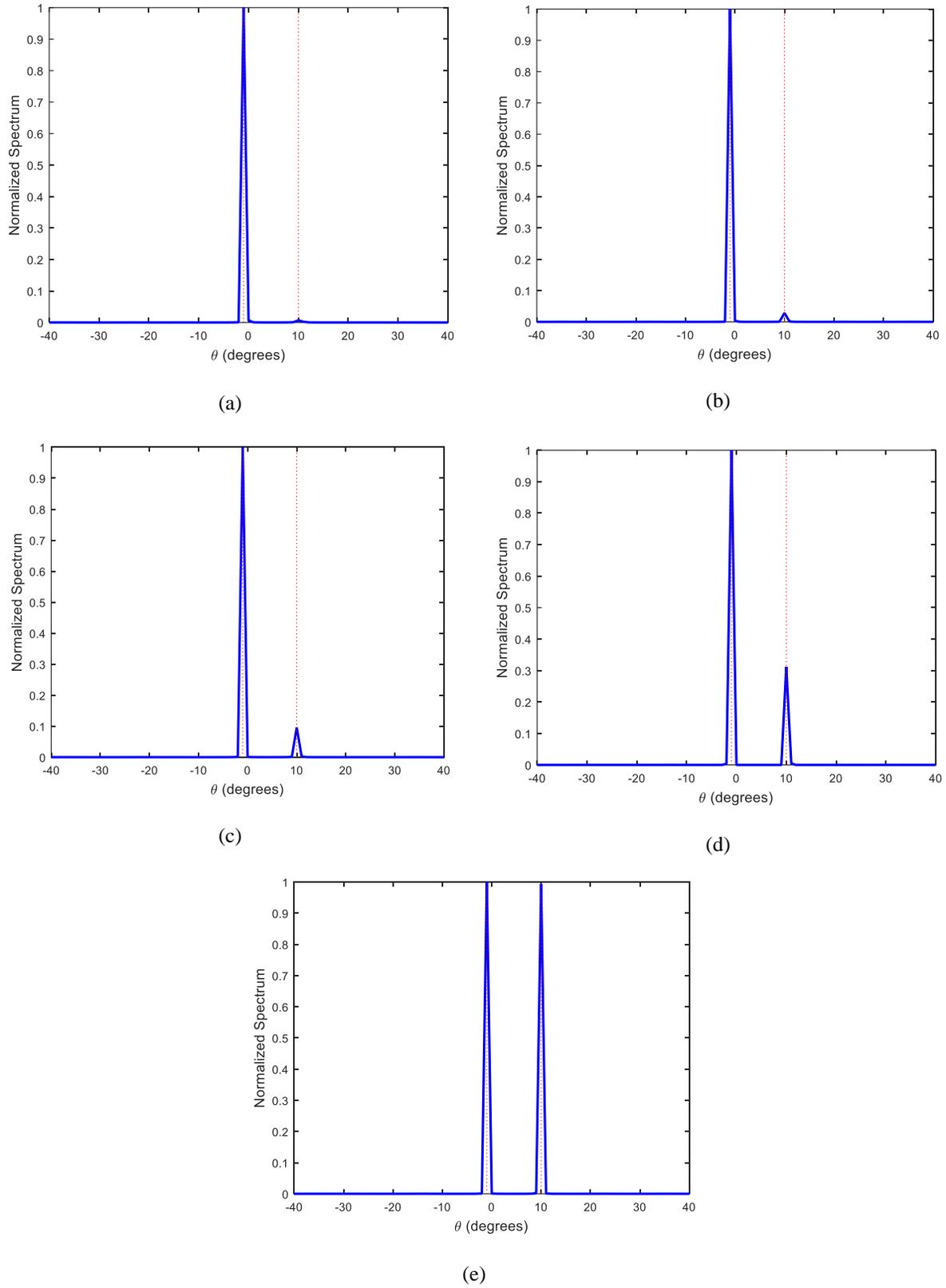

**Figure 6.11:** Co-prime array ($M = 2$, $N = 3$) with Configuration A, $D = 2$, $\Delta\theta = 11°$, (a) $\Delta$SNR = 20dB (b) $\Delta$SNR = 15dB (c) $\Delta$SNR = 10dB (d) $\Delta$SNR = 5dB (e) $\Delta$SNR = 0dB.

and $\Delta SNR = 5\ dB$; however, the estimates are biased. Third, when the source separation is increased to $11°$, the two targets are correctly estimated for all SNR levels.

Next, the same scenarios are repeated with a different co-prime array. A transmit/receive co-prime array with $M = 2$ and $N = 5$ and Configuration A is considered. The transmit array consists of three elements with positions $[5, 10, 15]d_0$ and the receive array consists of five elements with positions $[0, 2, 4, 6, 8]d_0$. These arrays along with the corresponding sum coarray and DCSC are shown in Fig. 6.12. The same simulation parameters a used, and the obtained results are grouped in Fig. 6.13, Fig. 6.14, and Fig. 6.15. It is evident that the considered array provides an improved performance over the previous array. For the case where $\Delta\theta = 2°$, the weak target is completely missed when $\Delta SNR = 20\ dB$, and the two targets are correctly estimated when they have the same power. In the remaining scenarios in Fig. 6.13, the two targets are resolved, however their estimates are biased. For the cases where $\Delta\theta = 6°$ and $\Delta\theta = 11°$ the two targets are correctly estimated for all SNR levels. Overall, the performance of the proposed method improves in three scenarios i) when a larger array is used, ii) when the source separation is increased, iii) when the dynamic range of the target powers is decreased.

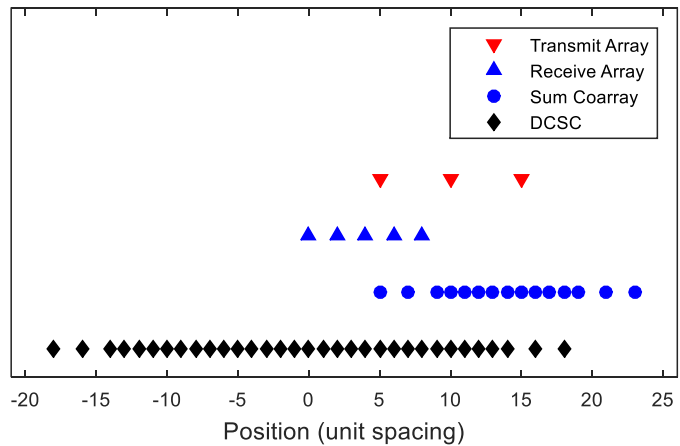

**Figure 6.12:** Co-prime array ($M = 2$, $N = 5$) with Configuration A, sum coarray, and DCSA.

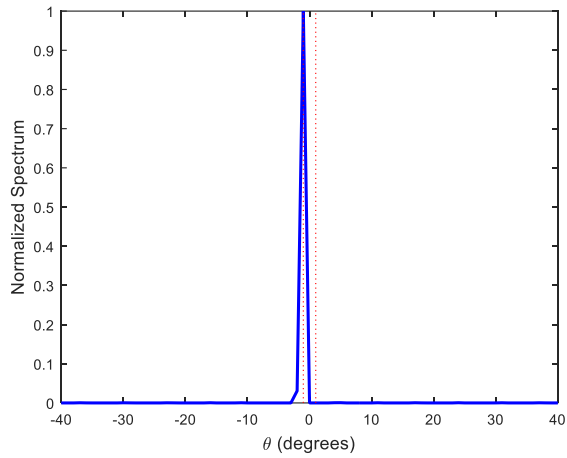

(a)

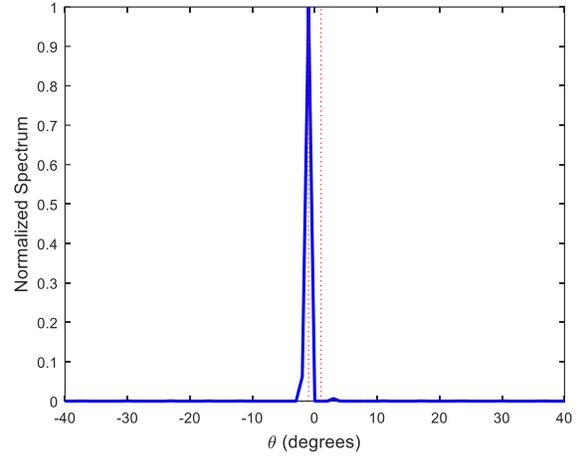

(b)

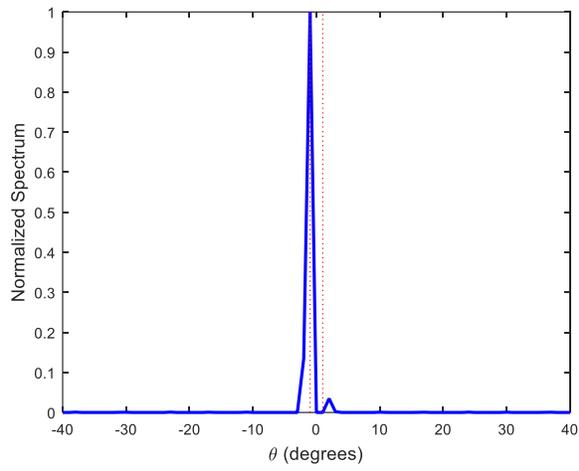

(c)

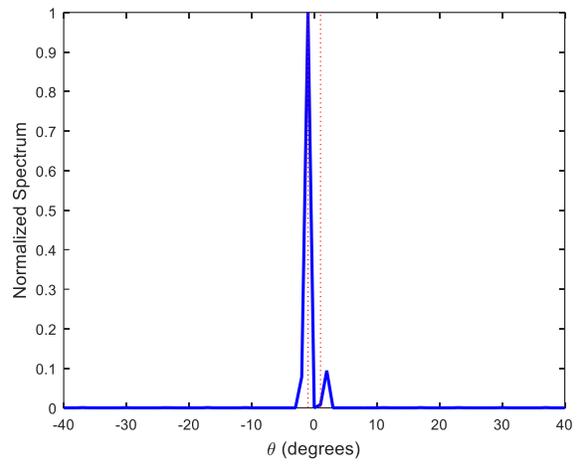

(d)

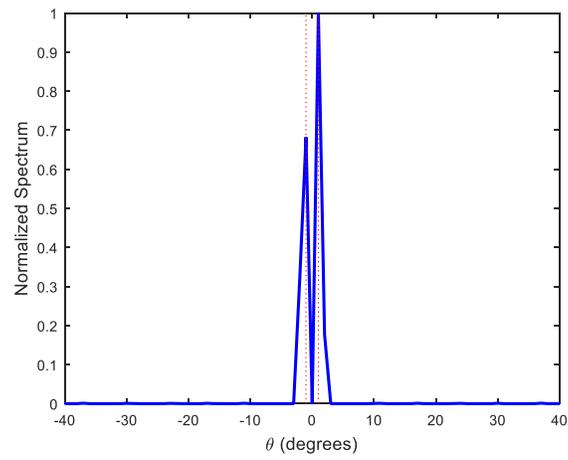

(e)

**Figure 6.13:** Co-prime array ($M = 2$, $N = 5$) with Configuration A, $D = 2$, $\Delta\theta = 2°$, (a) $\Delta$SNR = 20dB (b) $\Delta$SNR = 15dB (c) $\Delta$SNR = 10dB (d) $\Delta$SNR = 5dB (e) $\Delta$SNR = 0dB.

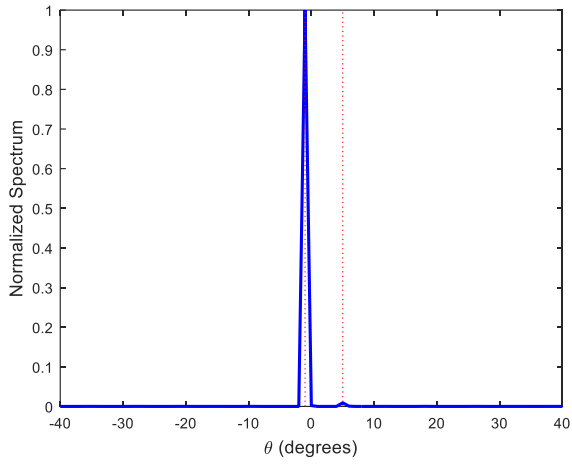

(a)

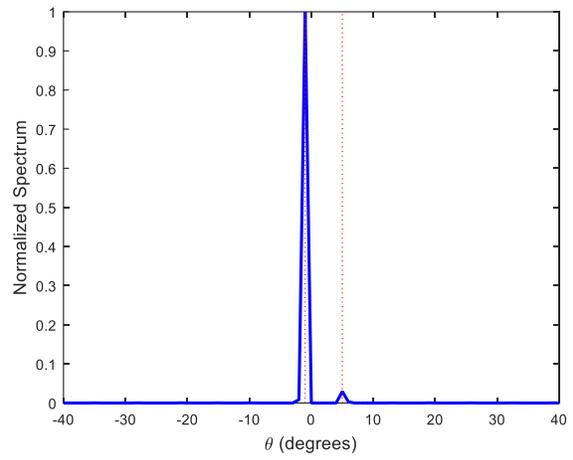

(b)

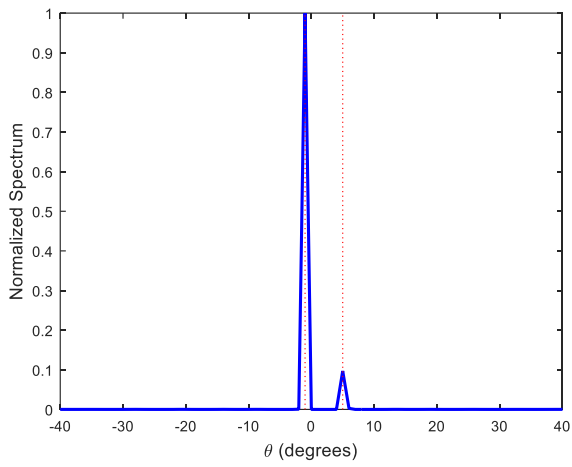

(c)

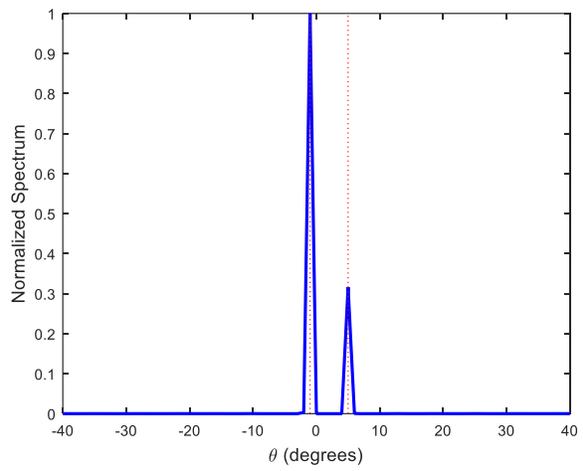

(d)

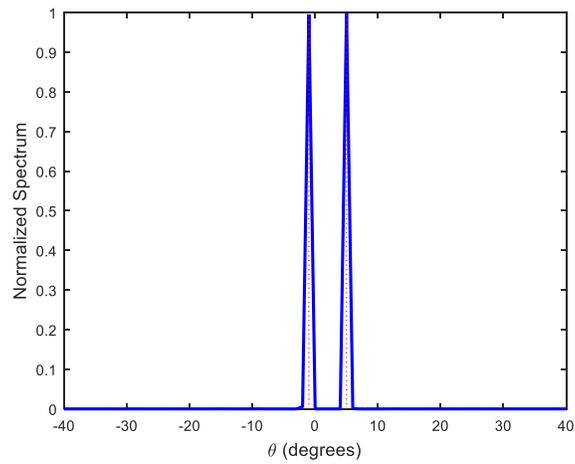

(e)

**Figure 6.14:** Co-prime array ($M = 2$, $N = 5$) with Configuration A, $D = 2$, $\Delta\theta = 6°$, (a) $\Delta$SNR = 20dB (b) $\Delta$SNR = 15dB (c) $\Delta$SNR = 10dB (d) $\Delta$SNR = 5dB (e) $\Delta$SNR = 0dB.

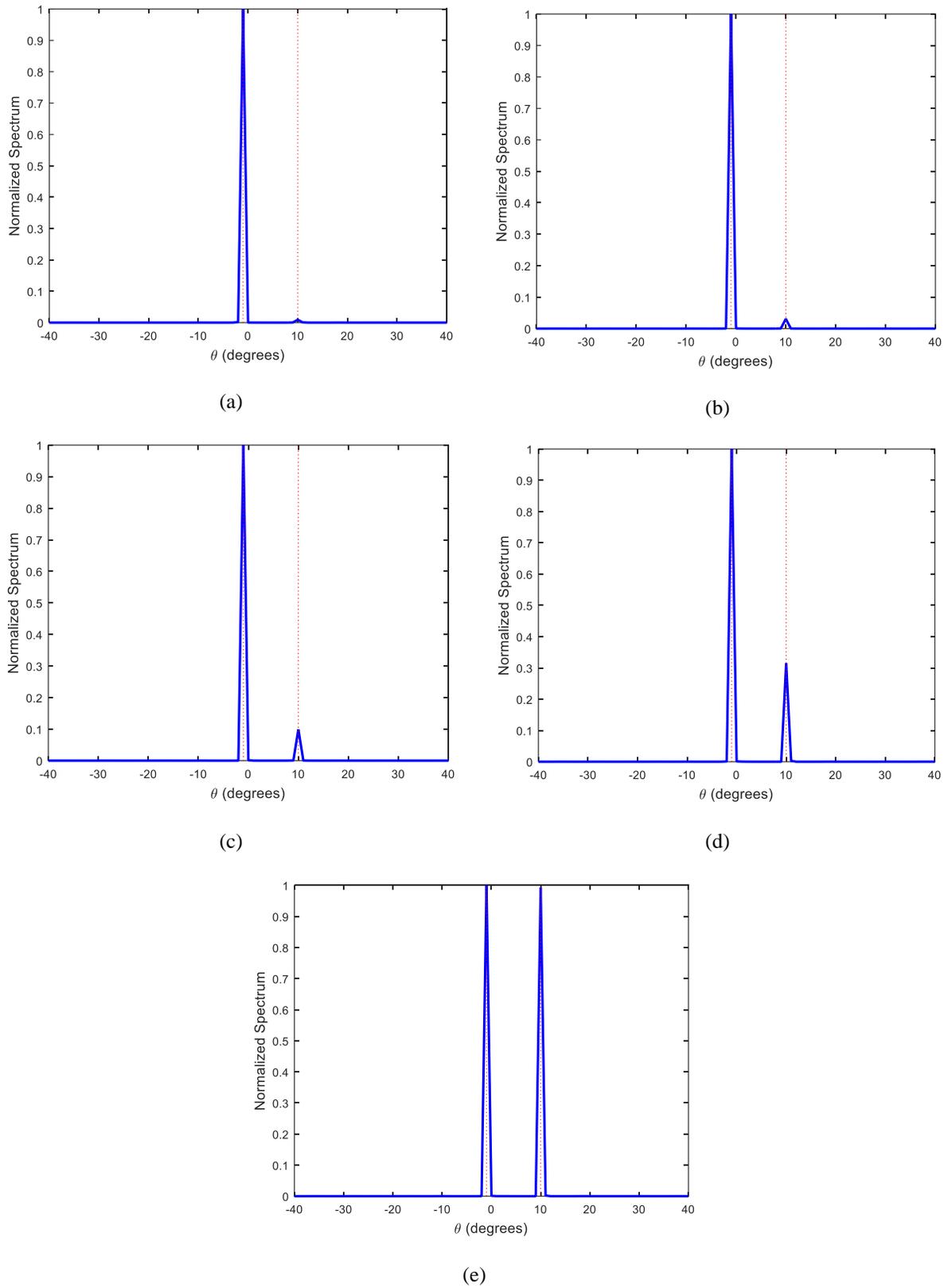

**Figure 6.15:** Co-prime array ($M = 2$, $N = 5$) with Configuration A, $D = 2$, $\Delta\theta = 11°$, (a) $\Delta$SNR = 20dB (b) $\Delta$SNR = 15dB (c) $\Delta$SNR = 10dB (d) $\Delta$SNR = 5dB (e) $\Delta$SNR = 0dB.

**6.5.    Concluding Remarks**

In this chapter, a sparsity-based method was proposed for DOA estimation using active non-uniform arrays. The proposed approach offers a significant enhancement in the DOFs over the currently employed methods by using the covariance matrix of the sum coarray measurements to emulate observations at the difference coarray of the sum coarray. The proposed method was tested using two non-uniform array configurations and was shown to successfully estimate the directions of a mixture of coherent and uncorrelated targets.

**6.5.1.  Contributions**

The main contributions in the chapter are the following.

1)  Exploiting the difference-of-sum coarray to enhance the DOFs for direction finding of a mixture of coherent and uncorrelated targets.

2)  Devising a sparsity-based reconstruction algorithm for direction finding of a mixture of coherent and uncorrelated targets.

3)  Determining a condition that must be satisfied in order to guarantee a unique solution.

4)  Examining the performance of the proposed method for different target separations and powers.

# CHAPTER VII

## CONCLUSIONS AND RECOMMENDATIONS

Antenna arrays are widely used in signal processing applications due to their multiple offerings. Non-uniform arrays provide an effective way to deal with the issue of increased hardware cost and complexity in large antenna arrays. These arrays deliver a similar performance to that of a uniform array with a reduced number of elements. Several non-uniform array configurations have been reported in the literature including minimum redundancy arrays, minimum hole arrays, nested arrays, and co-prime arrays, among many others. Each of these configurations provides certain advantages and few drawbacks over the others.

In this dissertation, the analysis and design of non-uniform arrays were investigated for direction-of-arrival estimation. Various methods were proposed to resolve the different challenges that are encountered by non-uniform arrays. The challenges include the reduction of the available degrees-of-freedom due to the presence of missing elements in the difference coarray, the mutual coupling effect in practical antenna arrays, and the presence of correlated or coherent targets in the field of view.

In Chapter 2, a brief review of the different DOA estimation techniques was provided. The various non-uniform array configurations were also defined in this chapter. In addition, the three main challenges that were treated in the dissertation were also explained.

In Chapter 3, multi-frequency operation was considered to perform DOA estimation using non-uniform arrays with increased DOFs. First, a multi-frequency technique was presented for high-resolution DOA estimation using non-uniform arrays with missing elements in their difference coarrays. A virtual augmented covariance matrix at the reference frequency was created using elements of the narrowband covariance matrices corresponding to the different employed frequencies. In the same chapter, a sparsity-based method was proposed for DOA estimation using multi-frequency non-uniform arrays. For sources with proportional spectra, all observations at the employed frequencies were combined to form a received signal vector at a larger virtual array, whose elements are given by the combination of the difference coarrays at the individual frequencies. For sources with non-proportional spectra, the common support that is shared by the observations at the employed frequencies was exploited through group sparse reconstruction. Supporting numerical examples, under both proportional and non-proportional spectra scenarios, were provided for the proposed approaches.

In Chapter 4, a sparsity-based interpolation technique was proposed to fill the missing coarray measurements and allow DOA estimation with increased degrees-of-freedom. The proposed method starts with the observations at the unique difference coarray locations and applies sparse reconstruction to interpolate the missing measurements. MUSIC with spatial smoothing is then applied to a combined measurements vector which comprises the actual measurements and the interpolated ones. The proposed method was successfully applied to co-prime arrays as well as minimum hole arrays. Numerical simulations validated the proposed method and evaluated its performance under different scenarios.

In Chapter 5, the mutual coupling effect on DOA estimation using non-uniform arrays was investigated. The impact of mutual on the estimation accuracy was compared for different array configurations and different antenna element types. In addition, two mutual coupling compensation methods which allow DOA estimation using non-uniform arrays in the presence of mutual coupling were proposed. The first method assumes imprecisely known mutual coupling and iteratively updates the mutual coupling matrix and the source directions. The second method simultaneously estimates the coupling matrix and the DOAs and is better suited to scenarios where no prior knowledge of the MCM is available. The performance of the proposed methods and their convergence were evaluated using numerical simulations.

In Chapter 6, a sparsity-based method was proposed for direction finding of a mixture of coherent and uncorrelated targets using transmit/receive non-uniform arrays. The proposed approach utilizes the vectorized covariance matrix of the sum coarray measurements which emulates observations at the difference coarray of the sum coarray. Two non-uniform array configurations, namely, co-prime arrays and MIMO MRAs, were used to test the proposed approach under different scenarios.

As for future recommendations, the various proposed techniques can be combined to tackle more than one challenge at a time. For instance, the multi-frequency approaches, presented in Chapter 3, can be combined with the mutual coupling compensation methods of Chapter 5 to allow multi-frequency DOA estimation in the presence of mutual coupling. In a similar fashion, the mutual coupling compensation methods can be combined with the sparsity-based active sensing method, presented in Chapter 6, to perform direction finding of a mixture of coherent and uncorrelated targets in practical arrays.

Other improvements can also be made to the current methods. In the sparsity-based interpolation technique of Chapter 4, only the interpolated observations at the missing elements in the coarray are kept, and the remaining ones are discarded. The discarded observations can be combined with the actual ones using a weighted average, which may result in an improved performance. The sparsity-based methods, presented throughout the dissertation, can undergo some improvements as well. First, the presented models can be modified to account for off-grid sources and allow DOA estimation with super-resolution [20, 27, 28]. Second, the choice of the regularization can be based on a modified generalized cross-validation instead of being chosen empirically [28]. Finally, the source power estimates can be improved by applying a least squares minimization after finding the directions of the sources [83].

# REFERENCES

[1]   D. H. Johnson and D. E. Dudgeon, *Array Signal Processing: Concepts and Techniques*. Englewood, NJ: Prentice Hall, 1993.

[2]   A. Moffet, "Minimum-redundancy linear arrays," *IEEE Trans. Antennas Propag.*, vol. AP-16, no. 2, pp. 172–175, Mar. 1968.

[3]   G. S. Bloom and S. W. Golomb, "Application of numbered undirected graphs," *Proc. IEEE*, vol. 65, no. 4, pp. 562–570, Apr. 1977.

[4]   P. Pal and P. P. Vaidyanathan, "Nested arrays: a novel approach to array processing with enhanced degrees of freedom," *IEEE Trans. Signal Process.*, vol. 58, no. 8, pp. 4167–4181, Aug. 2010.

[5]   P. P. Vaidyanathan and P. Pal, "Sparse sensing with co-prime samplers and arrays," *IEEE Trans. Signal Process.*, vol. 59, no. 2, pp. 573–586, 2011.

[6]   P. Pal and P. P. Vaidyanathan, "Coprime sampling and the MUSIC algorithm," in *IEEE Digital Signal Process. Workshop and IEEE Signal Process. Education Workshop*, Sedona, AZ, 2011, pp. 289–294.

[7]   C.-L. Liu and P. P. Vaidyanathan, "Cramér-Rao bounds for coprime and other sprase arrays, which find more sources than sensors," *Digital Signal Processing Special Issue on Co-prime Sampling and Arrays* (in press).

[8]   R. T. Hoctor and S. A. Kassam, "The unifying role of the coarray in aperture synthesis for coherent and incoherent imaging," *Proc. IEEE*, vol. 78, no. 4, pp. 735–752, Apr. 1990.

[9]   H. L. Van Trees, *Optimum Array Processing: Part IV of Detection, Estimation, and Modulation Theory*. New York, NY: John Wiley and Sons, 2002.

[10]  S. Chandran, *Advances in Direction-of-Arrival Estimation*. Norwood, MA: Artech House, 2006.

[11] T. E. Tuncer and B. Friedlander, *Classical and Modern Direction-of-Arrival Estimation*. Boston, MA: Academic Press (Elsevier), 2009.

[12] R. Schmidt, "Multiple emitter location and signal parameter estimation," *IEEE Trans. Antennas Propag.*, vol. 34, no. 3, pp. 276–280, Mar. 1986.

[13] S. U. Pillai, Y. Bar-Ness, and F. Haber, "A new approach to array geometry for improved spatial spectrum estimation," *Proc. IEEE*, vol. 73, pp. 1522–1524, Oct. 1985.

[14] Y. I. Abramovich, D. A. Gray, A. Y. Gorokhov, and N. K. Spencer, "Positive-definite Toeplitz completion in DOA estimation for nonuniform linear antenna arrays. I. Fully augmentable arrays," *IEEE Trans. Signal Process.*, vol. 46, pp. 2458–2471, Sep. 1998.

[15] Y. I. Abramovich, N. K. Spencer, and A. Y. Gorokhov, "Positive-definite Toeplitz completion in DOA estimation for nonuniform linear antenna arrays. II. Partially augmentable arrays," *IEEE Trans. Signal Process.*, vol. 47, pp. 1502–1521, Jun. 1999.

[16] T. J. Shan, M. Wax, and T. Kailath, "On spatial smoothing for direction-of-arrival estimation of coherent signals," *IEEE Trans. Acoust., Speech, Signal Process.*, vol. 33 no. 4, pp. 806–811, Aug. 1985.

[17] Q. Wu and Q. Liang, "Coprime sampling for nonstationary signal in radar signal processing," *EURASIP J. on Wireless Commun. Netw.*, 2013:58, 2013.

[18] K. Adhikari, J. R. Buck, and K. E. Wage, "Beamforming with extended co-prime sensor arrays," in *IEEE Int. Conf. Acoustics, Speech and Signal Process.*, Vancouver, BC, Canada, 2013, pp. 4183-4186.

[19] Y. Zhang, M. Amin, F. Ahmad, and B. Himed, "DOA estimation using a sparse uniform linear array with two CW signals of co-prime frequencies," in *IEEE 5th Int. Workshop Computational Advances in Multi-Sensor Adaptive Process.*, Saint Martin, French West Indies, France, 2013, pp. 404-407.

[20] Z. Tan and A. Nehorai, "Sparse direction of arrival estimation using co-prime arrays with off-grid targets," *IEEE Signal Process. Lett.*, vol. 21, no. 1, pp. 26–29, Jan. 2014.

[21] A. T. Pyzdek and R. L. Culver, "Processing methods for coprime arrays in complex shallow water environments," *J. Acoust. Soc. Am.*, vol. 135, no. 4, pp. 2392–2392, Apr. 2014.

[22] J. Chen and Q. Liang, "Rate distortion performance analysis of nested sampling and coprime sampling," *EURASIP J. Adv. Signal Process.*, 2014:18, 2014.

[23] J. Ramirez, J. Odom, and J. Krolik, "Exploiting array motion for augmentation of co-prime arrays," in *IEEE 8th Int. Sensor Array and Multichannel Signal Process. Workshop*, A Coruna, Spain, 2014, pp. 525–528.

[24] R. Roy and T. Kailath, "ESPRIT-Estimation of signal parameters via rotational invariance techniques," *IEEE Trans. Acoust., Speech, Signal Process.*, vol. 37, no. 7, pp. 984–995, Jul. 1989.

[25] A. J. Barabell, "Improving the resolution performance of eigenstructure-based direction-finding algorithms," in *IEEE Int. Conf. Acoustics, Speech and Signal Process.*, Boston, MA, 1983, pp. 336–339.

[26] C. El Kassis, J. Picheral, and C. Mokbel, "Advantages of nonuniform arrays using root-MUSIC," *Signal Processing*, vol. 90, no. 2, pp. 689–695, Feb. 2010.

[27] D. Malioutov, M. Cetin, and A. Willsky, "Sparse signal reconstruction perspective for source localization with sensor arrays," *IEEE Trans. Signal Process.*, vol. 53, no. 8, pp. 3010–3022, Aug. 2005.

[28] C.-Y. Hung and M. Kaveh, "Direction-finding based on the theory of super-resolution in sparse recovery algorithms," in *IEEE Int. Conf. Acoustics, Speech and Signal Process.*, Brisbane, Australia, 2015, pp. 2404–2408.

[29] Y. D. Zhang, M. G. Amin, and B. Himed, "Sparsity-based DOA estimation using co-prime arrays," in *IEEE Int. Conf. Acoustics, Speech and Signal Process.*, Vancouver, BC, Canada, 2013, pp. 3967–3971.

[30] B. Friedlander and A.J. Weiss, "Direction finding in the presence of mutual coupling," *IEEE Trans. Antennas Propag.*, vol. 39, no. 3, pp. 273–284, Mar. 1991.

[31] Q. Bao, C. C. Ko, and W. Zhi, "DOA estimation under unknown mutual coupling and multipath", *IEEE Trans. Aerosp. Electron. Syst.*, vol. 41, no. 2, pp. 565–573, Apr. 2005.

[32] J. Dai, D. Zhao, and X. Ji, "A sparse representation method for DOA estimation with unknown mutual coupling," *IEEE Antennas Wireless Propagat. Lett.*, vol. 11, pp. 1210–1213, 2012.

[33] E. BouDaher, Y. Jia, F. Ahmad, and M. Amin, "Direction-of-arrival estimation using multi-frequency co-prime arrays," in *22nd European Signal Process. Conf.*, Lisbon, Portugal, 2014, pp. 1034–1038.

[34] E. BouDaher, Y. Jia, F. Ahmad, and M. G. Amin, "Multi-frequency co-prime arrays for high-resolution direction-of-arrival estimation," *IEEE Trans. Signal Process.*, vol. 63, no. 14, pp. 3797–3808, Jul. 2015.

[35] J. L. Moulton, and S. A. Kassam, "Resolving more sources with multi-frequency coarrays in high-resolution direction-of-arrival estimation," in *43rd Annual Conference on Information Sciences and Systems*, Baltimore, MD, 2009, pp. 772–777.

[36] M. J. Hinich, "Processing spatially aliased arrays," *J. Acoust. Soc. Am.*, vol. 64, no. 3, pp. 793–795, Sep. 1978.

[37] M. G. Amin, "Sufficient conditions for aliased free direction of arrival estimation in periodic spatial spectra," *IEEE Trans. Antennas Propag.*, vol. 41, no. 4, pp. 508–511, Apr. 1993.

[38] Y. Yoon, L. M. Kaplan, and J. H. McClellan, "TOPS: New DOA estimator for wideband signals," *IEEE Trans. Signal Process.*, vol. 54, no. 6, pp. 1977–1989, Jun. 2006.

[39] H. Wang and M. Kaveh, "Coherent signal-subspace processing for the detection and estimation of angles of arrival of multiple wide-band sources," *IEEE Trans. Acoust., Speech, Signal Processing*, vol. 33, no. 4, pp. 823–831, Mar. 1985.

[40] F. Ahmad and S. A. Kassam, "Performance analysis and array design for wide-band beamformers," *J. Electron. Imaging*, vol. 7, no. 4, pp. 825–838, Oct. 1998.

[41] J. L. Moulton, "Enhanced high-resolution imaging through multiple-frequency coarray augmentation," Ph.D. dissertation, ESE Department, University of Pennsylvania, Philadelphia, PA, 2010.

[42] S. Holm, A. Austeng, K. Iranpour, and J.-F. Hopperstad, "Sparse sampling in array processing," in *Nonuniform Sampling – Theory and Practice (F. Marvasti Ed.)*, Plenum, NY: Springer, 2001, ch. 19, pp. 787–833.

[43] P. Pal and P. P. Vaidyanathan, "On application of LASSO for sparse support recovery with imperfect correlation awareness," in *Asilomar Conference on Signals, Systems and Computers*, Pacific Grove, CA, 2012, pp. 958–962.

[44] R. Tibshirani, "Regression shrinkage and selection via the Lasso," *J. R. Statist. Soc. B*, vol. 58, no. 1, pp. 267–288, 1996.

[45] J. A. Tropp and A. C. Gilbert, "Signal recovery from random measurements via orthogonal matching pursuit," *IEEE Trans. Inf. Theory*, vol. 53, no. 12, pp. 4655–4666, Dec. 2007.

[46] D. Needella and J. A. Tropp, "CoSaMP: iterative signal recovery from incomplete and inaccurate samples," *Appl. Comput. Harmon. Anal.*, vol. 26, no. 3, pp. 301–321, May 2009.

[47] V. A. Morozov, "On the solution of functional equations by the method of regularization," *Soviet Math. Dokl. 7*, pp. 414–417, 1966.

[48] M. Davenport, M. Duarte, Y. Eldar, and G. Kutyniok, *Compressed Sensing: Theory and Applications*. Cambridge, UK: Cambridge University Press, 2012.

[49] M. Yuan and Y. Lin, "Model selection and estimation in regression with grouped variables," *J. R. Statist. Soc. B*, vol. 68, no. 1, pp. 49–67, Feb. 2006.

[50] Y. Eldar, P. Kuppinger, and H. Bolcskei, "Block-sparse signals: Uncertainty relations and efficient recovery," *IEEE Trans. Signal Process.*, vol. 58, no. 6, pp. 3042–3054, Jun. 2010.

[51] C.-L. Liu, P. P. Vaidyanathan, and P. Pal, "Coprime coarray interpolation for DOA estimation via nuclear norm minimization," in *IEEE Int. Symp. Circuits and Syst.*, Montreal, Canada, 2016.

[52] J. F. Gemmeke and B. Cranen, "Using sparse representations for missing data imputation in noise robust speech recognition," *116th European Signal Processing Conference*, Lausanne, Switzerland, 2008, pp. 1–5.

[53] J. F. Gemmeke, H. Van Hamme, B. Cranen, and L. Boves, "Compressive sensing for missing data imputation in noise robust speech recognition," *IEEE J. Sel. Topics Signal Process.*, vol. 4, no. 2, pp. 272–287, Apr. 2010.

[54] Q. F. Tan, P. G. Georgiou, and S. Narayanan, "Enhanced sparse imputation techniques for a robust speech recognition front-end," *IEEE Trans. Audio, Speech, Language Process.*, vol. 19, no. 8, pp. 2418–2429, Nov. 2011.

[55] P. Shen, S. Tamura, and S. Hayamizu, "Feature reconstruction using sparse imputation for noise robust audio-visual speech recognition," in *Signal & Information Processing Association Annual Summit and Conference (APSIPA ASC)*, Hollywood, CA, 2012, pp. 1–4.

[56] L. Anitori, W. van Rossum, and A. Huizing, "Array aperture extrapolation using sparse reconstruction," in *IEEE Radar Conference (RadarCon)*, Arlington, VA, 2015, pp. 237–242.

[57] C.-C. Yeh, M.-L. Leou, and D. R. Ucci, "Bearing estimations with mutual coupling present," *IEEE Trans. Antennas Propag.*, vol. 37, no. 10, pp. 1332–1335, Oct. 1989.

[58] I. J. Gupta and A. A. Ksienski, "Effect of mutual coupling on the performance of adaptive arrays," *IEEE Trans. Antennas Propag.*, vol. 31, no. 5, pp. 785–791, Sep. 1983.

[59] H. Yamada, Y. Ogawa, and Y. Yamaguchi, "Mutual impedance of receiving array and calibration matrix for high-resolution DOA estimation," in *IEEE/ACES International Conference on Wireless Communications and Applied Computational Electromagnetics*, Honolulu, HI, 2005, pp. 361–364.

[60] H. T. Hui, "Improved compensation for the mutual coupling effect in a dipole array for direction finding," *IEEE Trans. Antennas Propag.*, vol. 51, no. 9, pp. 2498–2503, Sep. 2003.

[61] H. S. Lui and H. T. Hui, "Effective mutual coupling compensation for direction-of-arrival estimation using a new, accurate determination method for the receiving mutual impedance," *J. of Electromagn. Waves and Appl.*, vol. 24, pp. 271–281, 2010.

[62] R. S. Adve and T. K. Sarkar, "Compensation for the effects of mutual coupling on direct data domain adaptive algorithms," *IEEE Trans. Antennas Propag.*, vol. 48, no. 1, pp. 86–94, Jan. 2000.

[63] H. Singh, H. L. Sneha, and R. M. Jha, "Mutual coupling in phased arrays: A review," *International J. Antennas Propag.*, vol. 2013, Article ID 348123, 2013.

[64] C. A. Balanis, *Antenna Theory: Analysis and Design, 3rd Ed*. New York: John Wiley and Sons, 2005.

[65] J. Dai, D. Zhao, and Z. Ye, "DOA estimation and self-calibration algorithm for nonuniform linear array," in *International Symposium on Intelligent Signal Processing and Communications Systems*, Chengdu, China, 2010, pp. 1–4.

[66] FEKO Suite 6.3, EM Software & Systems – S.A. (Pty) Ltd.

[67] N. Hansen and A. Ostermeier, "Completely derandomized self-adaptation in evolutionary strategies," *Evolutionary Computation*, vol. 9, no. 2, pp. 159–195, 2001.

[68] D. E. Goldberg, *Genetic Algorithms in Search, Optimization and Machine Learning*. Reading, MA: Addison-Wesley, 1989.

[69] J. Kennedy and R. Eberhart, "Particle swarm optimization," in *IEEE International Conference on Neural Networks*, Perth, WA, 1995, pp. 1942–1948.

[70] G. B. Fogel and D. B. Fogel, "Continuous evolutionary programming: Analysis and experiments," *Cybern. Syst.*, vol. 26, no. 1, pp. 79–90, 1995.

[71] H. P. Schwefel, *Evolution and Optimum Seeking*, *Sixth Generation Computer Technology Series*. New York: Wiley, 1995.

[72] E. BouDaher and A. Hoorfar, "Electromagnetic optimization using mixed-parameter and multiobjective covariance matrix adaptation evolution strategy," *IEEE Trans. Antennas Propag.*, vol. 63, no. 4, pp. 1712–1724, Apr. 2015.

[73] E. BouDaher, F. Ahmad, M. Amin, and A. Hoorfar, "DOA estimation with co-prime arrays in the presence of mutual coupling," in *23^{rd} European Signal Processing Conference*, Nice, France, 2015, pp. 2830–2834.

[74] N. Hansen, "A CMA-ES for mixed integer nonlinear optimization," *Institut National de Recherche en Informatique et en Automatique*, no. 7751, 2011.

[75] M. M. Hayder and K. Mahata, "Direction-of-arrival estimation using a mixed $\ell_{2,0}$ norm approximation," *IEEE Trans. Signal Process.*, vol. 58 no. 9, pp. 4646–4655, Sep. 2010.

[76] D. D. Ariananda and G. Leus, "Direction of arrival estimation for more correlated sources than active sensors," *Signal Process.*, vol. 93, no. 12, pp. 3435–3448, Dec. 2013.

[77] R. T. Hoctor and S. A. Kassam, "High resolution coherent source location using transmit/receive arrays," *IEEE Trans. Image Process.*, vol. 1 no. 1, pp. 88–100, Jan. 1992.

[78] X. Wang, W. Wang, J. Liu, X. Li, and J. Wang, "A sparse representation scheme for angle estimation in monostatic MIMO radar," *Signal Process.*, vol. 104, pp. 258–263, Nov. 2014.

[79] C.-C. Weng and P.P. Vaidyanathan, "Nonuniform sparse array design for active sensing," in *Asilomar Conf. on Signals, Systems, and Computers*, Pacific Grove, CA, 2011, pp. 1062–1066.

[80] X. Zhang and D. Xu, "Low-complexity ESPRIT-based DOA estimation for colocated MIMO radar using reduced-dimension transformation," *Electr. Lett.*, vol. 47 no. 4, pp. 283–284, Feb. 2011.

[81] X. Zhang, Y. Huang, C. Chen, J. Li, and D. Xu, "Reduced-complexity Capon for direction of arrival estimation in a monostatic multiple-input multiple-output radar," *IET Radar Sonar Navig.*, vol. 8 no. 5, pp. 796–801, Oct. 2012.

[82] A. J. Laub, *Matrix Analysis for Scientists and Engineers*. Philadelphia, PA: SIAM, 2005.

[83] Y. Gu, N. A. Goodman, S. Hong, and Y. Li, "Robust adaptive beamforming based on interference matrix sparse reconstruction," *Signal Process.*, vol. 96, pp. 375–381, Mar. 2014.